\documentclass[5p]{elsarticle}
\usepackage{amssymb}
\usepackage{lineno}
\usepackage{color}
\usepackage{epsfig,latexsym}
\usepackage{float}
\usepackage{indentfirst}
\usepackage{amsmath}
\usepackage{amssymb}
\usepackage{times}
\usepackage{subfigure}
\usepackage{psfrag}
\usepackage{color}
\usepackage{algorithm}
\usepackage{setspace}
\usepackage{mathrsfs}
\usepackage{stfloats}
\usepackage{graphicx}
\usepackage{multirow}
\usepackage{array}
\usepackage{makecell}

\usepackage[colorlinks=true,linkcolor=black,citecolor=blue,urlcolor=blue,]{hyperref}

\usepackage[doipre={doi:~}]{uri}

\journal{Computer Communications}

\begin{document}
\begin{sloppypar}
\begin{frontmatter}
\title{\huge{A Comprehensive Survey on Aerial Mobile Edge Computing: Challenges, State-of-the-Art, and Future Directions}}

\author[address1]{Zhengyu Song}
\ead{songzy@bjtu.edu.cn}

\author[address1]{Xintong Qin}
\ead{20111046@bjtu.edu.cn}

\author[address2,address3]{Yuanyuan Hao}
\ead{yhao106@163.com}

\author[address1]{Tianwei Hou \corref{cor1}}
\ead{twhou@bjtu.edu.cn}

\cortext[cor1]{Corresponding author}

\author[address1]{Jun Wang}
\ead{wangjun1@bjtu.edu.cn}

\author[address1]{Xin Sun}
\ead{xsun@bjtu.edu.cn}

\address[address1]{School of Electronic and Information Engineering, Beijing Jiaotong University, Beijing 100044, China}
\address[address2]{Institute of Telecommunication and Navigation Satellites, China Academy of Space Technology, Beijing 100094, China}
\address[address3]{Innovation Center of Satellite Communication System, China National Space Administration, Beijing 100094, China}

\begin{abstract}
Driven by the visions of Internet of Things (IoT), there is an ever-increasing demand for computation resources of IoT users to support diverse applications. Mobile edge computing (MEC) has been deemed a promising solution to settle the conflict between the resource-hungry mobile applications and the resource-constrained IoT users. On the other hand, in order to provide ubiquitous and reliable connectivity in wireless networks, unmanned aerial vehicles (UAVs) can be leveraged as efficient aerial platforms by exploiting their inherent attributes, such as the on-demand deployment, high cruising altitude, and controllable maneuverability in three-dimensional (3D) space. Thus, the UAV-enabled aerial MEC is believed as a win-win solution to facilitate cost-effective and energy-saving communication and computation services in various environments. In this paper, we provide a comprehensive survey on the UAV-enabled aerial MEC. Firstly, the related advantages and research challenges for aerial MEC are discussed. Then, we provide a comprehensive review of the recent research advances, which is categorized by different domains, including the joint optimization of UAV trajectory, computation offloading and resource allocation, UAV deployment, task scheduling and load balancing, interplay between aerial MEC and other technologies, as well as the machine-learning (ML)-driven optimization. Finally, some important research directions deserved more efforts in future work are summarized.
\end{abstract}


\end{frontmatter}

\section{Introduction}
In recent years, the rapid proliferation of Internet of Things (IoT) has triggered the ever-increasing demand for various applications with diverse quality-of-service (QoS) requirements. Most of these applications, such as image/video processing, real-time online gaming, virtual/augmented reality, etc., are computation-intensive and latency-sensitive. However, the resource-constrained IoT users may not be able to execute these applications with devices that have limited energy and computation resources. Thus, how to tackle the conflict between the resource-hungry mobile applications and the resource-constrained users is an unprecedented challenge for IoT \cite{HGuo2018, CZhan2021, WFeng2021}. In this context, mobile edge computing (MEC) has been deemed a promising solution to address the above challenge. The MEC servers, deployed at the edge of wireless networks, can provide powerful computing services for IoT users with low transmission and execution latency. By offloading partial or all computation tasks to MEC servers, the IoT users can reduce their task execution latency and energy consumption, which significantly extends the capability to support various computation-intensive and latency-sensitive applications \cite{PWang2019}.

However, the traditional terrestrial infrastructure-based IoT and MEC networks are not applicable in remote regions or natural disaster areas, since deploying network facilities in these harsh environments is usually cost-inefficient or even infeasible. Fortunately, a new paradigm of UAV-enabled aerial MEC has been proposed recently and received increasing research attentions from both industry and academia \cite{FZhou2020}. Thanks to the inherent attributes of UAVs, including the on-demand deployment, low cost, controllable maneuverability, high cruising altitude, and line-of-sight (LoS) connectivity, etc., UAVs acting as aerial MEC platforms can be employed in a wide range of applications varying from civilians to the military for critical operations \cite{ZUllah2020}. Typically, the aerial MEC can work as a complement to the terrestrial MEC networks when the servers embedded in ground base stations (GBSs) are overloaded or unavailable \cite{SJoo2021, YXu2021-2}. In particular, the LoS connectivity and maneuverability of UAVs is able to significantly reduce the task offloading latency and energy consumption for MEC systems. As such, the aerial MEC integrating UAV communications and MEC is believed as a win-win technology for next generation wireless networks, which will play an indispensable role in providing flexible and ubiquitous communication and computing supports in diverse environments \cite{WZhang2020}.

Compared with conventional terrestrial infrastructure-based MEC systems, the aerial MEC has exhibited prominent advantages, which mainly stem from the unique features of UAVs. The representative advantages of aerial MEC can be summarized as follows:

\begin{itemize}
\item \textbf{Cost-effective and on-demand deployment:} Thanks to the high availability and easy deployment features of UAVs, the aerial MEC systems can be rapidly deployed with low cost according to the real-time demands and provide computation offloading opportunities to users with limited local computing capacities, especially in the areas where the network facilities are sparsely distributed or even fully destroyed.

\item \textbf{Coverage and computation capacity enhancement:} As UAVs usually have a higher cruising altitude, a large area can be effectively covered with a relatively small number of UAVs. More importantly, a swarm of UAVs formed a flying \emph{ad hoc} network (FANET) can collaboratively perform tasks in large-scale areas with the aid of inter-UAV links. As additional computation resource providers, the MEC servers carried by UAVs can also greatly enhance the computation capacity in hot-spot areas, such that more users can be accommodated with high-quality computing services.

  \item \textbf{Reliable LoS offloading links:} In addition to the coverage enhancement, another benefit of the higher cruising altitudes of UAVs in aerial MEC is the high possibility of LoS links. Compared with terrestrial fading channels, the LoS links can provide more reliable wireless connectivity for task offloading and computation result downloading, thus satisfying the stringent QoS requirements of MEC.

\item \textbf{Energy consumption and latency reduction:} In comparison to infrastructure-based MEC systems, the controllable maneuverability of UAVs introduces an additional design degree of freedom (DoF) for aerial MEC. By UAV trajectory optimization, better channel conditions can be experienced. Then, supplemented by appropriate resource allocation strategies, both the offloading energy consumption and task latency for aerial MEC can be significantly reduced.

\end{itemize}

Due to the aforementioned attractive characteristics, numerous research efforts have been dedicated to reap the benefits of UAV-enabled aerial MEC. As UAVs are imposed by stringent size, weight, and power (SWaP) constraints, their operational altitudes, coverage, computation capacities, and endurance are diversified. Nevertheless, since different types of UAVs mostly share similar characteristics from the perspectives of communication and computing, the aerial MEC can be investigated in a unified manner. For each specific application scenarios, the performance optimization while taking into account various constraints is indispensable. The key optimization considerations include trajectory design, resource allocation, optimal UAV deployment, cooperative aerial computing mechanisms, and so forth. In addition, there are also some research focusing on the interplay between aerial MEC and other advanced technologies like wireless power transfer (WPT), physical-layer security (PLS), and reconfigurable intelligent surface (RIS), etc., to further enhance the performances. Furthermore, with intensive efforts to optimize the aerial MEC by traditional optimization tools such as convex optimization and game theory, machine-learning (ML)-driven optimization has also been widely applied in aerial MEC, which exhibits huge potentials in solving complex control and resource allocation problems in highly dynamic aerial MEC environments \cite{FJiang2021}. All these efforts have effectively promoted the applications of UAV-enabled aerial MEC in critical areas.

\begin{table}[t]
\footnotesize
\caption{List of Abbreviations}
\centering
\begin{tabular}{|m{53pt}|m{170pt}|}
\hline
\bfseries Abbreviations &\bfseries {\centering}Full Name \\
\hline
ABS & Aerial Base Station \\
\hline
ADMM & Alternating Direction Method of Multipliers \\
\hline
AI & Artificial Intelligence \\
\hline
ANN & Artificial Neural Network \\
\hline
B\&B & Branch and Bound \\
\hline
BCD & Block Coordinate Descent \\
\hline
CTDE & Centralized Training and Decentralized Execution \\
\hline
DC & Difference of Convex \\
\hline
DDPG & Deep Deterministic Policy-Gradient \\
\hline
DNN & Deep Neural Network \\
\hline
DRL & Deep Reinforcement Learning \\
\hline
DVFS & Dynamic Voltage and Frequency Scaling \\
\hline
FANET & Flying \emph{ad hoc} Network \\
\hline
GBS & Ground Base Station \\
\hline
IoT & Internet of Things \\
\hline
IRS & Intelligent Reflecting Surface \\
\hline
LEO  & Low Earth Orbit \\
\hline
LoS  & Line-of-Sight \\
\hline
MADDPG  & Multi-Agent DDPG \\
\hline
MBS  & Macro Base Station \\
\hline
MDP  & Markov Decision Process \\
\hline
MEC  & Mobile Edge Computing \\
\hline
ML  & Machine Learning \\
\hline
NLoS  & Non-Line-of-Sight \\
\hline
NOMA  & Non-Orthogonal Multiple Access \\
\hline
OMA  & Orthogonal Multiple Access \\
\hline
PLS  & Physical-Layer Security \\
\hline
RL  & Reinforcement Learning \\
\hline
SADDPG  & Single-Agent DDPG \\
\hline
SCA  & Sequential Convex Approximation \\
\hline
SDN  & Software Defined Networking \\
\hline
SWaP  & Size, Weight, and Power \\
\hline
UAV  & Unmanned Aerial Vehicle \\
\hline
WPT  & Wireless Power Transfer \\
\hline
\end{tabular}
\end{table}

\subsection{Existing Surveys and Tutorials}
In recent years, a number of surveys and tutorials related to UAV communications have been published. For example, in \cite{BLi2019}, an exhaustive survey of various fifth-generation (5G) and beyond 5G (B5G) techniques based on UAV platforms was presented, which were categorized into three domains, i.e., physical layer, network layer, joint communication, computing and caching. From the perspective of game theory, M. E. Mkiramweni \emph{et al.} \cite{MEMkiramweni2019} surveyed the recent progress of modelling and analyzing the UAV-aided communication networks. Meanwhile, advanced distributed interference management schemes for large-scale UAV-assisted networks were introduced. In addition, Y. Zeng \emph{et al.} \cite{YZeng2019} provided a tutorial overview of the recent advances in the UAV-assisted communications and cellular-connected UAVs, where UAVs were integrated into the network and played as new aerial communication platforms and users, respectively. Similarly, M. Mozaffari \emph{et al.} \cite{MMozaffari2019} presented a comprehensive tutorial on the potential benefits and applications of UAVs in wireless communications. The fundamental tradeoffs as well as various analytical frameworks and mathematical tools for UAV-enabled communication networks were thoroughly investigated along with representative results. More recently, Q. Wu \emph{et al.} \cite{QWu2021} gave an overview of the latest research efforts on the integration of UAVs with cellular networks, where some advanced techniques, such as the RIS, short packet transmission, joint communication and radar sensing, as well as edge intelligence, were exploited to satisfy the diverse service requirements of next-generation wireless networks.

On the other hand, there have also been some representative surveys related to MEC systems. In particular, \cite{YMao2017} provided a survey focusing on the joint radio-and-computational resource management in MEC systems, including the cases of single user, multiple users and heterogeneous MEC servers.
T. Taleb \emph{et al.} \cite{TTaleb2017} introduced a survey on MEC orchestration, reference architecture and main deployment scenarios. Another survey by P. Porambage \emph{et al.} \cite{PPorambage2018} concentrated on the exploitation of MEC for IoT realization and their synergies, which greatly broadened the horizons of potential inter-dependencies of MEC and IoT technologies.
Besides, N. Abbas \emph{et al.} \cite{NAbbas2018} presented a comprehensive survey
on the definition, advantages, architectures, and potential applications of MEC. They also discussed the security and privacy issues and existing solutions.  In \cite{JMoura2019}, J. Moura \emph{et al.} concentrated on the challenges imposed by MEC over the limited wireless resources and surveyed the recent MEC-related research efforts from the viewpoint of game theory, where both classical and evolutionary games were presented. In a more recent survey, P. Ranaweera \emph{et al.} \cite{PRanaweera2021} addressed the security and privacy as a key aspect in the realization of MEC deployments, where they introduced a thorough investigation of the threat vectors and potential solutions.

In spite of the existing surveys and tutorials solely related to UAV communications or MEC systems, there is a lack of comprehensive survey focusing on the integration of UAV communications and MEC systems. To our best knowledge, W. Zhang \emph{et al.} \cite{WZhang2020} reviewed the applications, challenges, and state-of-the-art research on the air-ground integrated mobile edge networks from the aspects of UAV-assisted communications, computing and caching. Nevertheless, the UAV-assisted edge computing was not the main focus, where the security and privacy issues, and the recently emerging ML-driven optimization were not surveyed comprehensively. Besides, a magazine paper \cite{FZhou2020} presented a survey for the current research of UAV-enabled MEC networks. Although some of the recent advances and implementation issues were summarized, the related literature was not reviewed in an exhaustive way, especially for the most recent progress on the UAV deployment, applications of ML, as well as the security protection. Since the existing studies have not been systematically reviewed, in this paper, we attempt to bridge this gap by presenting an in-depth survey of aerial MEC with a vision towards a comprehensive computing infrastructure in future wireless networks.

\subsection{Paper Contributions and Organization}
As discussed above, while the UAV-enabled aerial MEC is expected as a promising solution to provide ubiquitous and reliable MEC services in the 5G-and-beyond networks, the successful realization is still in its infancy and demands for constant efforts from both academic and industry communities. Therefore, it is of great necessity to review the current studies on UAV-enabled aerial MEC. Motivated by this vision, we aim to present a comprehensive survey on the recent research advances in this domain, which are categorized by different topics, including the joint optimization of computation offloading, resource allocation and trajectory design, UAV deployment, task scheduling and load balancing, interplay between aerial MEC and other technologies, as well as the ML-driven optimization for aerial MEC. Moreover, we also provide an enlightening guidance for future research directions.

The rest of this survey is organized as follows. In Sec. II, we provide an overview of aerial MEC. In Sec. III, the joint optimization of UAV trajectory, computation offloading, and resource allocation is surveyed. Then, the UAV deployment and load balancing issues are presented in Sec. IV. In Sec. V, the interplay between aerial MEC and other technologies is introduced. The state-of-the-art studies dedicated to ML-driven optimization are reviewed in Sec. VI. Finally, a range of open problems for future researches in UAV-enabled aerial MEC are summarized in Sec. VII, followed by the conclusions in Sec. VIII. The organization of this survey is shown in Fig. 1. For facilitating reading, we give a list of abbreviations in Table 1.

\begin{figure}[!t]
\centering
\includegraphics[width=3.6in]{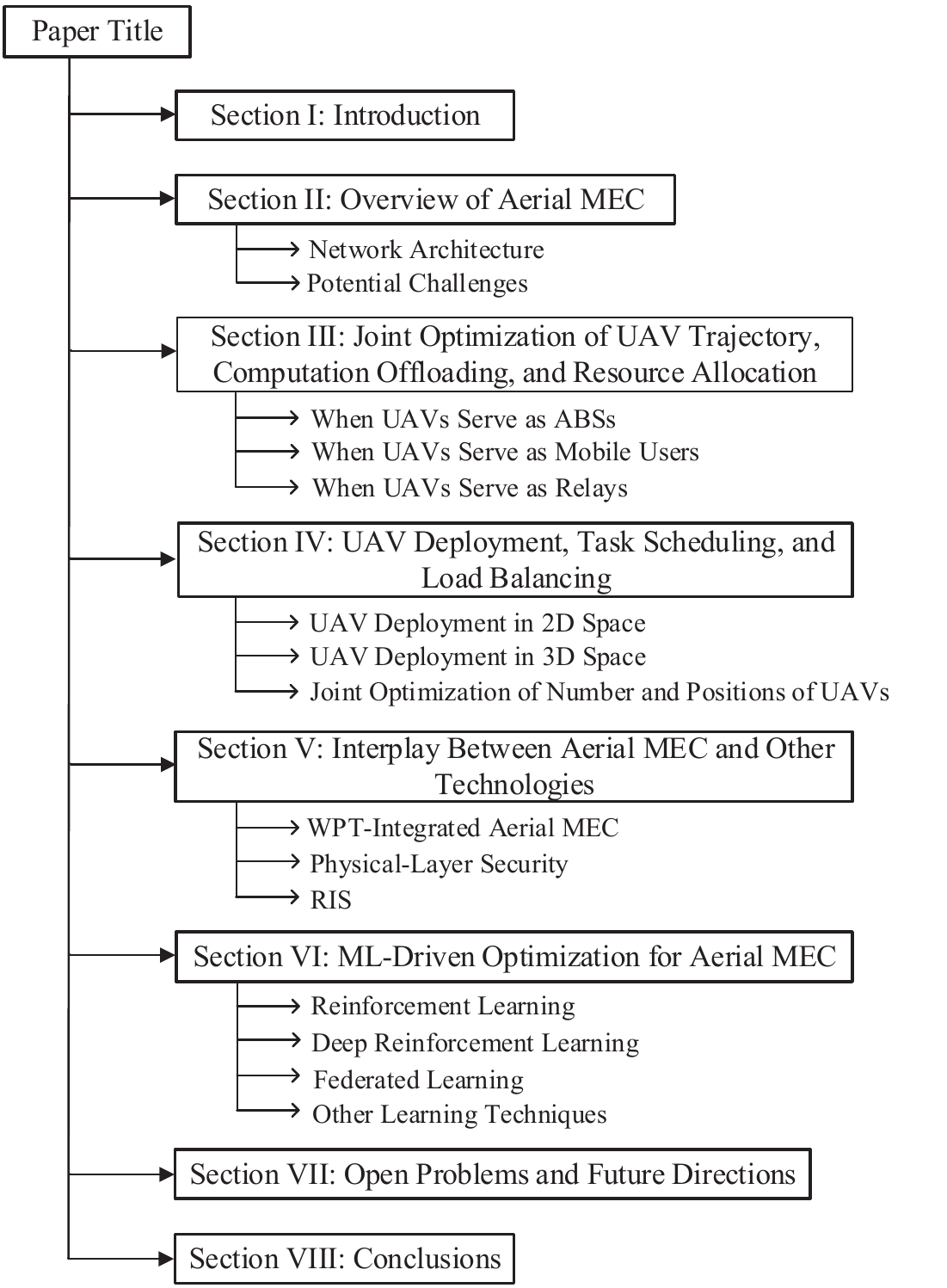}
\caption{The organization of this survey.}
\vspace{-0.1in}
\end{figure}

\section{Overview of Aerial MEC}
\subsection{Network Architecture}
Depending on the size, weight, wing configuration, flying duration and altitude, UAVs can be classified into different categories, such as large \emph{vs.} small/mini UAVs, fixed-wing \emph{vs.} rotary-wing UAVs, high altitude platforms (HAPs) \emph{vs.} low altitude platforms (LAPs), etc. \cite{MHassanalian2017, AFotouhi2019}. For aerial MEC, different types of UAVs can be deployed according to the diversified application scenarios. For example, the fixed-wing UAVs such as small aircrafts have higher flying speed, longer traveling distances, and can carry more payloads than rotary-wing UAVs, but they have to continuously move forward to remain aloft. Thus, the fixed-wing UAVs usually continuously fly forward and are used to cover a large area to provide computing services. In contrast, the rotary-wing UAVs such as quadrotor drones can remain stationary over a specific location. Besides, although the payloads of rotary-wing UAVs are relatively small due to their limited size and weight, the major advantages lie in that they are able to take off and land vertically without a runway or launcher, which make the deployment more rapidly and flexibly. HAPs usually have altitudes above 17km and are designed for long term operations (days or months) and large geographic areas \cite{YZeng2016}. For the LAPs, the advantage lies in the low-cost and rapid deployment or replacement, while their computing capacities are limited.

\begin{figure}[t]
\centering
\includegraphics[width=3.5in]{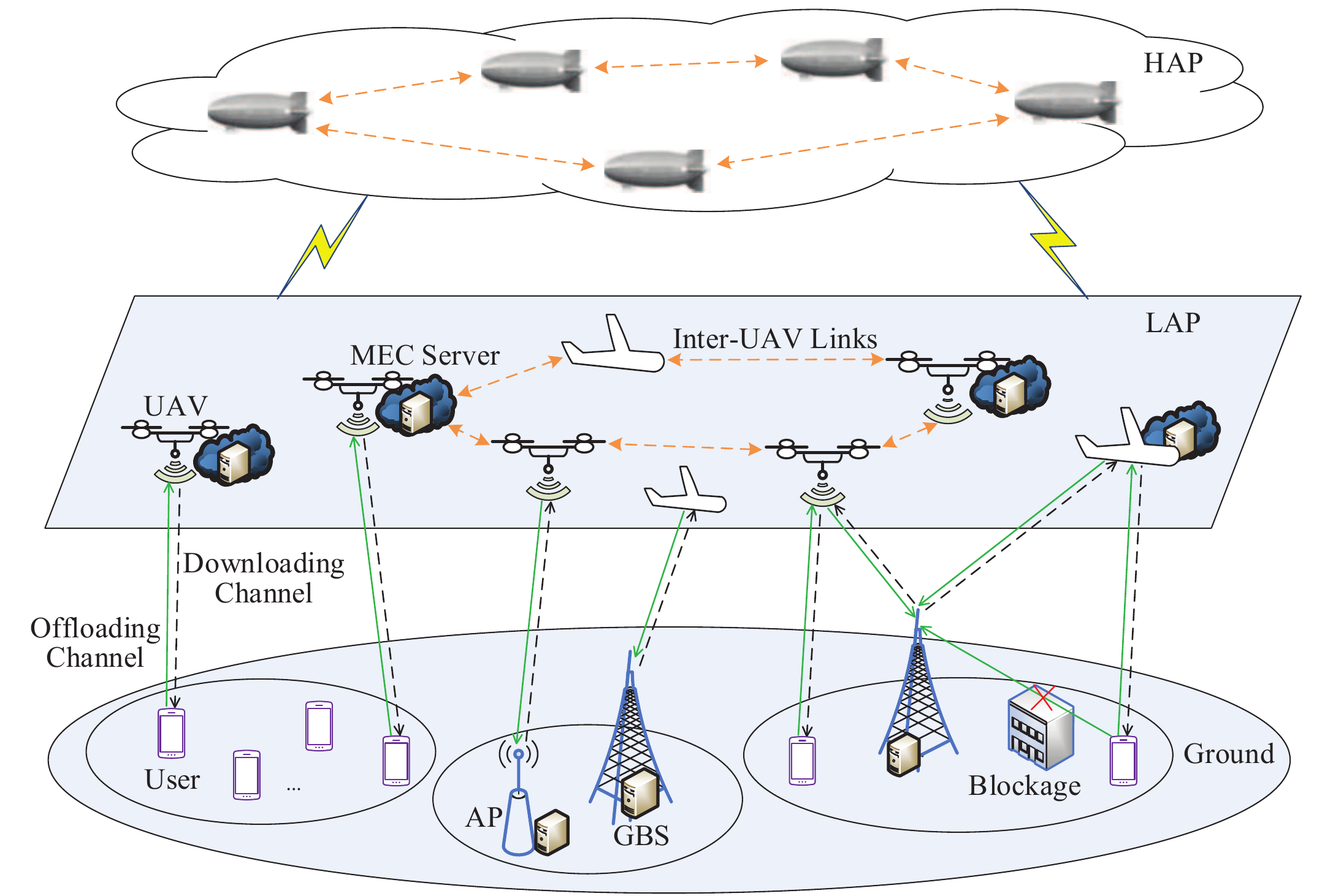}
\caption{Network architecture for aerial MEC.}
\end{figure}

An integrated network architecture for aerial MEC is shown in Fig. 2. As aforementioned, in the areas where the network facilities are sparsely distributed or even fully destroyed, UAVs can play as aerial base stations (ABSs) equipped with MEC servers to provide computing services to mobile users. Besides, as the infrastructure-based MEC deployed within a predefined region may not flexibly adapt to the time-varying distributions of service requirements, UAVs can be rapidly deployed to designated areas to meet the temporary or unexpected demands. On the other hand, UAVs can also act as relays to assist task offloading from users to more powerful remote MEC servers with two or more hops. Since UAVs can successively adjust their positions for experiencing good channel conditions, the UAV relays provide new opportunities for performance enhancement compared with conventional static relaying \cite{MHua2018, DZhai2021}. Due to the limited computation capacity of a single UAV, multiple UAVs can cooperate to enlarge the coverage area and increase the computation capacities, where the cooperative computing mechanisms require elaborate design. Furthermore, a swarm of UAVs can be constructed as a FANET via the inter-UAV links, where the task bits generated by small UAVs can be offloaded to a head UAV with rich computing resource for real-time processing, and the offloading tasks from ground users can be reallocated among multiple UAVs within the FANET. The LAPs can also be integrated with HAPs and terrestrial infrastructures to enable a comprehensive information network. Since HAPs have a global view for the whole system, they are responsible for providing global coverage and computing services, while the LAPs play as additional computation resources to supplement the terrestrial MEC networks and guarantee the stringent QoS requirements.

\subsection{Potential Challenges}
Incurred by the inherent dynamism and complexity of UAV communications and MEC systems, the design and optimization for aerial MEC are becoming more complicated. Consequently, despite the advantages of aerial MEC compared to terrestrial MEC systems, there are also some urgent technical challenges to be addressed, which are listed as follows.

\begin{itemize}
\item \textbf{Real-time trajectory design:} As one of the key features of UAVs, the controllable maneuverability brings new design DoF for aerial MEC. By the real-time trajectory design, better channel conditions for task offloading and result downloading can be experienced. Nevertheless, the onboard energy of an UAV is usually limited, while a significant proportion of energy is consumed by the flying. Besides, trajectory design is subject to the UAVs' wing configuration, maximum speed and acceleration, as well as the safe distance among multiple UAVs. Thus, how to design real-time trajectory for better channels without violating UAVs' constraints while saving the flying energy consumption is a big challenge for aerial MEC. Additionally, the real-time trajectory design becomes much more challenging when considering it is closely coupled with computation offloading and resource allocation strategies.

\item \textbf{Energy-efficient and latency-aware resource allocation:} Energy-efficient and latency-aware resource allocation plays an vital role in reaping the benefits of aerial MEC. However, due to the large number of closely-coupled parameters related to both UAVs and MEC, the formulated optimization problems are generally non-convex and it is rather challenging to derive optimal solutions with lower complexity. Besides, there is a fundamental tradeoff between the energy consumption and task latency, since the task latency can be generally reduced at the cost of consuming more energy. As a result, how to address the tradeoff by efficient resource allocation is an urgent challenge for aerial MEC.

\item \textbf{Optimal UAV deployment:} For the purpose of coverage and computation capacity enhancement, it is of necessity to optimize the UAV deployment for aerial MEC, including both the 3D placement and the number of UAVs. It is noteworthy that with higher altitudes of UAVs, the probability of LoS links increases, while the path loss is also more severe due to the increased transmission distance. Hence, the tradeoff between LoS connectivity and path loss needs to be carefully addressed by altitude optimization. Besides, depending on the geographical distribution of service requirements and computation capacities of UAVs, the horizontal deployment positions and minimum number of UAVs that can achieve full coverage while satisfying the service requirements is also a challenging optimization problem.

\item \textbf{Security protection and privacy-preserving:} Because of the open-access nature of wireless medium, aerial MEC is exposed to potential adversaries, leading to a high risk in data security and privacy. Especially when considering the high availability and easy deployment of small UAVs, a legitimate UAV in mission is very likely to be attacked by malicious or rogue UAVs, which is a particular threat that does not exist in the infrastructure-based MEC networks. Therefore, it is a huge challenge to develop comprehensive security protection and privacy-preserving mechanisms for aerial MEC.

\item \textbf{Advanced optimization tools:} Due to the large number of parameters, the optimization of aerial MEC usually suffers from exceedingly high levels of dimensionality. The resulting optimization problem can be handled by traditional approaches such as convex optimization, game theory, and heuristic algorithms, which requires expert knowledge on the environments. When encountering highly dynamic environments where expert knowledge is non-trivial to obtain, new ML tools with adaptive modelling and intelligent learning are usually superior. However, the optimality of ML approaches cannot be theoretically proved or strictly guaranteed. Hence, it is a challenge to identify heterogeneous scenarios and select appropriate optimization tools for aerial MEC.

\end{itemize}

Recently, numerous efforts have been dedicated to address these technical challenges. In the following sections, we will provide a comprehensive survey of the existing contributions to the UAV-enabled aerial MEC systems.

\section{Joint Optimization of UAV Trajectory, Computation Offloading, and Resource Allocation}
To realize energy-efficient and low-latency MEC, computation offloading and resource allocation strategies play a pivotal role. For the UAV-enabled aerial MEC, in spite of the stringent SWaP constraints, the controllable maneuverability of UAVs provides an additional DoF to further improve the performances. In this section, we will discuss the research progress on joint optimization of UAV trajectory, computation offloading and resource allocation for aerial MEC, which is categorized by three typical roles of UAVs, i.e., ABSs, mobile users, and relays.

\subsection{When UAVs Serve as ABSs}
By mounting the miniaturized MEC servers, the UAVs can serve as ABSs to provide timely communication and edge computing services for ground users. Compared with terrestrial infrastructure-based MEC systems, the LoS links between UAVs and ground users facilitates highly reliable air-ground transmissions.
However, the integration of UAVs and MEC also results in heterogeneity of the networks and the coupling among trajectory design and resource allocation. Therefore, it is of great necessity to jointly optimize the UAV trajectory, as well as the computation offloading and resource allocation strategies.

\textbf{\emph{1) Optimization for the Single-UAV Case:}}
At the early stage of optimization for aerial MEC, the single-UAV case is investigated to acquire some design insights. For example, in \cite{JHu2019}, a single UAV equipped with an MEC server is deployed in the sky to help ground users complete their computation tasks. It is assumed that the position of the UAV remains fixed during the whole mission period. Then, the energy consumption of all users is minimized by jointly optimizing the UAV position, time slot allocation and computation task partition. Although the globally optimal solution can be found in \cite{JHu2019} by the two-dimensional (2D) search over all possible UAV positions, the benefits of UAVs' mobility are not exploited.
Generally speaking, it is difficult to simultaneously optimize the UAV trajectory, computation offloading and resource allocation, especially for the large-scale multi-user scenarios, since the optimization variables are closely coupled and the formulated problems are usually highly non-convex. Decomposing the original problems into several subproblems to decouple the variables is an effective way to reduce the computation complexity \cite{XZhang2019}. For instance, in order to minimize the total energy consumption of users, M. Hua \emph{et al.} \cite{MHua2019} decompose the formulated problem into two subproblems, where the resource allocation is optimized by Lagrangian dual method, and the UAV trajectory is designed by leveraging the sequential convex approximation (SCA) technique. Similarly, the total energy consumption of all users is minimized in \cite{HGuo2020, JXiong2019, YLuo2020} by decomposing the original problems into several subproblems. Various optimization methods, such as the alternating direction method of multipliers (ADMM), Lagrange dual method, block coordinate descent (BCD), linear programming (LP) and SCA, are applied to solve the decomposed subproblems.

It is worth noting that the energy consumption optimization in the aforementioned literature is user-centric, where the UAV's energy consumption is not taken into account. Due to the limited on-board energy storage and considerable energy consumption for computing and propulsion, optimizing the UAV's energy consumption is of essential importance to prolong the service time \cite{YDu2018, MZhao2021}. Although the energy consumption of the UAV is imposed as a budget constraint in \cite{HGuo2020} when optimizing the energy consumption of users, there is still a gap between the minimum energy consumption and the energy budget for the UAV. Therefore, the energy consumption of both users and the single UAV is minimized in \cite{YTun2021, JZhang2019}. In particular, Y. K. Tun \emph{et al.} \cite{YTun2021} introduce a block successive upper-bound minimization (BSUM) algorithm to jointly optimize the task offloading decision, resource allocation mechanism as well as the UAV's trajectory, where the energy consumption of the UAV can be greatly reduced. Besides, in order to balance the energy consumption between users and the UAV, the average weighted energy consumption of users and the UAV is minimized in \cite{JZhang2019}. The Lyapunov optimization technique is applied to optimize the stochastic computation offloading, resource allocation, and trajectory scheduling. It is interestingly found that when only considering the energy consumption of the UAV, the optimal flying trajectory is a straight line, which can reduce the flying distance and save the UAV’s energy. On the contrary, if only the energy consumption of users is optimized, the UAV first flies close to users to improve the quality of communication channels. Then, the UAV has to return to the final position within the limited mission time. Such a trajectory definitely increases the flying distance and results in larger energy consumption for the UAV. As a comparison, under the proposed scheme in \cite{JZhang2019}, the energy consumption of users and the UAV can be well balanced.

In addition to the energy consumption optimization for aerial MEC, the latency problem has also drawn considerable attentions. In \cite{QHu2019}, a penalty dual decomposition (PDD)-based algorithm is proposed to minimize the sum of the maximum delay among all users by jointly optimizing the UAV trajectory, the ratio of offloading tasks, and the user scheduling variables. Simulations show that the proposed algorithm can always obtain a lower time delay compared to the scheme of fixed location and circular trajectory for the UAV. Intuitively, when the UAV flies to its served users at a higher speed, the task completion time can be further reduced. However, in this case, the propulsion power consumption of the UAV will be increased. Therefore, the task completion time and the energy consumption of the UAV are two opposite aspects related to UAV's flying speed and trajectory, which cannot be minimized at the same time. To address this challenge, C. Zhan \emph{et al.} \cite{CZhan2020} first investigate the completion time and the UAV's energy consumption minimization problem. A Pareto-optimal solution is derived to balance the tradeoff between the UAV's energy consumption and the task completion time.

\begin{figure}[t]
\centering
\includegraphics[width=3in]{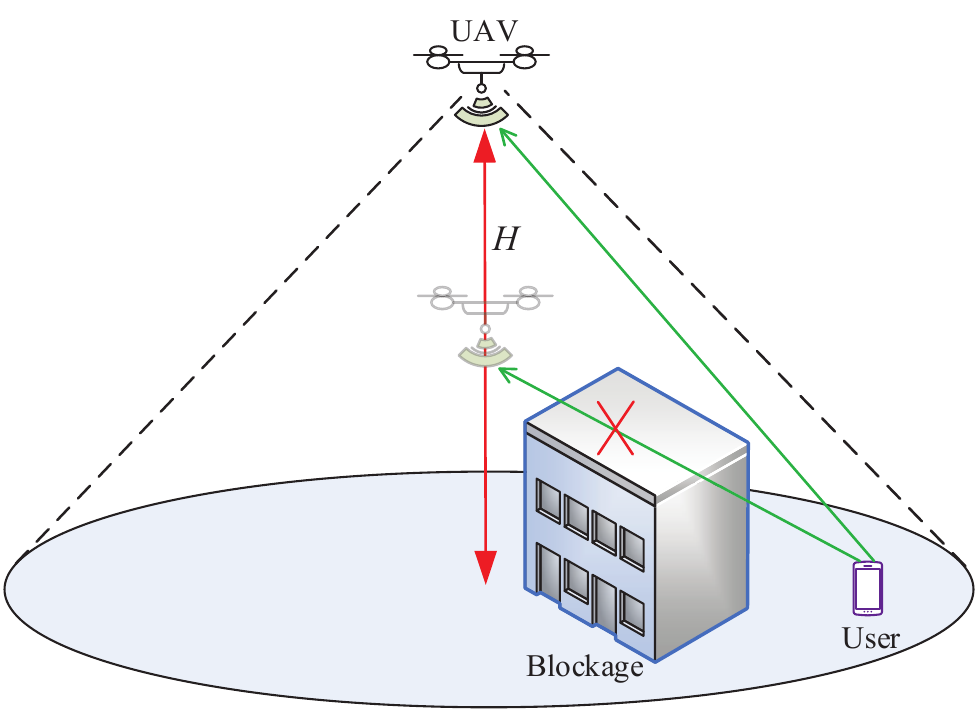}
\caption{Illustration of the height optimization for aerial MEC.}
\end{figure}

In most of the existing literature, the flying height of the UAV is assumed to be a constant. In practice, an UAV can freely fly in 3D space, and the flying height of the UAV has great impacts on the air-ground channel gains in terms of the probability of LoS links and path loss \cite{HMei2019}. Thus, F. Costanzo \emph{et al.} \cite{FCostanzo2020} investigate the height of the UAV in aerial MEC and propose a stochastic approximation method to select the UAV's flying height. As shown in Fig. 3, a reduction of the height can induce a positive impact on the overall energy consumption, because of the reduced path loss in LoS conditions. However, below a certain altitude, due to the presence of obstacles, some users begin to experience worse communication links, and thus the system energy consumption increases. Therefore, there is a fundamental tradeoff between the probability of LoS links and path loss when designing the 3D trajectory for the UAV. Focusing on this problem, H. Mei \emph{et al.} \cite{HMei2020} aim to jointly optimize the resource allocation and UAV trajectory in 3D space to minimize the overall energy consumption of the UAV. The trajectory of the rotary-wing/fixed-wing UAV is optimized in both vertical and horizontal dimensions by the quadratically constrained quadratically program (QCQP) and BCD algorithm. From the simulation results, it can be clearly observed that the UAV's flying height is constantly changing in order to save the UAV's service and propulsion energy while guaranteeing the service requirements of users.

Multiple access scheme is also a key consideration in the optimization of aerial MEC. While the orthogonal multiple access schemes, such as TDMA \cite{QHu2019, YQian2019}, FDMA \cite{JZhang2019}, and orthogonal frequency division multiple access (OFDMA) \cite{XDiao2019, SLiu2020}, are commonly adopted, non-orthogonal multiple access (NOMA) has also been introduced into the aerial MEC in order to accommodate massive connectivity, further reduce the transmission latency and improve the spectrum efficiency \cite{IBudhiraja2021}. NOMA allows multiple users to access the UAV and offload tasks at the same time and frequency. By applying the SIC technique, the inter-user interference can be mitigated and each user's offloading task data can be successively decoded by the UAV. X. Diao \emph{et al.} \cite{XbDiao2019} investigate the application of NOMA in aerial MEC, where all users simultaneously offload task bits to the UAV in the uplink via NOMA. The maximum energy consumption among users is minimized by jointly optimizing the UAV's trajectory, task data and computing resource allocation. In \cite{YYang2020}, the total energy consumption minimization problem is studied for NOMA-based aerial MEC systems. Simulation results demonstrate that the NOMA-based approach achieves higher energy efficiency than that based on orthogonal multiple access (OMA). Moreover, under both NOMA and OMA schemes, S. Jeong \emph{et al.} \cite{SJeong2018} minimize the total energy consumption of the users subject to the task tolerance latency and UAV’s energy budget by leveraging the SCA techinique. Similarly, the weighted sum energy consumption of the UAV and users is minimized in \cite{JJi2021}, by jointly optimizing the UAV trajectory and computation resource allocation, where both OMA and NOMA schemes are investigated based on the BCD method. Aiming to improve the user experience while optimizing the energy consumption of the UAV, M. Li \emph{et al.} \cite{MsLi2020} study the UAV's energy efficiency maximization problem for OMA and NOMA schemes by jointly optimizing the UAV trajectory, user transmit power, and computation load allocation, where the Dinkelbach algorithm and the SCA technique are applied to solve the formulated non-convex fractional programming.

A summary of contributions to the joint optimization for single-UAV case when UAVs serve as ABSs is provided in Table 2.

\renewcommand\arraystretch{1.25}
\begin{table*}[t]
\footnotesize
\caption{Summary of Contributions to the Optimization for Single-UAV Case When UAVs serve as ABSs}
\centering
\begin{tabular}{|m{20pt}<{\centering}|m{100pt}|m{185pt}|m{150pt}|}
\hline
\bfseries Refs. & \bfseries Optimization Objective & \bfseries Optimization Variables & \bfseries Optimization Methods\\
\hline
\cite{JHu2019} & User energy consumption & Computation task partition, time slot, UAV position & Augmented Lagrangian active set, 2D search \\
\hline
\cite{XZhang2019}& Computation efficiency  & CPU frequency, power, time slot, UAV trajectory,  & Dinkelbach’s method, Lagrange duality, SCA  \\
\hline
\cite{MHua2019} & Energy consumption of users & Bit allocation, connection scheme, CPU frequency, power, UAV trajectory & Lagrangian duality, SCA \\
\hline
\cite{HGuo2020} & Energy consumption of users & Bit allocation, offloading decision, UAV trajectory & BCD, LP, SCA \\
\hline
\cite{JXiong2019} & Energy consumption of users & Bit allocation, offloading decision, UAV trajectory & Greedy algorithm, Lagrangian duality, BCD\\
\hline
\cite{YLuo2020}& Energy consumption of users & Bit allocation, UAV trajectory & ADMM \\
\hline
\cite{YTun2021} & Energy consumption of both users and the UAV & Bandwidth, bit allocation, CPU frequency, power, UAV trajectory & BSUM \\
\hline
\cite{JZhang2019} & Energy consumption of both users and the UAV & Bit allocation, CPU frequency, UAV trajectory &  Lyapunov optimization, ADMM  \\
\hline
\cite{QHu2019} & Sum of the maximum delay among all users & Offloading decision, offloading ratio, UAV trajectory & PDD, CCCP  \\
\hline
\cite{CZhan2020} & Energy consumption and completion time of the UAV & CPU frequency, offloading decision, UAV trajectory & SCA, Pareto-optimal  \\
\hline
\cite{FCostanzo2020} & Long term system average energy consumption & CPU frequency, transmission rate, UAV height & Stochastic optimization, Kiefer-Wolfowitz algorithm \\
\hline
\cite{HMei2020} & UAV energy consumption & Bandwidth, power, CPU frequency, UAV's 3D trajectory & Lagrange duality, SCA, QCQP \\
\hline
\cite{XbDiao2019} & Maximum energy consumption among all users & Bit allocation, CPU frequency, UAV trajectory & SCA  \\
\hline
\cite{YYang2020} & Energy consumption for offloading and computing & Bit allocation, power, UAV trajectory & SCA, Lagrange duality\\
\hline
\cite{SJeong2018} & Energy consumption of users & Bit allocation, UAV trajectory & SCA  \\
\hline
\cite{JJi2021} & Weighted sum energy consumption of users and UAV & Bit allocation, CPU frequency, UAV trajectory & BCD, SCA, Lagrange duality  \\
\hline
\cite{MsLi2020}& Energy efficiency & CPU frequency, power, UAV trajectory & Dinkelbach’s method, SCA, ADMM  \\
\hline
\end{tabular}
\end{table*}
\renewcommand\arraystretch{1}

\textbf{\emph{2) Optimization for the Multi-UAV Case:}} In aerial MEC, if all the ground users rely on the single UAV for edge computing, the limited on-board energy storage will lead to the fact that the UAV cannot provide services for a long time. In addition, the insufficient computing capacity of the single UAV will increase the task completion time and thus may not meet the users' QoS requirements \cite{YQu2021}. Furthermore, the coverage area of a single UAV is limited, and the users out of the coverage area cannot enjoy the communication and computing services. To overcome these limitations, multiple UAVs can be deployed in the areas of interest to act as ABSs mounting MEC servers \cite{AANasir2021}. Compared with the single-UAV aerial MEC, multiple UAVs can provide more powerful and persistent edge computing services for ground users, and also enlarge the coverage area \cite{PAApostolopoulos2021}. However, to exploit the full benefits of multiple UAVs, there are several key challenges that need to be addressed. Firstly, each user needs to decide not only how much task bits it should offload, but also which UAV it should offload to, and the latter is called the user association problem. Secondly, the joint optimization of user association, UAVs' trajectories and resource allocation is much more complex than the aerial MEC systems with a single UAV, since the number of related variables increases and these variables are more closely coupled with each other. Last but not least, in order to avoid the collision of UAVs, the safe distance among multiple UAVs must be carefully considered, which results in more difficulties of trajectory design.

Recently, there has been some work providing solutions to the above-mentioned challenges for multi-UAV aerial MEC systems. For the user association, \cite{JiZhang2020, XDiao2020, ZYang2019} tackle this problem from different perspectives. Specifically, J. Zhang \emph{et al.} \cite{JiZhang2020} investigate the computation efficiency maximization for a multi-UAV-enabled aerial MEC system, where the user association, CPU frequency allocation, power and spectrum resources, as well as trajectory scheduling of multiple UAVs are jointly optimized by an iterative algorithm with a double-loop structure. The user association in \cite{JiZhang2020} is formulated as a standard integer linear programming (ILP) given the resource allocation and trajectories of UAVs. Then, existing algorithms, such as the branch-and-bound (B\&B) and cutting plane methods, can be applied to solve the ILP problem. Different from \cite{JiZhang2020}, X. Diao \emph{et al.} \cite{XDiao2020} propose a greedy-based offloading strategy variable rounding (GOSVR) algorithm to obtain a near-optimal integer solution for the user association. Then, the UAVs' trajectories are optimized to minimize the weighted sum of the maximum energy consumption among users and among UAVs. In \cite{ZYang2019}, after decomposing the sum power minimization problem into three subproblems, the user association subproblem is approximated as a sequence of weighted ${l_0}$-norm minimizations and a compressive sensing based algorithm is proposed to obtain the closed-form solution.

Note that the UAVs' location planning in \cite{ZYang2019} is obtained via the one-dimensional searching method, while the mobility of UAVs is not considered. In \cite{XZhang2020}, the trajectory design and resource allocation for NOMA-based multi-UAV aerial MEC are jointly optimized. The formulated problem is decomposed into two subproblems and an efficient iterative algorithm is proposed to minimize the weighted sum energy consumption of users and UAVs. During each iteration, the resource allocation subproblem is solved by SCA given the UAVs' trajectories, while based on the resource allocation schemes, the trajectory planning subproblem for multiple UAVs is addressed via the quadratic approximation. As a further advance, X. Qin \emph{et al.} \cite{XQin2021} focus on the multi-access feature in multi-UAV aerial MEC systems and propose a joint trajectory design and resource allocation algorithm. With the aid of multiple radio access, each user is allowed to offload task bits to multiple UAVs simultaneously for parallel computing. It is demonstrated that the proposed algorithm outperforms the fixed trajectory, fixed bandwidth allocation and single access schemes.

It is worth noting that the aforementioned literature assumes rational users in the multi-UAV aerial MEC systems can make offloading decisions in a risk-neutral manner for maximizing their perceived utility, such as minimizing the energy consumption \cite{ZYang2019}, maximizing the computation efficiency \cite{JiZhang2020}, etc. However, in the real life, the users tend to exhibit the risk-seeking or loss-aversion behavior under the presence of uncertainties introduced by the underlying computing resource availability of the servers \cite{PAApos2020, PAApos2020Access, PAApos2019}. To be more specific, since the energy storage of the UAV is limited, the energy that can be used for computing is correspondingly decreased over time when the UAV continuously consumes energy for flying. In this case, the computing uncertainty appears as the UAV's capability to process the offloading task bits from users is unknown in prior. Therefore, by taking both the risk-aware user behavior and the UAVs' computing resource availability into consideration, P. A. Apostolopoulos \emph{et al.} \cite{PAApostolopoulos2021} aim to maximize each user's satisfaction utility, where a linear probability of failure function is proposed to describe the computing resource availability of the UAVs. According to the simulation results, the proposed offloading approach can achieve the best performance compared to alternative approaches. Besides, the users’ physical and risk-aware characteristics have a significant impact on their task offloading decisions. For the UAVs, they will receive more task bits from users if they have higher computation capability and more energy storage.

\begin{figure}[t]
\centering
\includegraphics[width=3.5in]{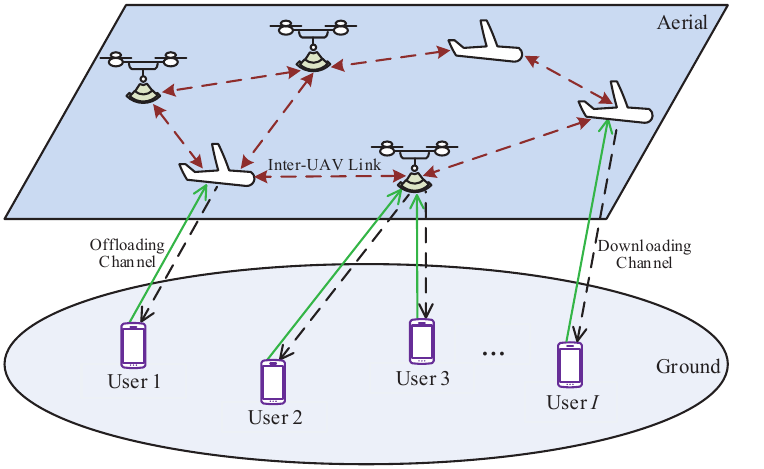}
\caption{Cooperative aerial MEC via the inter-UAV links.}
\end{figure}

Despite the intensive investigation on the optimization for multi-UAV aerial MEC, most of them assume each UAV operates in an independent way. In other words, each UAV computes users' tasks by their own computation capacity, where the UAVs taking too more tasks may overload. As a candidate solution, the UAVs' load can be balanced by adjusting the user association. However, the users within the coverage of one overloaded UAV may be far away from the other UAVs, which makes it impossible to change the user association. Actually, multiple UAVs can also cooperate in the air to share their capacities, where the task data offloaded to one UAV can be reallocated among multiple UAVs with the aid of inter-UAV links \cite{HShimada2021}. As shown in Fig. 4, via the inter-UAV links using various radio interfaces, such as WiFi, millimeter wave (mmWave) and the 5G New Radio (NR), a swarm of UAVs can be organized as a FANET and exchange information among each other, including the task data, control and coordination information \cite{NCheng2018}. Although additional transmission delay and overheads will be imposed due to the task reallocation, the computation capacity sharing exhibits a great potential to reduce the task processing delay and energy consumption for multi-UAV aerial MEC systems.

Nowadays, the investigations on the cooperative MEC among multiple UAVs are still in its infancy stage. In \cite{HDong2020}, a multi-UAV collaborative MEC model with inter-UAV links is investigated. When one UAV moving in a rectangular area obtains a sequence-dependent computing task, it can forward sub-tasks to other UAVs for computing with the aid of inter-UAV links. A spatial branch limiting algorithm is proposed to minimize the overall energy consumption of UAV cluster. It is found that when the number of tasks is less, the single-UAV scheme consumes lower energy than the proposed multi-UAV collaborative MEC. However, if the number of tasks increases to a certain extent, the single-UAV scheme cannot complete all tasks, while the multi-UAV collaborative MEC can still complete the tasks as the computation capacities of multiple UAVs are shared through the inter-UAV links. In \cite{WMa2019}, the tasks generated by UAV users can be offloaded to other UAVs via the FDMA-based inter-UAV links. A strategic game is utilized to analyze the interactions among UAVs which act in their own interests, and a decentralized strategic offloading algorithm is proposed to optimize the energy consumption of UAVs. Similarly, in \cite{WLiu2020}, the UAVs with low-latency and computation-intensive tasks can select to offload partial task data to a nearby UAV helper with more powerful computing ability. With the objective of minimizing the total energy consumption of all UAVs, a two-stage resource allocation scheme that exploits the convex optimization and stochastic learning automata (SLA) algorithm is proposed to optimize the UAV helper selection, transmission channel, and offloading rate. K. Yao \emph{et al.} \cite{KYao2020-C2} study the cooperation of two types of UAVs, namely, scout UAV (SU) and computing UAV (CU), in a UAV swarm. One SU and one CU are grouped into a pair and the SU can offload partial tasks to its CU for processing via non-overlapped sub-channels. A game-theoretic based solution is proposed to solve the computation offloading and variable-width channel access problem, where the overall energy consumption of UAVs is minimized. In addition to the energy consumption optimization, R. Chen \emph{et al.} \cite{RChen2020} focus on the delay minimization in cooperative UAV swarm-assisted MEC systems. It is assumed that the UAVs in a swarm are divided into several coalitions which consist of a leader and multiple members, where each member UAV needs to decide the offloading ratio and transmission channels to the leader UAV. In order to minimize the computing delay, an exact potential game (EPG) is formulated and a distributed offloading algorithm is designed to achieve the Nash equilibrium (NE).

For the trajectory design in multi-UAV aerial MEC systems, collision avoidance is also an essential consideration in real environment. Specifically, if there is a no-fly zone or obstacle in UAV’s flying range, the UAV needs to keep a safe distance with them \cite{JiZhang2020}. Moreover, the multiple UAVs are required to keep a safe distance with each other to avoid collision \cite{XDiao2020}. To avoid the collision between UAVs and obstacles, in \cite{YQu2021}, it is assumed that multiple UAVs fly at different fixed altitudes, and thus they could not interfere with each other. As a further advance, a conflict elimination strategy is applied in \cite{YLuo2021} for the trajectory design, which guarantees the minimum distances among potential conflict UAVs satisfy the safety constraints. Besides, a reinforcement learning (RL) approach is proposed in \cite{HChang2021} to design the UAVs' paths, where the collision risk is incorporated into the reward function to avoid the collision.

A summary of contributions to the joint optimization for multi-UAV case when UAVs serve as ABSs is provided in Table 3.

\renewcommand\arraystretch{1.25}
\begin{table*}[t]
\footnotesize
\caption{Summary of Contributions to the Optimization for Multi-UAV Case When UAVs serve as ABSs}
\centering
\begin{tabular}{|m{20pt}<{\centering}|m{135pt}|m{165pt}|m{135pt}|}
\hline
\bfseries Refs. & \bfseries Optimization Objective & \bfseries Optimization Variables & \bfseries Optimization Methods\\
\hline
\cite{JiZhang2020} & Computation efficiency & Bandwidth, CPU frequency, power, UAVs' trajectories, user association & Dinkelbach’s method, SCA, ILP \\
\hline
\cite{XDiao2020} & Sum of the maximum energy consumption among users and UAVs & UAVs' trajectories, user association &  Greedy algorithm, SCA \\
\hline
\cite{ZYang2019}& Weighted sum power of users and UAVs & Computation capacity, power, UAVs' locations, user association, &  Compressive sensing, Lagrangian duality  \\
\hline
\cite{XZhang2020}& Weighted sum energy consumption of users and UAVs & CPU frequency, power, UAVs' trajectories & Quadratic approximation, SCA  \\
\hline
\cite{XQin2021} & Weighted sum energy consumption of users and UAVs & Bandwidth, bit allocation, CPU frequency, power, UAVs' trajectories & BCD, SCA \\
\hline
\cite{HDong2020}& Energy consumption of UAVs & Bandwidth, bit allocation & B\&B \\
\hline
\cite{WMa2019} & Energy consumption of UAVs & Bit allocation & EPG \\
\hline
\cite{WLiu2020} & Energy consumption of UAVs & Channel, offloading decision and rate & SLA \\
\hline
\cite{KYao2020-C2} & Energy consumption of UAVs & Channel, offloading ratio & EPG \\
\hline
\cite{RChen2020} & Total delay of UAVs & Channel, offloading ratio & EPG \\
\hline
\cite{YLuo2021}& Energy consumption of users & Bit allocation, UAVs' trajectories, user association & Dynamic programming, ADMM   \\
\hline
\end{tabular}
\end{table*}
\renewcommand\arraystretch{1}

\subsection{When UAVs Serve as Mobile Users}
Characterized by the advantages of low cost, on-demand deployment and high maneuverability, UAVs can serve as mobile users and be dispatched to carry out critical tasks in various scenarios, such as target tracking \cite{XDeng2021}, emergency rescue \cite{ZNing2021}, smart delivery \cite{RLi2020}, and mapping \cite{MMessous2020}, etc. In these scenarios, UAVs may not be able to process the collected data in real time, due to the energy and computation capacity limitations. To address this problem, computation tasks can be offloaded to powerful MEC servers deployed at GBSs or other BS entities \cite{DCallegaro2018, JChen2021, MLiwang2021}. Compared with ground users, UAVs can flexibly adjust their positions in 3D space to find better channels to offload their task data.

In the context of UAVs serving as mobile users, a fixed-wing UAV is employed in \cite{LFan2019} to transmit part of the collected data to the ground MEC server for real-time processing. Under the constraints of computation capacities of the UAV and ground MEC server as well as the UAV's velocity and acceleration, the weighted sum energy consumption of the UAV and MEC server is minimized, where the UAV's trajectory, task assignment, and CPU's computational speed are jointly optimized. It is interestingly found that the transmit power of fixed-wing UAV has a significant impact on the trajectory and energy consumption. Specifically, if the transmit power of fixed-wing UAV is smaller, the UAV's trajectory has the shape of “8” approximately, which means the UAV has to continuously accelerate and decelerate in order to hover as closely as possible to the ground MEC servers to maintain a better channel and complete the task offloading with lower transmit power. In this case, the UAV's flight consumes more energy and it results in greater energy consumption for the system. On the contrary, if the transmit power is larger, the UAV can still complete the task offloading even if the channel is relatively worse. In this case, the UAV's trajectory follows a droplet shape with much lower average acceleration. Accordingly, the energy consumption for flying can be saved. Similarly, M. Hua \emph{et al.} \cite{MHua2018} investigate the aerial MEC systems where multiple UAVs serving as users are associated to an MEC server for computation offloading. Considering four access schemes, i.e, TDMA, OFDMA, one-by-one access and NOMA, the total energy consumption of all UAVs is minimized by SCA. From the simulation results, it can be found that NOMA has a better performance on energy consumption than OMA. Nevertheless, all UAVs superposing on the same resource block may cause severe decoding delay and co-channel interference. To deal with this problem, a multi-UAV grouping method is proposed in \cite{ZZhu2020} to reduce the number of UAVs on the same resource block. Then, the UAVs' transmit power and BSs' computation resources are optimized based on the Karush-Kuhn-Tucker (KKT) conditions to minimize the sum of energy consumption related to communication and computing.

Although less energy consumption is beneficial to the UAVs with limited battery storage, the QoS, especially the transmission and execution delay are also of great importance for the UAVs when they serve as users to perform critical tasks \cite{XCao2018}. Based on the transmission and execution delay, M. Dai \emph{et al.} \cite{MDai2020} propose an energy-efficient offloading scheme for UAVs. The matching scheme is designed to choose the optimal partners between UAVs and edge nodes, and then the offloading process of UAVs is modeled as a bargaining game to maximize the offloading utility. Although the transmission and execution delay can be reduced by improving the transmit power of UAVs, it will result in larger energy consumption \cite{YRen2020}. In this context, R. Duan \emph{et al.} \cite{RDuan2019} investigate the trade-off between the transmit energy of UAVs and the computation delay for heterogenous cloud aided multi-UAV systems. A power-vs-delay trade-off is struck by leveraging Lyapunov optimization, and the optimal task scheduling and resource allocation strategy is found.

Besides the MEC server in close proximity to UAVs, satellites or other BSs equipped with MEC servers can also provide additional offloading opportunities for UAVs. For instance, in \cite{MMessous2019}, a fleet of small UAVs performing an exploration mission can offload their tasks to a neighboring WiFi BS or a more powerful cellular-connected MEC server. A non-cooperative game theory based algorithm is proposed to address the offloading decision of UAVs. It is observed that a considerable amount of energy for the computation offloading can be saved with the right communication medium.
In \cite{PCao2019}, UAVs are used to detect the wind turbines. After detection, both the ground MEC server and satellites can provide edge computing services for the UAVs. Aiming to minimize the completion time, the UAVs' trajectories and computation strategy are jointly optimized. It is found that the proposed ground-satellite-integrated offloading scheme achieves a lower completion time than the scheme without satellites, which verifies the advantages of introducing satellites to provide offloading opportunities for UAVs.

A summary of contributions to the joint optimization when UAVs serve as mobile users is provided in Table 4.

\renewcommand\arraystretch{1.25}
\begin{table*}[t]
\footnotesize
\caption{Summary of Contributions to the Joint Optimization When UAVs Serve as Mobile Users}
\centering
\begin{tabular}{|m{22pt}<{\centering}|m{100pt}|m{125pt}|m{105pt}|m{90pt}|}
\hline
\bfseries Refs. & \bfseries Network Elements & \bfseries Optimization Objective  & \bfseries Optimization Variables & \bfseries Optimization Methods \\
\hline
\cite{MHua2018} & Multiple UAVs, one BS  & Energy consumption of UAVs &Bit allocation, power, UAV trajectory &SCA\\
\hline
\cite{LFan2019} & One UAV, one BS & Energy consumption of the UAV and BS & Offloading radio, CPU frequency, UAV trajectory &Discrete linear state-space approximation, CCCP  \\
\hline
\cite{ZZhu2020} & Multiple UAVs, Multiple BSs  &  Sum of transmission energy of UAVs and computation energy of BSs & CPU frequency, power, UAV grouping & Lagrangian duality\\
\hline
\cite{XCao2018} & One UAV, multiple BSs & UAV’s mission completion time & Offloading time, UAV trajectory& SCA \\
\hline
\cite{RDuan2019} & Multiple UAVs, an edge server, a remote server & Power-delay trade-off of UAVs & Bit allocation, CPU frequency & Queueing theory, Lyapunov optimization \\
\hline
\cite{MMessous2019} & Multiple UAVs, one WiFi BS, one MEC server & Energy consumption, delay and communication cost & User association & Non-cooperative game \\
\hline
\cite{PCao2019} & Multiple WTs, multiple UAVs, one BS, one satellite & Total completion time& Offloading mode, power, offloading time, CPU frequency, UAV trajectory & Lagrangian duailty\\
\hline
\end{tabular}
\end{table*}
\renewcommand\arraystretch{1}

\subsection{When UAVs Serve as Relays}
For local resource limited users who locate in remote areas or whose direct links to BSs have been blocked by obstacles, the UAVs can serve as relays to forward computation-intensive tasks to more powerful remote MEC servers by two or more hops \cite{YYu2021}. An illustrative network architecture when UAVs serve as relays can be seen in Fig. 5. Compared with conventional static relay, the UAV relay provides new opportunities for system performance enhancement, since a UAV can successively adjust its position for experiencing better channel conditions. However, due to the participation of UAV relays, the channel condition in each hop has a direct impact on the task offloading, which renders the UAV trajectory design much more complex. Meanwhile, the computation offloading and resource allocation optimization is also indispensable in order to fully unlock the potentials of UAV relays in aerial MEC.

\begin{figure}[t]
\centering
\includegraphics[width=3.65in]{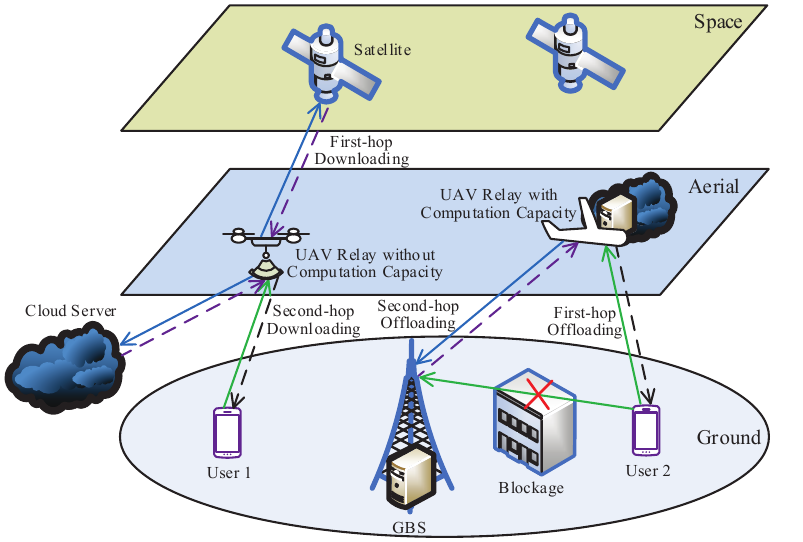}
\caption{Illustrative network architecture when UAVs serve as relays.}
\vspace{-0.1in}
\end{figure}

\textbf{\emph{1) UAV Relay without Computation Capacity:}} In \cite{XLiu2020}, as the ground channels from IoT nodes to the data center is in severe fading, the UAV serves as a relay to froward the information from IoT nodes to the data center controlled by the fog computing BS. In order to maximize the throughput of IoT nodes, the subcarrier allocation, transmit power of IoT nodes and UAV, as well as UAV trajectory are jointly optimized. Simulation results reveal that the throughput of IoT nodes under the mobile UAV relay is higher than that under the static UAV relay. In \cite{FGuo2019}, with the UAV acting as a mobile relay between users and the BS, UAV trajectory, power allocation, and user scheduling scheme are jointly optimized to minimize the total latency of all users. In \cite{QWang2021}, by utilizing the multiple input multiple output (MIMO) technique, users' tasks can be transmitted to multiple UAVs in parallel and next further offloaded to the BS by UAV relays. Then, a genetic based heuristic joint power and quality of experience (HJPQ) algorithm is proposed to minimize the weighted sum of energy consumption and delay. From the perspectives of revenue and operating expense (OPEX), S. Zheng \emph{et al.} \cite{SZheng2020} investigate the offloading and caching strategies, as well as bandwidth and computing resource allocation in aerial MEC when the UAV serves as a relay. A multi-dimensional hybrid adaptive particle swarm (MHAPSO) algorithm is proposed to maximize the net revenue. Simulation results show that the operator can earn more revenue when users have more task-input data, which coincides with the physical reality.

\textbf{\emph{2) UAV Relay with Computation Capacity:}} When the UAV serves as a relay to assist the task offloading, MEC servers can also be carried by the UAV to perform edge computing. Since the single-hop transmission delay from users to UAVs is less than the multi-hop delay from users to remote MEC servers, the users prefer to compute their tasks at UAVs. However, if the required computation resource exceed the UAV's capacity, part of the tasks will be further relayed to remote MEC servers for real-time processing \cite{LZhang2021}. Besides, for a specific amount of task bits, the computing process and relay transmission by the UAVs both consume considerable energy. Thus, from the perspectives of energy consumption and delay, the task allocation between relay UAVs and remote MEC servers should be carefully designed.

In \cite{TZhang2020}, a UAV is deployed as a MEC server to help terminal devices (TDs) compute their partial tasks. Meanwhile, the UAV can also act as a mobile decode-and-forward (DF) relay to further offload part of the TDs' task bits to the ground AP for computing. In order to minimize the total energy consumption, the task allocation, time slot scheduling, power allocation, and UAV trajectory design are jointly optimized. As expected, the proposed scheme consumes less energy compared with the “only relaying design” (i.e., the UAV does not possess computing capacity and can only act as a relay to assist task offloading from TDs to the ground AP), and the “no AP design” (i.e., the tasks of TDs are computed locally or by the UAV without the assistance of AP). Similarly, considering the UAV can simultaneously act as a MEC server and a relay, X. Hu \emph{et al.} \cite{XHu2019} assume the total bandwidth in each user equipment (UE)'s operation duration is separated into three independent parts for task reception from UEs, task offloading transmission to the AP, and task result downloading to the UEs, respectively. Then, in order to minimize the weighted sum energy consumption of the UAV and UEs, an alternating optimization algorithm is proposed to jointly optimize the bandwidth allocation, computation resource scheduling, and the UAV's trajectory.

It is worth remarking that the delay optimization problem is much more complex when the UAV acts as both a MEC server and a relay, as the multi-hop transmission delay and computation delay in the UAV and remote MEC servers, should be jointly considered. In order to reduce the total delay of all UEs, L. Zhang \emph{et al.} \cite{LZhang2020} decompose the formulated NP-hard delay minimization problem into three subproblem, and then an approximation algorithm is proposed to jointly optimize the UAV placement, user association, communication and computing resource assignment. Simulation results demonstrate that the proposed algorithm can achieve a lower delay compared with the scheme where the UAV can only act as a relay and the scheme where all users directly offload tasks to the MBS. This is because the UAV as a MEC server can directly execute the task computing for those users far away from the MBS and thus their task completion time may be decreased. In addition, since some users may experience poor channel conditions when they directly offload tasks to the MBS, the UAV serving as a relay can provide opportunities for users to improve their offloading channel gains.

Moreover, in order to address the delay and energy consumption tradeoff, a SCA-based algorithm is proposed in \cite{ZYu2020} to minimize the weighted sum of the delay and the UAV energy consumption. Simulation results show the weighted sum of delay and energy consumption is lower than other baseline schemes, such as the UAV-only scheme where all tasks are offloaded and processed at the UAV, the edge clouds (EC)-only scheme where all tasks are relayed by the UAV and processed at the ground ECs. Similarly, Y. Zhu \emph{et al.} \cite{YZhu2020} and L. Zhao \emph{et al.} \cite{LZhao2020} aim to optimize the task execution delay and energy consumption with the participation of SDN controller when the UAV serves as both a MEC server and a relay. In particular, based on the convex optimization and branch and bound (B\&B) method, a joint task computing mode selection and resource allocation algorithm is proposed in \cite{YZhu2020}. Simulation results show that the UAV relay scheme can significantly reduce the weighted sum of the delay and energy consumption of the UAV and all users. Differently, L. Zhao \emph{et al.} \cite{LZhao2020} solve the offloading mode selection problem by game theory, where vehicle users are regarded as multiple players and make offloading decisions according to their own interests. A decision-making algorithm based on the sequential game is proposed to optimize the task execution delay and system energy consumption.

A summary of contributions to the joint optimization when UAVs serve as relays is provided in Table 5.

\renewcommand\arraystretch{1.25}
\begin{table*}[t]
\footnotesize
\caption{Summary of Contributions to the Joint Optimization When UAVs Serve as Relays}
\centering
\begin{tabular}{|m{73pt}|m{22pt}<{\centering}|m{100pt}|m{160pt}|m{85pt}|}
\hline
\bfseries Categories &\bfseries Refs. & \bfseries Optimization Objective  & \bfseries Optimization Variables & \bfseries Optimization Methods \\
\hline
~ &\cite{XLiu2020} & Users’ throughput & Power, subcarrier, UAV trajectory & Water filling, SCA \\
\cline{2-5}
~ &\cite{FGuo2019}  & Total latency of users & Power, user scheduling, UAV trajectory & SCA, DC \\
\cline{2-5}
UAV Relay without Computation Capacity &\cite{QWang2021} & Weighted sum of energy consumption and
delay & Power, user association, UAV locations & HJPQ\\
\cline{2-5}
~ & \cite{SZheng2020} & Net revenue & Bandwidth, caching, CPU frequency, user association & MHAPSO\\
\hline
~ &\cite{TZhang2020} & Total energy consumption of UAV and users & Bit allocation, power, time slot, UAV trajectory & Lagrangian duality, SCA \\
\cline{2-5}
~ & \cite{XHu2019} & Weighted sum energy consumption of UAV and users & Bandwidth, CPU frequency, UAV trajectory & Lagrangian duality, SCA \\
\cline{2-5}

~ &\cite{LZhang2020} & Average latency of all users & Communication and computing resource assignment, UAV placement, user association & Linear optimization, exhaustive search  \\
\cline{2-5}
UAV Relay with Computation Capacity &\cite{ZYu2020} & Weighted sum of users' delay and UAV energy consumption & Communication and computing resource, task splitting decisions, UAV position & SCA  \\
\cline{2-5}
~ &\cite{YZhu2020} & Weighted sum of delay and energy consumption & Computing mode, CPU frequency & LP, B\&B \\
\cline{2-5}
~ & \cite{LZhao2020} & Weighted sum of delay and energy consumption & Offloading mode & Sequential game\\
\cline{2-5}
~ &\cite{XHu2020} & Weighted sum completed task-input bits of users & Task allocation, UAV's WPT power and trajectory, offloading and execution time & Lagrangian duality, SCA, BCD \\
\hline
\end{tabular}
\end{table*}
\renewcommand\arraystretch{1}

\section{UAV Deployment, Task Scheduling, and Load Balancing}
With the extensive employment of UAVs in civilian and military environments, effective UAV deployment for coverage and capacity maximization has become a challenging issue in aerial MEC \cite{JZhang2020}. Although there have already been much work dedicated to the UAV trajectory planning to provide high-quality edge computing services for users, UAVs, in these work, are generally deployed within a predefined region, and the locations of users are fixed. However, the users' distribution is usually time-varying in practice. For example, Fig. 1 in \cite{JrWang2020} demonstrates the Tencent service request distributions in Happy Valley, a theme park in Beijing, on October 1, 2018. It is shown that the service requests are highly nonuniformly distributed in the park, where users keep forming hot-spot areas at different places for different time in one day. Thus, the deployment of UAVs within a predefined region cannot satisfy the requirements of time-varying hot spot of user and service distributions. Besides, for the sake of saving energy and reducing the complexity of network management, the number of deployed UAVs should be as small as possible under the condition that all tasks can be executed within the tolerance latency \cite{YWang2020}. Nevertheless, the minimum number of deployed UAVs that can achieve full coverage while satisfying the task requirements cannot be obtained by trajectory planning. To address these problems, the deployment parameters of UAVs, including the horizontal position, flying height, and the number of UAVs, should be jointly optimized.

\vspace{-0.1in}
\subsection{UAV Deployment in 2D Space}
Generally speaking, multi-UAV deployment optimization is a NP-hard problem \cite{NHMotlagh2016}. The differential evolution (DE) is considered as an effective method to solve the multi-UAV deployment problem and can find a satisfactory solution \cite{LYang2020}. DE simulates the biological evolution. Through repeated iterations, the population that adapts to the environments can be retained. In \cite{LYang2020}, a DE-based multi-UAV deployment mechanism is proposed. The near-optimal 2D positions of UAVs can be obtained by iterations of the DE algorithm. By applying the proposed algorithm, the load balancing of UAVs can be guaranteed while the coverage constraint and the QoS requirements of IoT nodes are satisfied. Meanwhile, a deep reinforcement learning (DRL) algorithm is conceived for the task scheduling to improve the task execution efficiency. In addition, DE-based algorithm is also proposed in \cite{YWang2020, CJiang2020} to optimize the deployment of UAVs, with the objective to minimize the system energy consumption \cite{YWang2020}, and the load variance of UAVs \cite{CJiang2020}. The effectiveness of the DE-based algorithm in reducing the energy consumption and the number of deployed UAVs is demonstrated in \cite{YWang2020}.

In order to address the unexpected or temporary high traffic loads in hot-spot areas, H. El-Sayed \emph{et al.} \cite{HEl-Sayed2019} investigate the UAV deployment problem in vehicular networks, where UAVs are dynamically deployed to act as mobile edges. In the networks, a predefined group of UAVs is responsible for characterizing the traffic condition of each road link. By leveraging the bee swarm intelligence (BSI), a UAV deployment approach is proposed, which can achieve full network coverage without additional overhead or delay. Similarly, the on-demand UAV deployment for hot-spot areas is studied in \cite{JrWang2020}, with the aim to maximizing the number of served tasks. In order to optimize the hover locations of UAV-mounted edge servers among the dynamic hot-spot areas, a variable-sized bin-packing problem with geographic constraints is formulated. Since the bin-packing problem is NP-hard, an online dispatching scheme is proposed to find the hover locations of UAVs, and a greedy algorithm is developed to assign tasks to the UAV-mounted edge servers. Real-world experiments show that the proposed UAV dispatching scheme can provide timely services for more users in hot-spot areas while achieving a high resource utilization.
Different from the above literature, X. Wang \emph{et al.} \cite{XWang2021} focus on the economic viability of UAV-provided services (UPS) in hot-spot areas. It is found that if a UAV faces multiple hotspots, it should be deployed to only serve a single hotspot when considering the optimal pricing and energy allocation for each hotspot. For the case of multiple UAVs, a counterintuitive observation is that with UAVs' forking deployment to different hotspots, more UAVs may be deployed to the second-best hotspot instead of the expected first-best one for profit-maximizing purpose.

\subsection{UAV Deployment in 3D Space}
In the above-mentioned literature, the UAV deployment strategy is optimized in 2D space with fixed flying altitude. Actually, thanks to the maneuverability of UAVs, they can freely fly in 3D space. Although the higher altitude results in a stronger path-loss attenuation, the probability of LoS links between UAVs and ground devices increases with the UAVs' altitudes \cite{FCostanzo2020}. Besides, UAVs flying at higher altitude can cover larger ground areas. As a result, a DoF is lost if optimizing the UAV deployment only in 2D space.

As a further advance, S. Sun \emph{et al.} \cite{SSun2021} optimize the user association and 3D deployment of multiple UAVs, where the maximum operating time among all UAVs to complete the offloaded tasks is minimized. It is verified that the optimization of UAV deployment in 3D space can achieve better performance compared with the horizontal position only and vertical position only optimization. Afterwards, Z. Liao \emph{et al.} \cite{ZLiao2021} propose a novel UAV-assisted edge computing framework, named as HOTSPOT. In the framework, the UAV deployment positioning in 3D space is formulated as a maximum clique problem to minimize the average time delay, and an opportunistic offloading balanced scheme is also proposed based on the given UAV positions. It is demonstrated that the HOTSPOT framework with optimal UAV deployment in 3D space and opportunistic offloading balanced scheme can reduce the average offloading delay by up to 80\%. With a different objective from \cite{SSun2021} and \cite{ZLiao2021}, R. Islambouli \emph{et al.} \cite{RIslambouli2019} optimize the 3D deployment of UAVs to minimize the number of required UAVs. In the system, selected UAVs are equipped with MEC servers while the others serve as relays to help IoT devices offload tasks. The problem is formulated as a mixed integer program, and an efficient meta-heuristic solution is proposed. It is shown that the proposed solution achieves near-optimal performance in terms of the number of UAVs and can scale up to large IoT networks. In \cite{NAnsari202048}, the drone-mounted BS (DBS) aided heterogeneous network is investigated, where the free space optics communications are used for the tier-1 backhaul, and the in-band full-duplex communications are applied for the tier-2 backhaul. By decomposing the throughput maximization problem into three subproblems, the 3-D placement of UAVs, the user assignment and the resource allocation are optimized. It is found that the proposed algorithm is superior to that with fixed bandwidth and power assignment, and also outperforms the half-duplex DBSs in the tier-2 backhaul. In addition, the particle swarm optimization (PSO) \cite{LSun2021}, which is a population-based heuristic evolutionary algorithm and has shown impressive advantages compared to the genetic algorithm, is applied in \cite{CTang2020} to optimize the UAV placement in 3D space over the serving area. Simulations display that the proposed PSO-based algorithm can greatly reduce the number of deployed UAVs.

\renewcommand\arraystretch{1.25}
\begin{table*}[t]
\footnotesize
\caption{Summary of Contributions to UAV Deployment, Task Scheduling and Load Balancing}
\centering
\begin{tabular}{|m{0.75cm}<{\centering}|m{22pt}<{\centering}|m{4.0cm}|m{6.4cm}|m{3.5cm}|}
\hline
\bfseries 2D/3D Space & \bfseries Refs. & \bfseries Optimization Objective & \bfseries Optimization Variables & \bfseries Optimization Methods \\
\hline
~ &\cite{JrWang2020} & Number of served tasks & UAV positions and task assignment & Online dispatching and greedy algorithm \\
\cline{2-5}
~ &\cite{YWang2020} & System energy consumption & Number and positions of UAVs, task scheduling & DE, greedy algorithm\\
\cline{2-5}
~ &\cite{LYang2020} & Load balancing and execution latency & UAV positions, task scheduling & DE and DRL \\
\cline{2-5}
~ &\cite{CJiang2020} & Load balancing & Number and positions of UAVs, task offloading decision &	DE \\
\cline{2-5}
2D &\cite{HEl-Sayed2019} & QoS in vehicular networks & UAV positions & Swarm intelligence \\
\cline{2-5}
~ &\cite{XWang2021} & Total profit of UAVs & UAV positions, service price, hovering time and service capacity & Backward induction \\
\cline{2-5}
~ &\cite{YZheng2020} & Long-term profit of UAVs & UAV positions, communication and computation resources & Lyapunov optimization \\
\cline{2-5}
~ &\cite{XYu2020} & Throughput of task offloading & Coalition selection of ground nodes, UAV positions & Coalition formation game, Stackelberg game\\
\cline{2-5}
~ &\cite{WYou2021} & Coverage efficiency & Cluster head selection, UAV positions and transmit power & Penalty and BCD \\
\hline
~ &\cite{SSun2021} & Operating latency of UAVs & UAV positions and user association & SCA \\
\cline{2-5}
~ &\cite{ZLiao2021} & Average time delay of users & Number and positions of UAVs, task offloading decision & SGD \\
\cline{2-5}
3D &\cite{RIslambouli2019} & Number of deployed UAVs & Number and positions of UAVs, user association, task offloading decision & Meta-heuristic algorithm \\
\cline{2-5}
~ &\cite{CTang2020} & Coverage efficiency & Number and positions of UAVs & PSO \\
\hline
\end{tabular}
\end{table*}
\renewcommand\arraystretch{1}

\subsection{Joint Optimization of Number and Positions of UAVs}
As shown in \cite{RIslambouli2019} and \cite{CTang2020}, in addition to the horizontal and vertical positions of UAVs, the number of required UAVs is another big concern during the deployment, which should be as small as possible under the condition that the communication and computation requirements can be satisfied. Nevertheless, there have been limited literature optimizing the number of deployed UAVs. In \cite{RIslambouli2019}, the joint optimization of the number and positions of deployed UAVs in 3D space as well as the task offloading decision is studied, where the minimum number of deployed UAVs meeting various constraints is obtained. Afterwards, both the number and positions of UAVs are also considered in \cite{YWang2020}. With the objective to minimize the system energy consumption, a two-layer optimization method is proposed to jointly optimize the deployment of UAVs and task scheduling. Specifically, in the upper layer, a DE-based algorithm with an elimination operator is designed to optimize the deployment of UAVs, where the number of UAVs is adjusted by the elimination operator under the delay constraints, and the positions of UAVs are optimized by executing the DE. In the lower layer, the task scheduling is transformed as a large-scale 0-1 integer programming problem. A greedy algorithm is proposed to efficiently obtain the near-optimal solution. Besides, in \cite{CJiang2020}, the number and positions of UAVs are optimized by a method similar to \cite{YWang2020}, while its objective is to minimize the variance of the computation load of UAVs. Simulations show the proposed algorithm can effectively maintain the load balancing of UAVs.

Table 6 shows a summary of the major contributions to UAV deployment, task scheduling and load balancing.

\section{Interplay between Aerial MEC and Other Technologies}
\subsection{WPT-Integrated Aerial MEC}
Nowadays, WPT has been considered as a promising technology to provide stable and controllable energy supplies for users with energy harvesting (EH) modules \cite{JWang2020, LXie2021} to prolong their lifetime. In this context, WPT-integrated aerial MEC, as shown in Fig. 6, has received ever-increasing attentions from both industry and academia. For example, in \cite{FZhou2018-C}, an radio-frequency (RF) energy transmitter and an MEC server are implemented in a single UAV to provide wireless energy supplies and computing services for ground users, where the energy consumption of the UAV is minimized by an alternative algorithm based on SCA and Lagrange duality method. Afterwards, in \cite{FZhou2018}, the computation rate maximization for WPT-integrated aerial MEC is investigated under the energy-harvesting causal and the UAV's speed constraints. Simulation results reveal that the weighted sum computation bits of all users increase with the UAV transmit power. This is because with more transmit power from the UAV, users can harvest more energy to perform local computing and task offloading. In \cite{YDu2019}, a new TDMA-based workflow model is proposed for the wirelessly-powered aerial MEC systems. By applying the flow-shop scheduling techniques, the user association, computing resource allocation, UAV hovering time, wireless powering duration and the service sequence of users are jointly optimized to minimize the energy consumption of UAVs.

\begin{figure}[t]
\centering
\includegraphics[width=3.65in]{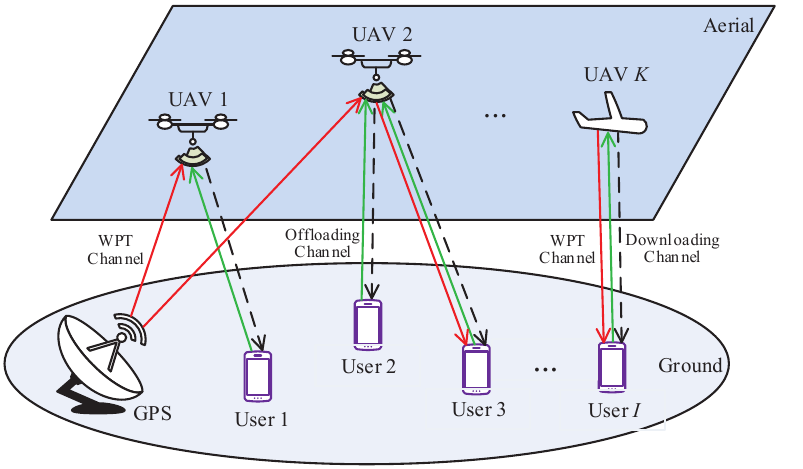}
\caption{Illustration of the WPT-integrated aerial MEC.}
\vspace{-0.1in}
\end{figure}

Thanks to the broadcast nature of wireless links, the transmit power from the UAV can be used to charge not only active users that have computation tasks, but also the idle users. Thus, in order to avoid the waste of UAV's transmit power, Y. Liu \emph{et al.} \cite{YLiu2020} assume that the idle users can also harvest energy from the UAV and cooperatively assist active users to compute their tasks. By jointly optimizing the resource allocation and the UAV's trajectory using SCA and a decomposition and iteration (DAI)-based algorithm, the energy consumption of the UAV is effectively reduced while the active users' tasks can be completed within the given tolerance latency. Additionally, the broadcast nature of wireless links also results in the near-far effect, which means those users located far away from the UAV harvest less energy while they need to communicate with the UAV from a farther distance. Aiming to alleviate this effect, W. Lu \emph{et al.} \cite{WLu2020, WLu2020IET} utilize the user which is closer to the UAV to act as a relay to assist the other user to offload tasks to the UAV. Accordingly, the communication distance of the farther user can be decreased, which dramatically saves the transmission energy harvested from the UAV.

Besides the low-power ground users, the UAV may also face battery capacity limitation. Therefore, the EH module can also be deployed at the UAV for harvesting energy from the ground power station (GPS) or other power suppliers \cite{YLiu2020I, YLiu2021-1}. Under this circumstance, Y. Liu \emph{et al.} \cite{YLiu2020I} investigate the service utility maximization problem by jointly optimizing the UAV's trajectory, computation offloading decisions and offloading duration. It is found that the service utility first increases sharply and then slow down with the increase of the transmit power of the GPS. This is because the increase of the GPS's transmit power at the beginning can lead to a higher level of harvested energy for the UAV. Thus, the UAV can use the harvested energy to adjust its trajectory to find a better channel to undertake more offloading tasks from users. Nevertheless, when the suboptimal offloading duration and UAV's trajectory have been achieved, the increase of the UAV's harvested energy will not have a distinctive impact on the service utility.

Interestingly, while the UAV acts as an information relay and a MEC server, it can also serve as an energy relay to broadcast energy harvested from the AP to UEs. Based on these roles of an UAV, the weighted sum completed task-input bits of UEs is maximized in \cite{XHu2020}. A three-step BCD algorithm is proposed to optimize the task allocation and UAV's WPT power, offloading and execution time, as well as UAV's trajectory. It is demonstrated that the UAV's trajectory highly depends on the location of the AP. This is because the AP plays the roles of both a powerful MEC server and the original energy source, which incentivizes the UAV to move closer to the AP in order to harvest more energy and reduce the energy consumption for relaying task bits to the AP.

\subsection{Physical-Layer Security}
During the offloading process in aerial MEC systems, there may be malignant potential eavesdroppers to intercept the communication information \cite{WLu2021}, leading to a high risk in data security and privacy. To prevent such risks, physical-layer security technology has been widely applied, which can provide a confidential data transmission by exploiting the physical-layer features without using secret keys. In \cite{TBai2019}, a physical-layer security model for aerial MEC is proposed, where the AP relying on full-duplex technique receives the offloading task from the single UAV, and also acts as a jamming source to impose artificial noise on the eavesdroppers. The injection of artificial noise makes the eavesdroppers' SINR lower than the AP receiver's SINR. Thus, the eavesdroppers is unable to decode the transmitted confidential messages from the UAV to the AP. Then, energy-efficient computation offloading schemes are proposed for both active and passive eavesdroppers to minimize the energy consumption of UAVs while satisfying the security requirements. In \cite{DHan2020} and \cite{YZhou2020}, a full-duplex legitimate UAV server with double antennas is deployed to receive the offloading task bits from the ground users and simultaneously send jamming signals to the eavesdropper UAVs, as shown in Fig. 7 (a). In order to further enhance the security, the non-offloading users also send jamming signals to the eavesdropper UAVs. The security capacity is optimized with joint considerations of the UAV positions, transmit power of UAV and users, task offloading ratio, and user association. With a similar model, the secure computation efficiency is maximized in \cite{PAmos2021} by a two-stage alternative optimization algorithm, where the UAV's trajectory and transmit power of ground users are jointly optimized.
It is worth noting that the implementations of physical-layer security in \cite{TBai2019, DHan2020, YZhou2020, PAmos2021} all rely on the full-duplex technique, where the self-interference (SI) cancellation is a big challenge. It has been shown in \cite{YZhou2020} that the secrecy capacity keeps decreasing with the increase of SI efficiency, since a higher SI efficiency results in more residual SI power.

In order to overcome the limitations of full-duplex technique, Y. Xu \emph{et al.} \cite{YXu2021} investigate the security of aerial MEC systems where dual UAVs are deployed. As shown in Fig. 7 (b), it is assumed that one of the dual UAVs helps the ground users to compute the offloaded tasks and the other one acts as a jammer to suppress the vicious eavesdroppers. The minimum secure computing capacity are maximized for both TDMA and NOMA schemes by jointly optimizing the communication resources, computation resources, and UAVs’ trajectories. Besides, in \cite{SLai2021}, when a eavesdropper UAV tries to intercept the offloading transmission from MDs to computational AP, it is assumed there is no jammer and the channel condition of eavesdropper is naturally worse than that of the offloading links. Thus, the offloading transmission has a secrete data rate. After deriving the secrete offloading rates of MDs, the weighted sum of latency and energy consumption is minimized by applying DQN.

\begin{figure}[!t]
\centering
\subfigure[]{
\includegraphics[width=1.58in]{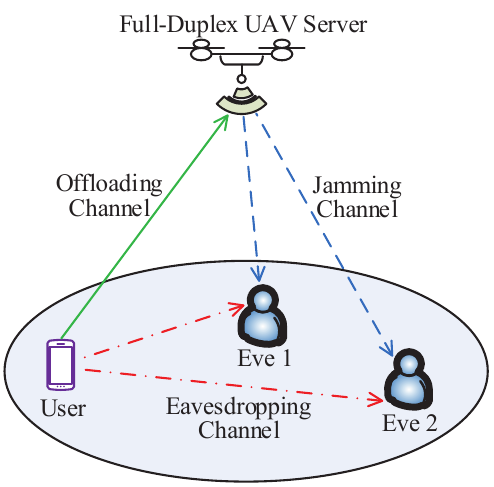}}
\subfigure[]{
\includegraphics[width=1.72in]{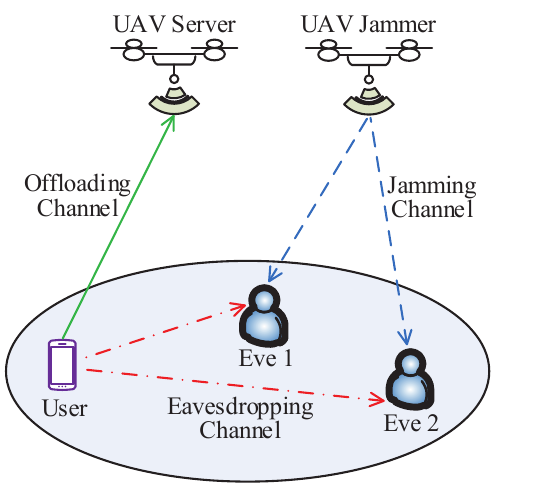}}
\caption{Illustrative model of physical-layer security in aerial MEC (a) with a full-duplex UAV server; (b) with dual UAVs (one UAV server and one UAV jammer).}
\vspace{-0.1in}
\end{figure}

\subsection{RIS}
In the current paradigm of aerial MEC optimization, the random radio environment is uncontrollable and thus it is not considered in the problem formulation. Recently, a novel concept of reconfigurable intelligent surface (RIS) has been proposed to construct a smart radio environment. RIS comprise a large array of passive scattering elements. By a joint phase control of all scattering elements, the reflecting phases and angles of the incident signals are tuned in a customizable way such that the reflected signals can be added coherently to improve the received signal power. For aerial MEC, in spite of the high possibility of LoS links due to the higher cruising altitudes of UAVs, the air-ground connections might be frequently blocked by ground obstacles in practical environments like urban areas. In order to further improve the propagation environment, RIS-assisted UAV communications and aerial MEC have been recently proposed. RIS can be deployed on the surface of buildings to reflect the signals between UAV and ground users \cite{SLi2021}. It can also be equipped on the UAV to reflect the signals between BS and users. Nevertheless, an urgent challenge is the continually moving of UAVs may render the IRS passive beamforming ineffective, since the phase-shift design of IRS is coupled with UAVs' trajectories \cite{SLi2021}. To address this challenge, in \cite{SLi2020}, the joint UAV trajectory and RIS passive beam-forming optimization is divided into two subproblems to maximize the data transmission rate in RIS-assisted UAV communications. A closed-form phase-shift solution for any given UAV trajectory is derived to achieve the phase alignment of the received signals, and a suboptimal UAV trajectory is designed by SCA with the optimal phase-shift solution.

Nowadays, there has been few work studying the RIS-assisted aerial MEC yet. In \cite{HMei2021}, one rotary-wing UAV is deployed to work as a MEC server, and a RIS is deployed on the the building surface to reflect the signals from users to the UAV during the task offloading. In order to minimize the energy consumption of the UAV while satisfying the QoS requirement of tasks, the UAV trajectory, task offloading and cache, as well as the phase-shift design of the RIS are jointly optimized by SCA. It is verified by simulations that RIS passive phase-shift can improve the transmission environment and thus reduce the energy consumption of the UAV.

\section{ML-Driven Optimization for Aerial MEC}
In previous sections, the optimization for aerial MEC is mostly based on traditional approaches, such as the convex optimization, game theory, and heuristic algorithm. However, these optimization problems usually suffers from the exceedingly high levels of dimensionality, due to the large number of parameters. Thus, it is generally difficult to derive the optimal solution with lower complexity. Although the originally formulated problems can be decomposed into several subproblems to reduce the computation complexity, the system performances are heavily limited \cite{XWang2020}. More importantly, traditional optimization approaches rely on expert knowledge on the environments to model the dynamic systems. Unfortunately, in practice, it is difficult to acquire accurate knowledge due to the complexity and uncertainty of dynamic systems. When considering 5G and IoT networks with massive amounts of devices, these conventional approaches are particularly inefficient or even infeasible. To address this concern, ML has been considered as a promising technique to solve the wireless control and resource management problems by adaptive modelling and intelligent learning without any manual manipulations \cite{FJiang2021, HChang2021}. Different from conventional optimization methods, ML has continuous and powerful learning ability, which can derive real-time and near-optimal decisions with lower complexity even for highly dynamic environments and large-scale networks. Actually, for the optimization of complex systems with high-dimensional parameters, ML-based algorithms are often the only feasible alternative \cite{TRodrigues2020}.

In general, ML techniques can be categorized into several tiers, i.e., supervised learning, unsupervised learning, reinforcement learning (RL), deep learning (DL), deep reinforcement learning (DRL), and federated learning (FL) \cite{ZUllah2020, NCLuong2019}. Supervised and unsupervised learning have been widely applied for the computation offloading optimization in MEC systems \cite{XDiao2018, AKSangaiah2019}. However, supervised learning can only be employed when there are sufficient labeled data. In contrast, the unsupervised learning is able to extract key features of the data for better prediction, where there is not labeled output or target values in the dataset used for training. In the field of aerial MEC, supervised/unsupervised learning can be used for user clustering to provide better service, while the RL, DRL, and FL based approaches are particularly popular for the resource allocation \cite{AShakarami2020}. In Table 7, we summarize the major contributions to the ML-driven optimization for aerial MEC systems.

\subsection{Reinforcement Learning}
Under the uncertain and stochastic environments of UAV-enabled aerial MEC systems, the decision-making problems can be typically modeled as a Markov Decision Process (MDP), where the next state of the system is only related to the current state, regardless of the historical states \cite{MWang2020}. As an important branch of ML, RL is an effective tool to address the MDPs \cite{SZhu2020}. RL is a control-theoretic trial-and error learning method with rewards and punishments of a sequence of actions \cite{AShakarami2020}. There are five core components in the RL: environment, agent, state, action, and reward. During the process of RL, the agent interacts with the environment and learn to make decisions, seeking to find optimal actions to maximize the immediate reward given the current state. Note that the reward in the RL is not acquired through the training set directly. Instead, it is generated dynamically from the feedback of the interaction between the agent and the environment \cite{SLai2021}. In \cite{AAsheralieva2019}, MEC servers are installed at terrestrial stationary BSs and UAVs, where UAVs are deployed as quasi-statinary BSs. Given the high randomness of MEC environment, the computational offloading problem cannot be addressed well by centralized optimization. Thus, the problem is formulated as a two-level hierarchical game model. Then, a MDP is defined and the exact and approximate RL algorithms are developed to find the optimal strategy for each BS. By applying the proposed algorithms, the long-term payoff of each BS is maximized and the strategies of BSs are in the mixed-strategy NE. In \cite{GFaraci2020}, the computation offloading management of multiple UAVs are modeled as a MDP to enable RL and optimize the selection of system parameters. Besides, in \cite{GFragkos2020}, RL is introduced to determine the amount of offloading data from UAVs to MEC servers, where the UAVs are utilized as mobile users, and the task offloading is modeled as a non-cooperative game to maximize the utility of each UAV.

In the field of RL, Q-learning is the most effective tool and widely used in the literature \cite{NCLuong2019}. For instance, in \cite{YLiao2020}, multiple UAVs and terrestrial BSs are deployed to provide edge computing services for industrial IoT (IIoT). In order to maximize the revenue of mobile network operators (MNOs) and decrease the IIoT operators' economic cost, a dynamic pricing problem where each UAV interacts with IIoT MDs and adaptively announces the service price is modeled as a discrete finite MDP. Thanks to the model-free feature of Q-learning, it is utilized to efficiently solve the decision-making problem, where the reward and transition probability from one state to another state are not required \cite{AFeriani2021}.
Similarly, in \cite{MWang2020}, with UAVs acting as flying edge servers, a MUs-Edge-Cloud network architecture is designed to provide computing services for MUs. The problem which minimizes the weighted sum of task completion time and MUs' energy consumption is formulated as a MDP, and then an efficient Q-learning based computation offloading algorithm (QCOA) is proposed. The proposed algorithm can decrease the task completion time and energy consumption of MUs, while significantly reducing the complexity of solving the optimization problem.

Besides, in \cite{XMa2020}, a rotary-wing UAV provides edge computing services for IoT devices and plays the role of agent. A single-agent Q-learning algorithm is proposed to jointly optimize the UAV's trajectory and the ratio of offloading tasks to minimize the total system delay. In \cite{MWang2020-C}, the power and computation resource allocation problem for multi-UAV-enabled MEC network is formulated as a MDP. Then, a fully decentralized multi-agent Q-learning algorithm is proposed to solve the formulated problem in a distributed manner, where UAVs are taken as the learning agents, and all agents share the common Q-learning structure. For this fully decentralized algorithm, since each UAV make decisions only based on its local observations while the information is not exchanged among different UAVs, the complexity of the proposed algorithm can be significantly reduced and is easily scalable to a high number of agents. However, the absence of cooperation between the UAV agents may bring convergence and stability problems.

\subsection{Deep Reinforcement Learning}
For large-scale systems with high-dimensional state and action spaces, the learning process for RL may take a long time to converge to the best policy. To overcome the limitations of RL, DRL, which leverage the advantages of deep neural networks (DNNs) \cite{FJiang2020, MMukherjee2020} to train the learning process, has been proposed. A DNN which is referred to as an artificial neural network (ANN) with two or more hidden layers is widely used for optimization problems to find the solutions by manipulating mathematics appropriately \cite{AShakarami2020}. Thanks to the power of DNNs for estimating the associated functions in RL, DRL can provide faster learning speed and is more efficient for large-scale systems with high-dimensional state-action spaces \cite{LWang20211}. In the following, we will discuss the representative literature whose approaches fall into the field of DRL in the optimization of aerial MEC systems.

\renewcommand\arraystretch{1.3}
\begin{table*}[t]
\footnotesize
\caption{Summary of Contributions to the ML-Driven Optimization for Aerial MEC}
\centering
\begin{tabular}{|m{22pt}<{\centering}|m{1.2cm}|m{3.7cm}|m{4.2cm}|m{4.0cm}|m{1.4cm}|}
\hline
\bfseries Refs. & \bfseries Agents & \bfseries States & \bfseries Actions & \bfseries Rewards & \bfseries Learning Techniques \\
\hline
\cite{AAsheralieva2019} & MEC servers & Task parameters of users and backlog of MEC server queue & Coalition formation, task offloading and processing decisions & Long-term payoffs of stationary BSs and UAVs & ~ \\
\cline{1-5}
\cite{GFaraci2020} & System controller & Behavior of the areas controlled by UAVs, number of jobs in UAVs' queues & Number of active computing elements and job allocation between UAVs & Weighted sum of power consumption, job loss and incurred delay & RL \\
\cline{1-5}
\cite{GFragkos2020} & UAVs & Task profile, computation capacities of UAVs and servers & Amount of offloading task data & Utility of UAVs & ~ \\
\hline
\cite{YLiao2020} & Edge servers & Offloading decisions of MDs & Service price & Weighted sum of MNO' and IIoT operators' revenue & ~ \\
\cline{1-5}
\cite{MWang2020} & MUs & MUs' weighted time and energy consumption & Offloading decisions & Weighted sum of completion time and MUs' energy consumption & Q-Learning \\
\cline{1-5}
\cite{XMa2020} & UAV & Processing status of IoT devices waiting to be served by the UAV & Service order of the UAV and the offloading ratio & Total system delay  & ~\\
\cline{1-5}
\cite{MWang2020-C} & UAVs & Task property, UAV positions, SINR & Energy and computation resource allocation & Power and computation resource consumption & ~ \\
\hline
\cite{FXu2020} & Central controller & Channel quality, computation capability & User association, computation resource allocation & Resource utilizing efficiency & ~ \\
\cline{1-5}
\cite{RKrisna2020} & UAV & Device condition, task position and priority & UAV trajectory & Communication and computation capacities for static devices & DQN \\
\cline{1-5}
\cite{SLai2021} & System controller & Offloading strategies, bandwidth allocation, users' transmit power & Offloading ratio, bandwidth allocation, users' transmit power & Weighted sum of latency and energy consumption & ~ \\
\hline
\cite{HWang2020} & Cloud controller & Task arrival, channel conditions, energy buffer queue & Offloading ratio to the UAV, offloading decision to the MEC server & Weighted sum of delay, energy consumption, and bandwidth cost & DDQN \\
\cline{1-5}
\cite{QLiu2021} & UAV & Locations of users and UAV, channel gains, energy of UAV & User association and UAV trajectory & The difference of system utility and UAV energy consumption & ~ \\
\hline
\cite{LWang2021} & Central controller & Coordinates of all UAVs & UAVs' trajectories & Energy consumption of all UEs & SADDPG \\
\hline
\cite{HPeng2021} & MeNB, UAVs & Coordinates of vehicles and UAVs, task property & Vehicle association, spectrum, computing, and caching resource allocation & Satisfaction degrees of delay and caching resources requested by vehicles & ~ \\
\cline{1-5}
\cite{LWang20211} & UAVs & Coordinates of UAVs, UAV distances, accumulated served time of UEs and UE-load of UAVs & UAVs' trajectories & Fairness of UAVs' load and the served time of UEs, overall energy consumption of UEs & MADDPG \\
\cline{1-5}
\cite{AMSeid2021} & UAV cluster heads & Task profile and queue buffer of IoT devices, SINR, UCH status & Device association, computation resource and transmit power allocation & Weighted sum of task processing delay and energy consumption & ~ \\
\hline
\cite{YNie2021} & UEs & Offloading decisions, transmit power of UEs, system power consumption & Offloading decisions, transmit power of UEs & Weighted sum energy consumption of UEs and UAVs & FL \\
\cline{1-5}
\cite{DRahbari2021} & UAVs & Computation resources & CPU frequency & Weighted sum of energy consumption and computation time & ~ \\
\hline
\end{tabular}
\end{table*}
\renewcommand\arraystretch{1}

\textbf{\emph{1) DQN:}} Although Q-learning has been applied in the above literature to address the MDPs, it is not well-suited to complicated system models with high-dimensional state-action spaces. This is because Q-learning algorithms make decisions based on the Q-table. When the state-action spaces in practice are large, the cost of managing the Q-table is unaffordable. As a result, the Q-learning algorithm may not be able to find the optimal policy in high-dimensional systems, which is known as the curse of dimensionality \cite{TRen2021}. In order to address this problem, deep versions of the Q-Learning algorithm such as Deep Q-Network (DQN) have been introduced, which falls into the field of DRL \cite{AFeriani2021}. DQN embraces the advantages of DNNs to estimate the approximate value of tabular items, thereby accelerating the learning speed and outperforming the traditional tabular based Q-learning algorithms. Furthermore, a pair of additional features, i.e., experience replay and target deep networks, make DQN more powerful and versatile \cite{FXu2020, VMnih2015}.

Recently, DQN and its variants have become the most popular online Q-learning algorithms with impressive performance in many communication applications. For instance, in \cite{FXu2020}, a space-air-ground-sea integrated network is introduced to facilitate the edge computing for maritime service. A DRL-based solution, i.e., DQN, is developed to solve the joint user association and computation resource allocation problem, and the effectiveness of DQN in improving the resource utilizing efficiency is shown by simulations. In \cite{RKrisna2020}, a single UAV is deployed as an edge node to provide service to the edge network. The trajectory of UAV is optimized by DQN to maximize the communication and computation capacities for static devices. In addition, S. Lai \emph{et al.} \cite{SLai2021} focus on the application of DQN in physical-layer security, where a UAV is the eavesdropper in the MEC systems to overhear the secure offloading tasks from the user to the computational AP. In order to decrease the latency and energy consumption while ensuring the physical-layer security, a DQN-based strategy is proposed to address the high-dimensional state-action spaces and automatically solve the optimization problem.

\textbf{\emph{2) DDQN:}} Although DQN has shown impressive performance in various communication applications, its main disadvantage is the overestimation problem. Due to the inherent defects of overestimation on the true Q-values, DQN suffers from the maximization bias which may lead to an unstable learning process. To eliminate the typical overestimation problem caused by DQN, double learning technique, i.e., double DQN (DDQN), has been proposed. Unlike DQN which directly chooses the maximum Q-value of each action in the target network, DDQN selects the action corresponding to the maximum Q-value for the next state in the predicted network, and then calculates the Q-value in the target network according to the selected action \cite{HWang2020}. In other words, DDQN decouples the action selection from the action evaluation to generate a more accurate state-action value function than DQN, which can effectively address the overestimation problem \cite{QLiu2021}.

Due to the effectiveness of DDQN, some applications of DDQN have been introduced recently to address the trajectory design and resource allocation problems for aerial MEC systems. For example, in \cite{QLiu2021}, DDQN with a single agent is applied to optimize the UAV trajectory and user association, where the average long-term system reward is maximized and the QoS constraint is satisfied. Simulations reveal that the proposed DDQN-based algorithm converges much faster and obtains a higher throughput than conventional RL and DQN algorithms. Similarly, H. Wang \emph{et al.} \cite{HWang2020} adopt DDQN to investigate the computation offloading strategy for multi-UAV aerial MEC systems. To design the DDQN-based algorithm, the task offloading ratios of users are first discretized. Then, the K-Means algorithm is used for task category classification to reduce the dimension of action space and accelerate the learning process. Finally, with the Follow-Me Cloud (FMC) controller as the agent which knows the global information of the MEC model, DDQN is applied to find the near-optimal offloading policy to minimize the weighted sum of delay, energy consumption, and bandwidth cost.

\textbf{\emph{3) Single and Multi-Agent DDPG:}} Despite the powerful ability of DQN to solve complex problems with high-dimensional state spaces, it cannon handle the continuous or high-dimensional action space because choosing the best action that maximizes the Q-function is quite difficult \cite{AMSeid2021}. It is worth noting that when studying the optimization of aerial MEC systems, some literature only consider the discrete action space. For example, in \cite{FXu2020}, the considered action space is which access point and MEC server should be assigned to a specific user to provide network access and computation services. Apparently, the action space in this paper is discrete and DQN can be readily applied.
Unfortunately, in the UAV-enabled aerial MEC systems, there are often many optimization problems with continuous variables, such as the UAV trajectory control, computation and communication resource allocation, etc. To deal with these kinds of problems, some literature try to first discretize the action space, and then develop DQN/DDQN-based algorithms. In \cite{XMa2020, MWang2020-C}, and \cite{QLiu2021}, they all assume the UAV can only hover over several pre-defined fixed positions, thereby discretizing the UAV's trajectory. Likewise, in \cite{HWang2020}, the offloading ratio of computation tasks is discretized in order to apply DDQN to optimize the partial offloading problem. However, disassembling the action space into discrete and efficient action vectors is often difficult, especially when the size of action vectors is large \cite{HPeng2021}. Besides, such quantization also throws away some useful information that can be critical in finding the optimal policy \cite{XWang2020}.

\begin{figure}[t]
\centering
\includegraphics[width=3.5in]{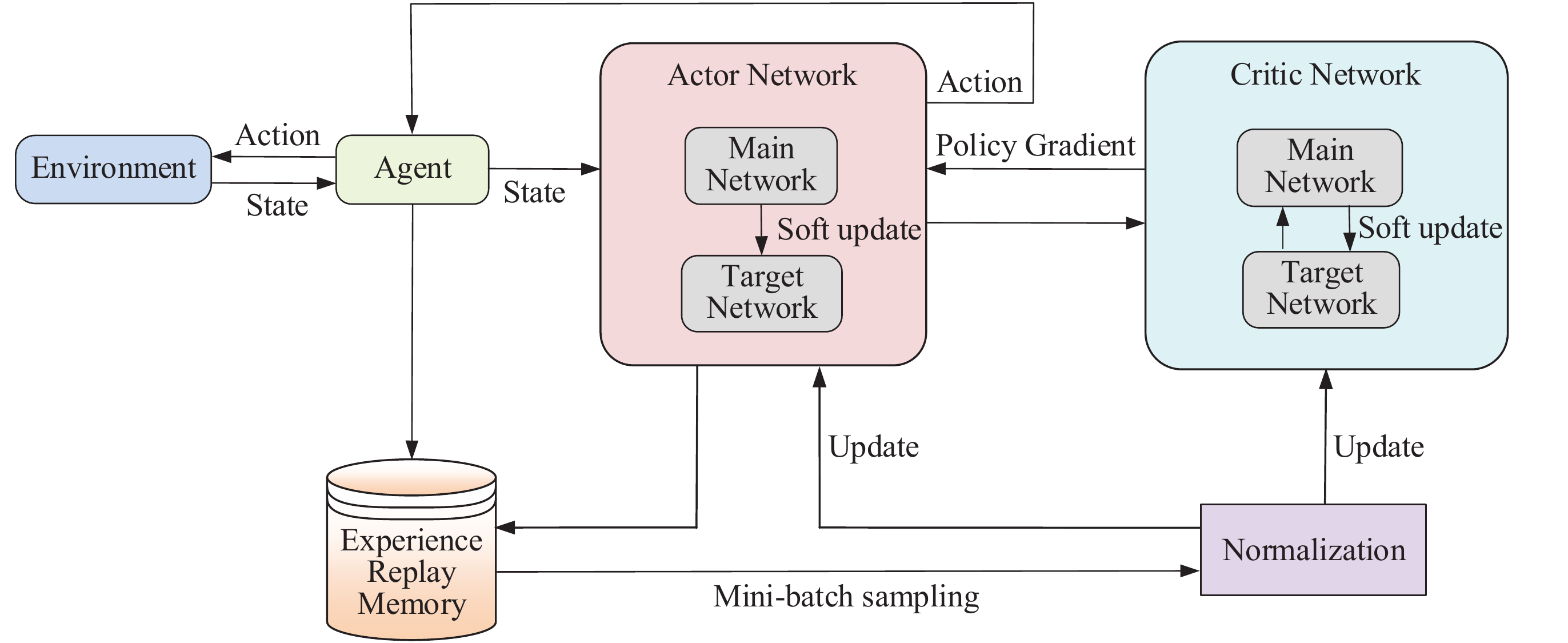}
\caption{Illustration of the DDPG model.}
\vspace{-0.1in}
\end{figure}

Therefore, in order to solve the continuous control problem without descretizing the action space, deep deterministic policy-gradient (DDPG), which is an enhanced version of the deterministic policy-gradient (DPG) algorithm, has been proposed by DeepMind \cite{TPLillicrap2015}. As illustrated in Fig. 8, DDPG integrates the actor-critic approach \cite{VRKonda2000} into DRL and include two DQNs, i.e., the actor network and critic network. Similar to DQN, target networks and experience replay are also adopted to improve the stability of DDPG. Recently, DDPG has been widely applied to optimize the aerial MEC with continuous action space.

In \cite{LWang2021}, UAVs are equipped with MEC servers to provide computation resource for ground UEs. The user association, resource allocation and UAVs' trajectories are optimized to minimize the energy consumption of all UEs. After addressing the user association and resource allocation problems by B\&B method, DDPG is applied to optimize the UAVs' trajectories, and a deep Reinforcement leArning based Trajectory control algorithm (RAT) is proposed. In RAT, a single agent is deployed in the control center to interact with the environment and train the actor and critic networks. After adequate training, the central agent sends the actions generate by actor network to all UAVs in each time slot. The proposed RAT derived from DDPG is insusceptible to the initial points of UAVs and can execute fast decision-making once the training has been completed.

The single anent learning as in \cite{LWang2021} relies on the central controller, which collects the global information and perform the training process. In practice, however, the global information collection is either costly or infeasible for large-scale and real-time systems. Without the intervention of a central controller, multi-agent learning plays an important role in the performance optimization. Multiple agents can either compete or collaborate to accomplish tasks with accuracy and efficiency \cite{GFragkos2020}. Nevertheless, for multi-agent learning, if each agent decides its action independently of the other agents, the resultant joint action may not be optimal. Hence, the coordination among multiple agents is the biggest concern in order to make sure the same joint action profile can be chosen \cite{AFeriani2021}. To establish collaboration among agents, one of the effective schemes is the centralized training and decentralized execution (CTDE). For CTDE, there is a central controller which gathers information about the agents to perform centralized training, while each agent chooses the best action only based on its local observation on the environment to conduct decentralized execution. More details on the coordination of multi-agent learning can be found in \cite{AFeriani2021}. With the framework of CTDE scheme, multi-agent DDPG (MADDPG) has been widely applied in the optimization of aerial MEC systems.

In the context of MEC- and UAV-assisted vehicular networks, MADDPG is applied in \cite{HPeng2021} to address the vehicle association and resource allocation problems with continuous action space. It is assumed that MEC servers are mounted at a macro eNodeB (MeNB) and multiple UAVs, where UAVs fly over the coverage area of the MeNB and collaborate with MeNB to provides computing resources to vehicles. The CTDE scheme is adopted in the framework of MADDPG as in \cite{LWang20211}. To be specific, the MeNB and UAV agents first centrally train the MADDPG model with aware of the other participants' actions. Then, during the online decentralized execution stage, each agent can obtain its action by the local observation only. Simulation results show that the convergence speed of the MADDGP-based scheme is comparable to that based on single-agent DDPG (SADDPG). However, in terms of the QoS satisfaction ratios, the MADDGP-based scheme is superior to the SADDPG-based one, which verifies the performance advantages of MADDGP.

In \cite{AMSeid2021}, the mobile UAVs work in clusters autonomously to act as ABSs and provide computing resource to IoT devices during emergencies. Each UAV cluster is controlled by a UAV cluster head (UCH), which communicates with the central network controller (CNC) and IoT devices. With CNC and UCHs acting as the agents, MADDGP is used to design the collaborative computation offloading and resource allocation scheme to minimize the energy consumption and computational delay.
To promote the coordination among agents, this paper adopts the CTDE scheme. Precisely, the centralized agent (i.e., CNC) first collects global information from the environment and trains the global observations in a centralized manner. Then, the trained data is distributed to UCHs. During the testing process, each UCH agent independently finds the optimal action based on its local observations only and sends the optimal action to IoT devices. Simulations reveals that the proposed MADDPG based scheme outperforms the DQN and greedy based offloading in stochastic dynamic environments.

As the trajectories of UAVs are not considered in \cite{AMSeid2021}, L. Wang \emph{et al.} \cite{LWang20211} focus on the trajectory optimization by MADDPG in UAV-enabled MEC systems. In this paper, it is assumed that each UAV acts as an agent, and all agents can exchange their observations and actions simultaneously with each other. Then the CTDE scheme is applied to promote the coordination of agents. After the centralized training, both the model and the network parameters are saved in UAV agents for testing. Given the trajectories of UAVs obtained by MADDPG, a low-complexity approach is proposed to optimize the offloading decisions of UEs. It is shown in simulations that the UAV trajectories designed by MADDPG outperform the random and circle trajectories.

\subsection{Federated Learning}
FL is an AI approach that enables users to collaboratively learn a shared model with data maintained on their own devices \cite{MChen2021}. Since FL enables agents to train their data locally and transmit the training parameters instead of sense
data among agents, it addresses concerns such as the privacy of agents and limited radio resources. As a contrast, for centralized learning approaches, the large-scale communications among agents are fully exposed to each other without privacy protecting. To exploit the benefits of FL, a multi-agent federated reinforcement learning (MAFRL) algorithm with integration of FL and DRL is proposed in \cite{YNie2021} to optimize the resource allocation and user association in UAV-aided MEC systems. The UEs fast learn models by keeping their data training locally, which can reduce the time cost of the data transmission and protect the privacy compared with the centralized MARL framework. Similarly, a federated DRL-based online task offloading and resource allocation algorithm is proposed in \cite{LZang2022} for MEC networks, where the DRL is executed in users, and FL uses the distributed architecture of MEC to aggregate and update the training parameters. Note that the major disadvantage of FL lies in local network failure affects the global network \cite{ZYang2021}. Thus, performing FL tasks may not be feasible due to the unavailability of terrestrial communications. To address this issue, the UAVs are used to perform FL tasks in \cite{QVPham2022} and provide sustainable FL solutions by leveraging the controllable maneuverability of UAVs. In addition, collaborative computing in a swarm of drones is considered in \cite{DRahbari2021}. A FL-based fast and fair offloading strategy is proposed, where the data transfer among agents in dynamic networks is not required. Thus, considerable latency and energy consumption can be reduced when designing the computation offloading strategy.

\subsection{Other Learning Techniques}
\textbf{\emph{1) Active Learning:}} Active learning (AL) is a form of semi-supervised ML where the algorithm can choose which data it wants to learn from \cite{SLi2020-C}. With this approach, the program can actively query an authority source, either the programmer or a labeled dataset, to learn the correct prediction for a given problem. The goal of AL is to speed along the learning process, especially when there is not a large labeled dataset to practice traditional supervised learning methods. One of the typical applications for AL is the object detection, which can produce similar results as supervised learning, with a fraction of the human involvement. For example, as the failure of UAV control system may cause abnormal flight of the UAV, the use of abnormal data from UAV sensors for fault detection and diagnosis has very important practical significance. With those less labeled and more unlabeled data collected by the UAV sensors, a classification method that combining AL and improved semi-supervised support vector machine is proposed in \cite{DPan2020}. Experiments show that the classification method will achieve the ideal classification effect. Besides, in \cite{BKellenberger2019}, given a set of images acquired with an UAV, an animal detector is trained by AL such that it can be reused for repeated acquisitions. It is found that less than half a percent of oracle-provided labels are enough to find almost 80\% of the animals in challenging sets of UAV images.

\textbf{\emph{2) Online Learning:}} Online learning is a method of ML where data becomes available in a sequential order and is used to update the best predictor for future data at each step, as opposed to batch learning techniques which generate the best predictor by learning on the entire training data set at once \cite{XWang2022}. To cope with the time-varying problems with unknown statistics on-the-go, the online learning algorithm has been applied to the task offloading decisions, server selection, and service caching in MEC \cite{ZSun2020, DRen2022}. Besides, in \cite{QCui2019}, stochastic online learning, and its promising applications to MEC are discussed. Stochastic online learning learns from the changes of dynamic systems rather than training data, decouples tasks between time slots and edge devices. By leveraging the stochastic online learning, the time-averaged operational cost of MEC can be asymptotically minimized with the increase of learning time.

\section{Open Problems and Future Directions}
In spite of the potentials of UAV-enabled aerial MEC systems, there are still many open problems to be addressed. In this section, we discuss the research opportunities and highlight interesting directions deserved more efforts for future work.

\subsection{Space-Air-Ground Integrated MEC}
Flying at high altitudes, UAVs have a better coverage and can provide reliable and seamless services with LoS connections. However, UAVs may not be able to perform critical tasks alone in some harsh environments. Thus, the assistance from other networks like satellite networks is indispensable \cite{NCheng2019, BMao2021}. As the coverage areas of satellites are usually large, they can relay data and control information between UAVs and remote ground networks. Besides, the satellites can be equipped with MEC servers to enhance the edge computing capability.

There are several challenges to construct a space-air-ground integrated MEC system. The most significant disadvantages of satellites are the propagation loss and delay. Meanwhile, the high operational cost of satellite communications may prevent their wide applications. Fortunately, recent improvements of satellite technologies make the LEO satellites much more economical, and the propagation delay can be reduced to 1-4 ms due to the low orbit altitude \cite{ZZhang2019}. Nevertheless, the high mobility of satellites and UAVs changes the channel state continuously and also results in frequent handover, which makes the optimization of space-air-ground integrated MEC systems difficult. More importantly, it is challenging to operate UAVs and satellites in a heterogeneous networks to ensure reliable signaling exchange and data transmission among different participants \cite{NCheng2018}. In order to cope with the interworking issues and achieve the expected benefits of space-air-ground integrated MEC systems, a comprehensive mechanism for the cooperative communication and computing, resource allocation, and cost-efficient protocol design deserves further studies.

\subsection{Interference Management}
Due to the high UAV altitude, the air-ground channels typically experience limited scattering, and thus have a higher-probability LoS link compared to the terrestrial channels. Although the LoS-dominant air-ground channels can provide more reliable wireless connectivity for task offloading and computation result downloading, they also cause strong air-ground interference, especially for the networks with a high density of UAVs. For example, when the UAV serves as a user, the task offloading from the UAV to its associated GBS may pose severe interference to other GBSs occupying the same spectrum due to the strong signals brought by LoS channels, and thereby result in the probabilities of error in signal detection. To cope with the performance degradation, interference management becomes one of the most critical issues in aerial MEC systems.

The dynamic and complexity nature of aerial MEC calls for new techniques to deal with the interference challenge. As a large number of UAVs is expected to be deployed in the future networks, there is an urgent need for distributed interference management schemes, in order to reduce the control and computational overhead. Mean field game (MFG) \cite{PSemasinghe2017} and RL are powerful candidate techniques to implement the distributed interference management in aerial MEC systems, where the interference management operation can be conducted by UAVs and other network elements independently, allowing the networks to have self-organizing capability.

In addition, some advanced physical-layer techniques, such as the directional antenna, and full dimension MIMO (FD-MIMO) (or 3D beamforming), also greatly contribute to the air-ground interference mitigation \cite{ASharma2020}. Specifically, UAVs can be equipped with directional antennas instead of omnidirectional antennas, where the antenna direction is aligned with the LoS direction to amplify the useful signal power and decrease the interference power coming from a broad range of angles. The FD-MIMO offers flexible beamforming in both azimuth and elevation dimensions. By exploiting the elevation angle separation, the interference mitigation capability can be significantly improved. However, due to the highly mobile structures of aerial MEC systems, some challenging problems arise with the application of directional antennas or 3-D beamforming, such as the LoS direction tracking, as well as the joint optimization of resource allocation and UAV trajectory, which are worthwhile for further investigations.

\subsection{ML-Driven Optimization for Aerial MEC}
As a promising technique, ML has been extensively applied to solve the wireless control and resource management problems by leveraging its adaptive modelling and intelligent learning capabilities. In spite of the existing contributions to ML-driven optimization for aerial MEC, there are still some issues in need of further investigations.

Multi-agent DRL is a popular approach for the optimization of aerial MEC systems. For multi-agent DRL, since each agent only has its local observation, coordination is the most important issue. Most of the existing literature in aerial MEC systems adopts the CTDE scheme to establish collaboration among multiple agents. Although CTDE provides an efficient solution to the partial observability, a central unit is still required to collect the observations and joint actions from the agents during training. The information collection leads to a considerable communication cost, especially for large-scale networks with a large amount of agents. Another coordination scheme among agents is the decentralized over networked agents scheme, where neighboring agents can exchange their local information over an existing communication structure to learn cooperative behaviors. However, subject to the high mobility and diversity constraints of UAVs, as well as the limited radio resources and frequently changing topologies, it is rather challenging to design cost-efficient communication protocols for learning coordination in aerial MEC systems. Although some methods such as pruning, and gating mechanisms have been applied to reduce the number of exchanged messages \cite{AFeriani2021}, there are still strong demands for more efficient solutions to the coordination issues in multi-agent environments.

In addition, security and privacy concerns will arise with the information exchanging among agents. For example, private information sharing for coordination over a wireless network is vulnerable to adversarial attacks. A possible solution to security and privacy concerns is the fully decentralized training algorithms, where each agent optimizes its policy based on local observation without exchanging information with the other agents. However, the fully decentralized algorithms suffer from convergence problems. Therefore, it deserve more efforts to develop more efficient security and privacy protection mechanisms for multi-agent learning.

\subsection{Security Protection and Privacy Preserving}
Due to the criticality of the services provided by UAVs and MEC, such as the sensitive data collection, public safety monitoring and target surveillance, the security protection and privacy preserving have become the major concerns for UAV-enabled aerial MEC. Although MEC can provide better security than cloud computing owing to the short distance between users and the MEC server, the features of aerial MEC make it vulnerable to potential attacks. For example, the public and open-access channels that UAVs use for communications may cause passive eavesdropping, active interfering, information leakage or manipulations, and denial of service (DoS) attacks, etc., while the detection of malicious nodes is difficult due to the dynamic network topology \cite{HSedjelmaci2019}. Existing solutions, such as the physical-layer security mechanisms and NFV/SDN-based security orchestration, have tried to address the eavesdropping and DoS attacks. In addition, when legitimate UAVs are used to collect sensing data, the malicious UAVs may provide false information and destroy the data collection services. In this case, introducing the trust threshold and designing trust evaluation schemes to filter the malicious UAVs out from the systems are effective solutions \cite{YSu2021}. Meanwhile, by implementing SDN in aerial MEC, the network traffic and malicious data can be isolated to satisfy the security requirements \cite{NAbbas2018}. Considering the open nature of wireless medium, all parties involved in the UAV-related communications have to be authenticated for security protection and privacy preserving. Thus, as a mechanism to control the accessibility, various efficient privacy aware authenticated key agreement (AKA) schemes have been proposed to facilitate the security \cite{PGope2020}. Since the UAVs are usually resource-constrained, lightweight authentication schemes with minimal memory overhead as well as computational and communication costs are essential to balance the tradeoff between security and lightweight features \cite{MYahuza2021}.

Besides, in the large-scale IoT systems, massive sensing data must be gathered, validated and stored in a secure and decentralized way to protect data from cyber-attacks, where the centralized ledger approaches cannot meet these requirements \cite{RGupta2021}. In this context, as a decentralized infrastructure rising with digital crypto currencies like Bitcoin, blockchain has gradually become a popular solution for the scalable, secure, and transparent data management \cite{ZGuan2020}. As blockchain is a data structure that supports features like pseudonymity, data integrity, etc., the data gathered by UAVs can be securely kept in blockchain at the MEC server, where the blockchain is used to prevent the unauthorised modifications and thus guarantee the security and integrity of the collected data.

Furthermore, some other external attacks to the components of sensing, computation, data storage and UAV control can also paralyze the UAV-enabled aerial MEC systems. For instance, the sensor spoofing attacks \cite{DHeeger2020} towards the on-board sensors can blind the sensors deployed at UAVs and send fake information to mislead the decision-makers. For the computation component in UAVs, an attacker can launch the sleep deprivation attack that drains the UAV's battery and computation resources \cite{AIHentati2020}. Thus, intrusion detection and prevention is a crucial prerequisite before designing the aerial MEC infrastructure. Concerning those diverse threats faced by aerial MEC systems, there must be comprehensive security and privacy mechanisms for forecasting, protecting, and recovering the system from catastrophic situations, while there are still not sufficient studies. Hence, the security protection and privacy preserving can be treated as an open issue with significant necessity in aerial MEC systems.

\subsection{Wireless Charging for UAVs}
Despite the attractive features of UAVs, the highly constrained onboard energy limits their flight time, typically less than 30 minutes. Although the UAVs can obtain energy replenishment via battery swapping \cite{PKChittoor2021}, it is time-consuming and may cause the service interruption. In order to continuously serve users, it is essential to prolong the lifetime of UAVs. In this context, extensive research efforts have been devoted to the wireless charging technology for UAVs \cite{WLiu2021, WLiu2022, NAnsari2019, JOuyang2018}. For example, N. Ansari \emph{et al.} \cite{NAnsari2019} propose a novel wireless charging architecture based on the free space optics (FSO), where the energy and data streams can be transferred from the AP to UAVs by utilizing the laser beams via the FSO links. Furthermore, for the laser-powered UAVs subject to the energy harvesting constraint, a downlink throughput maximization problem is investigated in \cite{JOuyang2018}, where the UAV trajectory and transmit power are jointly optimized. Besides the energy transferred from the power station, the UAV equipped with sensors and photo-voltaic (PV) cells can also harvest the solar energy from the environment. Boris Galkin \emph{et al.} \cite{BGalkin2019} investigate the impact of the PV cell area on the UAV charging performance. Simulation results demonstrate that the harvested energy increases with the cell area, but the additional weight of the cells also increases the power consumption of the UAVs.

Although the above-mentioned studies have provided some design insights for the development of UAVs' charging technology, most of them are deduced in theory or at the stage of simulations. Many external factors, such as the atmospheric attenuation, turbulence and weather conditions, need to be taken into account in the practical deployment \cite{NAnsari2019}. In addition, when the wireless charging technology is applied in aerial MEC, some challenging problems like resource allocation and trajectory design need to be jointly considered. Thus, to overcome the UAVs' energy limitation, the wireless charging techniques for UAVs are still worthy of further investigations.

\subsection{Cache-Enabled Aerial MEC}
In the last couple of years, as a promising solution to avoid frequent replication for the same content, wireless caching has attracted extensive attentions from both academia and industry. By proactively caching the popular contents at the APs or other storage devices located in close proximity to users, the content acquisition latency of users can be effectively reduced and the burden of the network backhaul can also be alleviated. When introducing wireless caching into the aerial MEC, new opportunities for performance enhancement are brought. For example, when the UAVs are deployed as ABSs to provide the air-to-ground wireless connectivity, the contents requested by ground users may be provided by the UAVs' local caches \cite{ABera2020}. Meanwhile, it is possible to effectively save the network storage resources by caching the popular contents at UAVs to serve the moving users, since the UAVs can be used as flying BSs to track the users' mobility patterns and thus improve the caching efficiency \cite{MChen2017}. In addition, from the perspective of computing, if parts of the database for completing the tasks or the computation results that may be reused have been cached in the UAVs, the computing latency of aerial MEC can be greatly reduced.

Despite the significant benefits of content caching in aerial MEC, there are still some urgent challenges to be addressed. For instance, considering the contradiction between the massive contents and finite storage capacity as well as the stringent SWaP constraints of UAVs, the caching mechanisms, i.e., what types of contents should be cached, require careful design. One widely accepted solution is to cache contents based on the users' interests \cite{JJi2020}. However, the users' interests may change over time and the content popularity distribution is often not available in advance. Besides, in the cache-enabled aerial MEC systems, the UAV will be simultaneously equipped with communication circuits, computing processors, and cache units. Accordingly, it is of great significance to consider the synergy of these components. To this end, the heterogeneous resource allocation including communication, computing, and caching along with the UAV trajectory design must be jointly optimized. However, the formulated joint optimization problems with highly-coupled variables are usually non-convex and challenging to be solved. Hence, in order to seamlessly integrate the wireless caching and aerial MEC, it still requires intensive studies to address the above-mentioned challenges.

\subsection{Integration with Recent Promising Techniques}
Recently, some promising techniques, such as the radio-based sensing, RIS, and edge intelligence, have emerged for supporting the 5G-and-beyond communications. The radio-based sensing is able to achieve contactless and privacy-preserving object detection and environmental monitoring, while RIS with the phase-shift design is a cost-effective solution to improve the propagation environment and suppress the interference. Besides, the edge intelligence, which integrates artificial intelligence (AI) and MEC capabilities into the network edge, enables various intelligent computation-intensive applications, and can dramatically reduce the end-to-end latency. However, these advanced techniques may not be directly applicable to aerial MEC systems, due to the unique features and operating environments of UAVs \cite{QWu2021}. For example, since the phase-shift design of RIS is coupled with UAVs' trajectories, the continually moving of UAVs may render the RIS passive beamforming ineffective \cite{SLi2020}. Additionally, when applying the edge intelligence, the edge devices like UAVs collaboratively train the AI model by their locally distributed data and computation capabilities, where the UAVs' trajectory design as well as the communication and computation resource scheduling have great impacts on the exchange of AI model parameters, and they are challenging issues to be addressed. There have been some research efforts on integrating these techniques with UAV and MEC systems \cite{YLiu2021, CDong2021, HMei2021, ZLi2021}, but they are still in an early stage. Therefore, along with the great opportunities brought by promising techniques, how to overcome the new challenges and fully exploit their potential benefits of aerial MEC systems requires careful investigations.

\section{Conclusions}
The UAV-enabled aerial MEC has been considered as a promising solution to provide ubiquitous and reliable MEC services in the current and future wireless networks, but the successful realization still demands for constant efforts from both academia and industry. In this paper, we have provided a comprehensive survey on the research progress of aerial MEC. We first gave an overview of aerial MEC, starting from the network architecture to the potential challenges. Then, based on three typical roles of UAVs, i.e., ABSs, mobile users, and relays, the joint optimization of UAV trajectory, computation offloading, and resource allocation was discussed, where the stringent SWaP constraints and the controllable maneuverability of UAVs were fully taken into account. Next, we reviewed the UAV deployment, task scheduling and load balancing strategies, which were designed to satisfy the time-varying requirements of users and services while reducing the deployment cost. We have also discussed the interplay between aerial MEC and other technologies, such as the WPT, PLS, and RIS. Furthermore, considering the huge potentials of ML in solving complex control and resource allocation problems in highly dynamic environments, the recent progress of ML-driven optimization for aerial MEC was comprehensively reviewed. Finally, we highlighted open problems and promising future directions pertaining to aerial MEC, including the space-air-ground integrated MEC, interference management, ML-driven optimization, security protection and privacy preserving, wireless charging for UAVs, and cache-enabled aerial MEC. Meanwhile, the integration of aerial MEC with other promising techniques, such as the radio-based sensing, RIS, and edge intelligence, was discussed.

\section*{Acknowledgement}
This work was supported by the National Natural Science Foundation of China under Grant 61901027.

\bibliographystyle{elsarticle-num}
\bibliography{refs}

\begin{thebibliography}{100}
\expandafter\ifx\csname url\endcsname\relax
  \def\url#1{\texttt{#1}}\fi
\expandafter\ifx\csname urlprefix\endcsname\relax\def\urlprefix{URL }\fi
\expandafter\ifx\csname href\endcsname\relax
  \def\href#1#2{#2} \def\path#1{#1}\fi

\bibitem{HGuo2018}
H.~Guo, J.~Liu, J.~Zhang, W.~Sun, N.~Kato, Mobile-edge computation offloading
  for ultradense {I}o{T} networks, IEEE Internet of Things Journal 5~(6) (2018)
  4977--4988.
\newblock \href {https://doi.org/10.1109/JIOT.2018.2838584}
  {\path{doi:10.1109/JIOT.2018.2838584}}.

\bibitem{CZhan2021}
C.~Zhan, H.~Hu, Z.~Liu, Z.~Wang, S.~Mao, Multi-{UAV}-enabled mobile-edge
  computing for time-constrained {I}o{T} applications, IEEE Internet of Things
  Journal 8~(20) (2021) 15553--15567.
\newblock \href {https://doi.org/10.1109/JIOT.2021.3073208}
  {\path{doi:10.1109/JIOT.2021.3073208}}.

\bibitem{WFeng2021}
W.~Feng, J.~Tang, N.~Zhao, X.~Zhang, X.~Wang, K.-K. Wong, J.~A. Chambers,
  Hybrid beamforming design and resource allocation for {UAV}-aided
  wireless-powered mobile edge computing networks with {NOMA}, IEEE Journal on
  Selected Areas in Communications 39~(11) (2021) 3271--3286.
\newblock \href {https://doi.org/10.1109/JSAC.2021.3091158}
  {\path{doi:10.1109/JSAC.2021.3091158}}.

\bibitem{PWang2019}
P.~Wang, C.~Yao, Z.~Zheng, G.~Sun, L.~Song, Joint task assignment,
  transmission, and computing resource allocation in multilayer mobile edge
  computing systems, IEEE Internet of Things Journal 6~(2) (2019) 2872--2884.
\newblock \href {https://doi.org/10.1109/JIOT.2018.2876198}
  {\path{doi:10.1109/JIOT.2018.2876198}}.

\bibitem{FZhou2020}
F.~Zhou, R.~Q. Hu, Z.~Li, Y.~Wang, Mobile edge computing in unmanned aerial
  vehicle networks, IEEE Wireless Communications 27~(1) (2020) 140--146.
\newblock \href {https://doi.org/10.1109/MWC.001.1800594}
  {\path{doi:10.1109/MWC.001.1800594}}.

\bibitem{ZUllah2020}
Z.~Ullah, F.~Al-Turjman, U.~Moatasim, L.~Mostarda, R.~Gagliardi, {UAV}s joint
  optimization problems and machine learning to improve the 5{G} and beyond
  communication, Computer Networks 182 (2020) 107478.
\newblock \href {https://doi.org/https://doi.org/10.1016/j.comnet.2020.107478}
  {\path{doi:https://doi.org/10.1016/j.comnet.2020.107478}}.

\bibitem{SJoo2021}
S.~Joo, H.~Kang, J.~Kang, {C}o{SM}o{S}: Cooperative sky-ground mobile edge
  computing system, IEEE Transactions on Vehicular Technology 70~(8) (2021)
  8373--8377.
\newblock \href {https://doi.org/10.1109/TVT.2021.3094584}
  {\path{doi:10.1109/TVT.2021.3094584}}.

\bibitem{YXu2021-2}
Y.~Xu, T.~Zhang, Y.~Liu, D.~Yang, L.~Xiao, M.~Tao, {UAV}-assisted {MEC}
  networks with aerial and ground cooperation, IEEE Transactions on Wireless
  Communications 20~(12) (2021) 7712--7727.
\newblock \href {https://doi.org/10.1109/TWC.2021.3086521}
  {\path{doi:10.1109/TWC.2021.3086521}}.

\bibitem{WZhang2020}
W.~Zhang, L.~Li, N.~Zhang, T.~Han, S.~Wang, Air-ground integrated mobile edge
  networks: A survey, IEEE Access 8 (2020) 125998--126018.
\newblock \href {https://doi.org/10.1109/ACCESS.2020.3008168}
  {\path{doi:10.1109/ACCESS.2020.3008168}}.

\bibitem{FJiang2021}
F.~{Jiang}, K.~{Wang}, L.~{Dong}, C.~{Pan}, W.~{Xu}, K.~{Yang}, {AI} driven
  heterogeneous {MEC} system with {UAV} assistance for dynamic environment:
  Challenges and solutions, IEEE Network 35~(1) (2021) 400--408.
\newblock \href {https://doi.org/10.1109/MNET.011.2000440}
  {\path{doi:10.1109/MNET.011.2000440}}.

\bibitem{BLi2019}
B.~Li, Z.~Fei, Y.~Zhang, {UAV} communications for 5{G} and beyond: Recent
  advances and future trends, IEEE Internet of Things Journal 6~(2) (2019)
  2241--2263.
\newblock \href {https://doi.org/10.1109/JIOT.2018.2887086}
  {\path{doi:10.1109/JIOT.2018.2887086}}.

\bibitem{MEMkiramweni2019}
M.~E. Mkiramweni, C.~Yang, J.~Li, W.~Zhang, A survey of game theory in unmanned
  aerial vehicles communications, IEEE Communications Surveys \& Tutorials
  21~(4) (2019) 3386--3416.
\newblock \href {https://doi.org/10.1109/COMST.2019.2919613}
  {\path{doi:10.1109/COMST.2019.2919613}}.

\bibitem{YZeng2019}
Y.~Zeng, Q.~Wu, R.~Zhang, Accessing from the sky: A tutorial on {UAV}
  communications for 5{G} and beyond, Proceedings of the IEEE 107~(12) (2019)
  2327--2375.
\newblock \href {https://doi.org/10.1109/JPROC.2019.2952892}
  {\path{doi:10.1109/JPROC.2019.2952892}}.

\bibitem{MMozaffari2019}
M.~Mozaffari, W.~Saad, M.~Bennis, Y.-H. Nam, M.~Debbah, A tutorial on {UAV}s
  for wireless networks: Applications, challenges, and open problems, IEEE
  Communications Surveys \& Tutorials 21~(3) (2019) 2334--2360.
\newblock \href {https://doi.org/10.1109/COMST.2019.2902862}
  {\path{doi:10.1109/COMST.2019.2902862}}.

\bibitem{QWu2021}
Q.~Wu, J.~Xu, Y.~Zeng, D.~W.~K. Ng, N.~Al-Dhahir, R.~Schober, A.~L.
  Swindlehurst, A comprehensive overview on 5{G}-and-beyond networks with
  {UAV}s: From communications to sensing and intelligence, IEEE Journal on
  Selected Areas in Communications 39~(10) (2021) 2912--2945.
\newblock \href {https://doi.org/10.1109/JSAC.2021.3088681}
  {\path{doi:10.1109/JSAC.2021.3088681}}.

\bibitem{YMao2017}
Y.~Mao, C.~You, J.~Zhang, K.~Huang, K.~B. Letaief, A survey on mobile edge
  computing: The communication perspective, IEEE Communications Surveys \&
  Tutorials 19~(4) (2017) 2322--2358.
\newblock \href {https://doi.org/10.1109/COMST.2017.2745201}
  {\path{doi:10.1109/COMST.2017.2745201}}.

\bibitem{TTaleb2017}
T.~Taleb, K.~Samdanis, B.~Mada, H.~Flinck, S.~Dutta, D.~Sabella, On
  multi-access edge computing: A survey of the emerging 5{G} network edge cloud
  architecture and orchestration, IEEE Communications Surveys \& Tutorials
  19~(3) (2017) 1657--1681.
\newblock \href {https://doi.org/10.1109/COMST.2017.2705720}
  {\path{doi:10.1109/COMST.2017.2705720}}.

\bibitem{PPorambage2018}
P.~Porambage, J.~Okwuibe, M.~Liyanage, M.~Ylianttila, T.~Taleb, Survey on
  multi-access edge computing for internet of things realization, IEEE
  Communications Surveys \& Tutorials 20~(4) (2018) 2961--2991.
\newblock \href {https://doi.org/10.1109/COMST.2018.2849509}
  {\path{doi:10.1109/COMST.2018.2849509}}.

\bibitem{NAbbas2018}
N.~Abbas, Y.~Zhang, A.~Taherkordi, T.~Skeie, Mobile edge computing: A survey,
  IEEE Internet of Things Journal 5~(1) (2018) 450--465.
\newblock \href {https://doi.org/10.1109/JIOT.2017.2750180}
  {\path{doi:10.1109/JIOT.2017.2750180}}.

\bibitem{JMoura2019}
J.~Moura, D.~Hutchison, Game theory for multi-access edge computing: Survey,
  use cases, and future trends, IEEE Communications Surveys \& Tutorials 21~(1)
  (2019) 260--288.
\newblock \href {https://doi.org/10.1109/COMST.2018.2863030}
  {\path{doi:10.1109/COMST.2018.2863030}}.

\bibitem{PRanaweera2021}
P.~Ranaweera, A.~D. Jurcut, M.~Liyanage, Survey on multi-access edge computing
  security and privacy, IEEE Communications Surveys \& Tutorials 23~(2) (2021)
  1078--1124.
\newblock \href {https://doi.org/10.1109/COMST.2021.3062546}
  {\path{doi:10.1109/COMST.2021.3062546}}.

\bibitem{MHassanalian2017}
M.~Hassanalian, A.~Abdelkefi, Classifications, applications, and design
  challenges of drones: A review, Progress in Aerospace Sciences 91 (2017)
  99--131.
\newblock \href {https://doi.org/10.1016/j.paerosci.2017.04.003}
  {\path{doi:10.1016/j.paerosci.2017.04.003}}.

\bibitem{AFotouhi2019}
A.~Fotouhi, H.~Qiang, M.~Ding, M.~Hassan, L.~G. Giordano, A.~Garcia-Rodriguez,
  J.~Yuan, Survey on {UAV} cellular communications: Practical aspects,
  standardization advancements, regulation, and security challenges, IEEE
  Communications Surveys \& Tutorials 21~(4) (2019) 3417--3442.
\newblock \href {https://doi.org/10.1109/COMST.2019.2906228}
  {\path{doi:10.1109/COMST.2019.2906228}}.

\bibitem{YZeng2016}
Y.~Zeng, R.~Zhang, T.~J. Lim, Wireless communications with unmanned aerial
  vehicles: opportunities and challenges, IEEE Communications Magazine 54~(5)
  (2016) 36--42.
\newblock \href {https://doi.org/10.1109/MCOM.2016.7470933}
  {\path{doi:10.1109/MCOM.2016.7470933}}.

\bibitem{MHua2018}
M.~Hua, Y.~Huang, Y.~Wang, Q.~Wu, H.~Dai, L.~Yang, Energy optimization for
  cellular-connected multi-{UAV} mobile edge computing systems with
  multi-access schemes, Journal of Communications and Information Networks
  3~(4) (2018) 33--44.
\newblock \href {https://doi.org/10.1007/s41650-018-0035-0}
  {\path{doi:10.1007/s41650-018-0035-0}}.

\bibitem{DZhai2021}
D.~Zhai, H.~Li, X.~Tang, R.~Zhang, Z.~Ding, F.~R. Yu, Height optimization and
  resource allocation for {NOMA} enhanced {UAV}-aided relay networks, IEEE
  Transactions on Communications 69~(2) (2021) 962--975.
\newblock \href {https://doi.org/10.1109/TCOMM.2020.3037345}
  {\path{doi:10.1109/TCOMM.2020.3037345}}.

\bibitem{JHu2019}
J.~Hu, M.~Jiang, Q.~Zhang, Q.~Li, J.~Qin, Joint optimization of {UAV} position,
  time slot allocation, and computation task partition in multiuser aerial
  mobile-edge computing systems, IEEE Transactions on Vehicular Technology
  68~(7) (2019) 7231--7235.
\newblock \href {https://doi.org/10.1109/TVT.2019.2915836}
  {\path{doi:10.1109/TVT.2019.2915836}}.

\bibitem{XZhang2019}
X.~Zhang, Y.~Zhong, P.~Liu, F.~Zhou, Y.~Wang, Resource allocation for a
  {UAV}-enabled mobile-edge computing system: Computation efficiency
  maximization, IEEE Access 7 (2019) 113345--113354.
\newblock \href {https://doi.org/10.1109/ACCESS.2019.2935217}
  {\path{doi:10.1109/ACCESS.2019.2935217}}.

\bibitem{MHua2019}
M.~Hua, Y.~Wang, C.~Li, Y.~Huang, L.~Yang, {UAV}-aided mobile edge computing
  systems with one by one access scheme, IEEE Transactions on Green
  Communications and Networking 3~(3) (2019) 664--678.
\newblock \href {https://doi.org/10.1109/TGCN.2019.2910590}
  {\path{doi:10.1109/TGCN.2019.2910590}}.

\bibitem{HGuo2020}
H.~Guo, J.~Liu, {UAV}-enhanced intelligent offloading for internet of things at
  the edge, IEEE Transactions on Industrial Informatics 16~(4) (2020)
  2737--2746.
\newblock \href {https://doi.org/10.1109/TII.2019.2954944}
  {\path{doi:10.1109/TII.2019.2954944}}.

\bibitem{JXiong2019}
J.~Xiong, H.~Guo, J.~Liu, Task offloading in {UAV}-aided edge computing: Bit
  allocation and trajectory optimization, IEEE Communications Letters 23~(3)
  (2019) 538--541.
\newblock \href {https://doi.org/10.1109/LCOMM.2019.2891662}
  {\path{doi:10.1109/LCOMM.2019.2891662}}.

\bibitem{YLuo2020}
Y.~Luo, W.~Ding, B.~Zhang, W.~Huang, C.~Liu, Optimization of bits allocation
  and path planning with trajectory constraint in {UAV}-enabled mobile edge
  computing system, Chinese Journal of Aeronautics 33~(10) (2020) 2716--2727.
\newblock \href {https://doi.org/10.1016/j.cja.2020.04.014}
  {\path{doi:10.1016/j.cja.2020.04.014}}.

\bibitem{YDu2018}
Y.~Du, K.~Wang, K.~Yang, G.~Zhang, Energy-efficient resource allocation in
  {UAV} based {MEC} system for {I}o{T} devices, in: 2018 IEEE Global
  Communications Conference (GLOBECOM), 2018, pp. 1--6.
\newblock \href {https://doi.org/10.1109/GLOCOM.2018.8647789}
  {\path{doi:10.1109/GLOCOM.2018.8647789}}.

\bibitem{MZhao2021}
M.~Zhao, W.~Li, L.~Bao, J.~Luo, Z.~He, D.~Liu, Fairness-aware task scheduling
  and resource allocation in uav-enabled mobile edge computing networks, IEEE
  Transactions on Green Communications and Networking 5~(4) (2021) 2174--2187.
\newblock \href {https://doi.org/10.1109/TGCN.2021.3095070}
  {\path{doi:10.1109/TGCN.2021.3095070}}.

\bibitem{YTun2021}
Y.~K. Tun, Y.~M. Park, N.~H. Tran, W.~Saad, S.~R. Pandey, C.~S. Hong,
  Energy-efficient resource management in {UAV}-assisted mobile edge computing,
  IEEE Communications Letters 25~(1) (2021) 249--253.
\newblock \href {https://doi.org/10.1109/LCOMM.2020.3026033}
  {\path{doi:10.1109/LCOMM.2020.3026033}}.

\bibitem{JZhang2019}
J.~Zhang, L.~Zhou, Q.~Tang, E.~C.-H. Ngai, X.~Hu, H.~Zhao, J.~Wei, Stochastic
  computation offloading and trajectory scheduling for {UAV}-assisted mobile
  edge computing, IEEE Internet of Things Journal 6~(2) (2019) 3688--3699.
\newblock \href {https://doi.org/10.1109/JIOT.2018.2890133}
  {\path{doi:10.1109/JIOT.2018.2890133}}.

\bibitem{QHu2019}
Q.~Hu, Y.~Cai, G.~Yu, Z.~Qin, M.~Zhao, G.~Y. Li, Joint offloading and
  trajectory design for {UAV}-enabled mobile edge computing systems, IEEE
  Internet of Things Journal 6~(2) (2019) 1879--1892.
\newblock \href {https://doi.org/10.1109/JIOT.2018.2878876}
  {\path{doi:10.1109/JIOT.2018.2878876}}.

\bibitem{CZhan2020}
C.~{Zhan}, H.~{Hu}, X.~{Sui}, Z.~{Liu}, D.~{Niyato}, Completion time and energy
  optimization in the {UAV}-enabled mobile-edge computing system, IEEE Internet
  of Things Journal 7~(8) (2020) 7808--7822.
\newblock \href {https://doi.org/10.1109/JIOT.2020.2993260}
  {\path{doi:10.1109/JIOT.2020.2993260}}.

\bibitem{HMei2019}
H.~Mei, K.~Wang, D.~Zhou, K.~Yang, Joint trajectory-task-cache optimization in
  {UAV}-enabled mobile edge networks for cyber-physical system, IEEE Access 7
  (2019) 156476--156488.
\newblock \href {https://doi.org/10.1109/ACCESS.2019.2949032}
  {\path{doi:10.1109/ACCESS.2019.2949032}}.

\bibitem{FCostanzo2020}
F.~Costanzo, P.~D. Lorenzo, S.~Barbarossa, Dynamic resource optimization and
  altitude selection in {UAV}-based multi-access edge computing, in: ICASSP
  2020 - 2020 IEEE International Conference on Acoustics, Speech and Signal
  Processing (ICASSP), 2020, pp. 4985--4989.
\newblock \href {https://doi.org/10.1109/ICASSP40776.2020.9053594}
  {\path{doi:10.1109/ICASSP40776.2020.9053594}}.

\bibitem{HMei2020}
H.~Mei, K.~Yang, Q.~Liu, K.~Wang, Joint trajectory-resource optimization in
  {UAV}-enabled edge-cloud system with virtualized mobile clone, IEEE Internet
  of Things Journal 7~(7) (2020) 5906--5921.
\newblock \href {https://doi.org/10.1109/JIOT.2019.2952677}
  {\path{doi:10.1109/JIOT.2019.2952677}}.

\bibitem{YQian2019}
Y.~Qian, F.~Wang, J.~Li, L.~Shi, K.~Cai, F.~Shu, User association and path
  planning for {UAV}-aided mobile edge computing with energy restriction, IEEE
  Wireless Communications Letters 8~(5) (2019) 1312--1315.
\newblock \href {https://doi.org/10.1109/LWC.2019.2913843}
  {\path{doi:10.1109/LWC.2019.2913843}}.

\bibitem{XDiao2019}
X.~Diao, J.~Zheng, Y.~Cai, Y.~Wu, A.~Anpalagan, Fair data allocation and
  trajectory optimization for {UAV}-assisted mobile edge computing, IEEE
  Communications Letters 23~(12) (2019) 2357--2361.
\newblock \href {https://doi.org/10.1109/LCOMM.2019.2943461}
  {\path{doi:10.1109/LCOMM.2019.2943461}}.

\bibitem{SLiu2020}
S.~Liu, T.~Yang, Delay aware scheduling in {UAV}-enabled {OFDMA} mobile edge
  computing system, IET Communications 14~(18) (2020) 3203--3211.
\newblock \href {https://doi.org/10.1049/iet-com.2020.0274}
  {\path{doi:10.1049/iet-com.2020.0274}}.

\bibitem{IBudhiraja2021}
I.~Budhiraja, N.~Kumar, S.~Tyagi, S.~Tanwar, Energy consumption minimization
  scheme for {NOMA}-based mobile edge computation networks underlaying {UAV},
  IEEE Systems Journal 15~(4) (2021) 5724--5733.
\newblock \href {https://doi.org/10.1109/JSYST.2021.3076782}
  {\path{doi:10.1109/JSYST.2021.3076782}}.

\bibitem{XbDiao2019}
X.~Diao, J.~Zheng, Y.~Wu, Y.~Cai, A.~Anpalagan, Joint trajectory design, task
  data, and computing resource allocations for {NOMA}-based and {UAV}-assisted
  mobile edge computing, IEEE Access 7 (2019) 117448--117459.
\newblock \href {https://doi.org/10.1109/ACCESS.2019.2936437}
  {\path{doi:10.1109/ACCESS.2019.2936437}}.

\bibitem{YYang2020}
Y.~Yang, M.~Cenk~Gursoy, Energy efficiency optimization in {UAV}-assisted
  communications and edge computing, in: 2020 IEEE 21st International Workshop
  on Signal Processing Advances in Wireless Communications (SPAWC), 2020, pp.
  1--5.
\newblock \href {https://doi.org/10.1109/SPAWC48557.2020.9154211}
  {\path{doi:10.1109/SPAWC48557.2020.9154211}}.

\bibitem{SJeong2018}
S.~Jeong, O.~Simeone, J.~Kang, Mobile edge computing via a {UAV}-mounted
  cloudlet: Optimization of bit allocation and path planning, IEEE Transactions
  on Vehicular Technology 67~(3) (2018) 2049--2063.
\newblock \href {https://doi.org/10.1109/TVT.2017.2706308}
  {\path{doi:10.1109/TVT.2017.2706308}}.

\bibitem{JJi2021}
J.~Ji, K.~Zhu, C.~Yi, D.~Niyato, Energy consumption minimization in
  {UAV}-assisted mobile-edge computing systems: Joint resource allocation and
  trajectory design, IEEE Internet of Things Journal 8~(10) (2021) 8570--8584.
\newblock \href {https://doi.org/10.1109/JIOT.2020.3046788}
  {\path{doi:10.1109/JIOT.2020.3046788}}.

\bibitem{MsLi2020}
M.~Li, N.~Cheng, J.~Gao, Y.~Wang, L.~Zhao, X.~Shen, Energy-efficient
  {UAV}-assisted mobile edge computing: Resource allocation and trajectory
  optimization, IEEE Transactions on Vehicular Technology 69~(3) (2020)
  3424--3438.
\newblock \href {https://doi.org/10.1109/TVT.2020.2968343}
  {\path{doi:10.1109/TVT.2020.2968343}}.

\bibitem{YQu2021}
Y.~Qu, H.~Dai, H.~Wang, C.~Dong, F.~Wu, S.~Guo, Q.~Wu, Service provisioning for
  {UAV}-enabled mobile edge computing, IEEE Journal on Selected Areas in
  Communications 39~(11) (2021) 3287--3305.
\newblock \href {https://doi.org/10.1109/JSAC.2021.3088660}
  {\path{doi:10.1109/JSAC.2021.3088660}}.

\bibitem{AANasir2021}
A.~A. Nasir, Latency optimization of {UAV}-enabled {MEC} system for virtual
  reality applications under rician fading channels, IEEE Wireless
  Communications Letters 10~(8) (2021) 1633--1637.
\newblock \href {https://doi.org/10.1109/LWC.2021.3075762}
  {\path{doi:10.1109/LWC.2021.3075762}}.

\bibitem{PAApostolopoulos2021}
P.~A. Apostolopoulos, G.~Fragkos, E.~E. Tsiropoulou, S.~Papavassiliou, Data
  offloading in {UAV}-assisted multi-access edge computing systems under
  resource uncertainty, IEEE Transactions on Mobile Computing (2021) 1--16\href
  {https://doi.org/10.1109/TMC.2021.3069911}
  {\path{doi:10.1109/TMC.2021.3069911}}.

\bibitem{JiZhang2020}
J.~{Zhang}, L.~{Zhou}, F.~{Zhou}, B.~{Seet}, H.~{Zhang}, Z.~{Cai}, J.~{Wei},
  Computation-efficient offloading and trajectory scheduling for multi-{UAV}
  assisted mobile edge computing, IEEE Transactions on Vehicular Technology
  69~(2) (2020) 2114--2125.
\newblock \href {https://doi.org/10.1109/TVT.2019.2960103}
  {\path{doi:10.1109/TVT.2019.2960103}}.

\bibitem{XDiao2020}
X.~Diao, M.~Wang, J.~Zheng, Y.~Cai, Fairness-aware offloading and trajectory
  optimization for multi-{UAV} enabled multi-access edge computing, IEEE Access
  8 (2020) 124359--124370.
\newblock \href {https://doi.org/10.1109/ACCESS.2020.3006112}
  {\path{doi:10.1109/ACCESS.2020.3006112}}.

\bibitem{ZYang2019}
Z.~Yang, C.~Pan, K.~Wang, M.~Shikh-Bahaei, Energy efficient resource allocation
  in {UAV}-enabled mobile edge computing networks, IEEE Transactions on
  Wireless Communications 18~(9) (2019) 4576--4589.
\newblock \href {https://doi.org/10.1109/TWC.2019.2927313}
  {\path{doi:10.1109/TWC.2019.2927313}}.

\bibitem{XZhang2020}
X.~Zhang, J.~Zhang, J.~Xiong, L.~Zhou, J.~Wei, Energy-efficient
  multi-{UAV}-enabled multiaccess edge computing incorporating {NOMA}, IEEE
  Internet of Things Journal 7~(6) (2020) 5613--5627.
\newblock \href {https://doi.org/10.1109/JIOT.2020.2980035}
  {\path{doi:10.1109/JIOT.2020.2980035}}.

\bibitem{XQin2021}
X.~Qin, Z.~Song, Y.~Hao, X.~Sun, Joint resource allocation and trajectory
  optimization for multi-{UAV}-assisted multi-access mobile edge computing,
  IEEE Wireless Communications Letters 10~(7) (2021) 1400--1404.
\newblock \href {https://doi.org/10.1109/LWC.2021.3068793}
  {\path{doi:10.1109/LWC.2021.3068793}}.

\bibitem{PAApos2020}
P.~A. Apostolopoulos, E.~E. Tsiropoulou, S.~Papavassiliou, Risk-aware data
  offloading in multi-server multi-access edge computing environment, IEEE/ACM
  Transactions on Networking 28~(3) (2020) 1405--1418.
\newblock \href {https://doi.org/10.1109/TNET.2020.2983119}
  {\path{doi:10.1109/TNET.2020.2983119}}.

\bibitem{PAApos2020Access}
P.~A. Apostolopoulos, E.~E. Tsiropoulou, S.~Papavassiliou, Cognitive data
  offloading in mobile edge computing for internet of things, IEEE Access 8
  (2020) 55736--55749.
\newblock \href {https://doi.org/10.1109/ACCESS.2020.2981837}
  {\path{doi:10.1109/ACCESS.2020.2981837}}.

\bibitem{PAApos2019}
P.~A. Apostolopoulos, E.~E. Tsiropoulou, S.~Papavassiliou, Risk-aware social
  cloud computing based on serverless computing model, in: 2019 IEEE Global
  Communications Conference (GLOBECOM), 2019, pp. 1--6.
\newblock \href {https://doi.org/10.1109/GLOBECOM38437.2019.9013182}
  {\path{doi:10.1109/GLOBECOM38437.2019.9013182}}.

\bibitem{HShimada2021}
H.~Shimada, Y.~Kawamoto, N.~Kato, Novel computation and communication resources
  allocation using relay communications in {UAV}-mounted cloudlet systems, IEEE
  Transactions on Network Science and Engineering 8~(4) (2021) 3140--3151.
\newblock \href {https://doi.org/10.1109/TNSE.2021.3105455}
  {\path{doi:10.1109/TNSE.2021.3105455}}.

\bibitem{NCheng2018}
N.~Cheng, W.~Xu, W.~Shi, Y.~Zhou, N.~Lu, H.~Zhou, X.~Shen, Air-ground
  integrated mobile edge networks: Architecture, challenges, and opportunities,
  IEEE Communications Magazine 56~(8) (2018) 26--32.
\newblock \href {https://doi.org/10.1109/MCOM.2018.1701092}
  {\path{doi:10.1109/MCOM.2018.1701092}}.

\bibitem{HDong2020}
H.~Dong, N.~Wu, G.~Feng, X.~Gao, Research on computing task allocation method
  based on multi-{UAV}s collaboration, in: 2020 IEEE International Conference
  on Smart Internet of Things (SmartIoT), 2020, pp. 86--93.
\newblock \href {https://doi.org/10.1109/SmartIoT49966.2020.00022}
  {\path{doi:10.1109/SmartIoT49966.2020.00022}}.

\bibitem{WMa2019}
W.~Ma, X.~Liu, L.~Mashayekhy, A strategic game for task offloading among
  capacitated {UAV}-mounted cloudlets, in: 2019 IEEE International Congress on
  Internet of Things (ICIOT), 2019, pp. 61--68.
\newblock \href {https://doi.org/10.1109/ICIOT.2019.00022}
  {\path{doi:10.1109/ICIOT.2019.00022}}.

\bibitem{WLiu2020}
W.~Liu, Y.~Xu, N.~Qi, K.~Yao, Y.~Zhang, W.~He, Joint computation offloading and
  resource allocation in {UAV} swarms with multi-access edge computing, in:
  2020 International Conference on Wireless Communications and Signal
  Processing (WCSP), 2020, pp. 280--285.
\newblock \href {https://doi.org/10.1109/WCSP49889.2020.9299713}
  {\path{doi:10.1109/WCSP49889.2020.9299713}}.

\bibitem{KYao2020-C2}
K.~Yao, J.~Chen, Y.~Zhang, L.~Cui, Y.~Yang, Y.~Xu, Joint computation offloading
  and variable-width channel access optimization in {UAV} swarms, in: GLOBECOM
  2020 - 2020 IEEE Global Communications Conference, 2020, pp. 1--6.
\newblock \href {https://doi.org/10.1109/GLOBECOM42002.2020.9322587}
  {\path{doi:10.1109/GLOBECOM42002.2020.9322587}}.

\bibitem{RChen2020}
R.~Chen, L.~Cui, Y.~Zhang, J.~Chen, K.~Yao, Y.~Yang, C.~Yao, H.~Han, Delay
  optimization with {FCFS} queuing model in mobile edge computing-assisted
  {UAV} swarms: A game-theoretic learning approach, in: 2020 International
  Conference on Wireless Communications and Signal Processing (WCSP), 2020, pp.
  245--250.
\newblock \href {https://doi.org/10.1109/WCSP49889.2020.9299801}
  {\path{doi:10.1109/WCSP49889.2020.9299801}}.

\bibitem{YLuo2021}
Y.~Luo, W.~Ding, B.~Zhang, Optimization of task scheduling and dynamic service
  strategy for multi-{UAV}-enabled mobile-edge computing system, IEEE
  Transactions on Cognitive Communications and Networking 7~(3) (2021)
  970--984.
\newblock \href {https://doi.org/10.1109/TCCN.2021.3051947}
  {\path{doi:10.1109/TCCN.2021.3051947}}.

\bibitem{HChang2021}
H.~Chang, Y.~Chen, B.~Zhang, D.~Doermann, Multi-{UAV} mobile edge computing and
  path planning platform based on reinforcement learning, IEEE Transactions on
  Emerging Topics in Computational Intelligence (2021) 1--10\href
  {https://doi.org/10.1109/TETCI.2021.3083410}
  {\path{doi:10.1109/TETCI.2021.3083410}}.

\bibitem{XDeng2021}
X.~Deng, J.~Li, P.~Guan, L.~Zhang, Energy-efficient {UAV}-aided target tracking
  systems based on edge computing, IEEE Internet of Things Journal 9~(3) (2022)
  2207--2214.
\newblock \href {https://doi.org/10.1109/JIOT.2021.3091216}
  {\path{doi:10.1109/JIOT.2021.3091216}}.

\bibitem{ZNing2021}
Z.~Ning, P.~Dong, M.~Wen, X.~Wang, L.~Guo, R.~Y.~K. Kwok, H.~V. Poor,
  5{G}-enabled {UAV}-to-community offloading: Joint trajectory design and task
  scheduling, IEEE Journal on Selected Areas in Communications 39~(11) (2021)
  3306--3320.
\newblock \href {https://doi.org/10.1109/JSAC.2021.3088663}
  {\path{doi:10.1109/JSAC.2021.3088663}}.

\bibitem{RLi2020}
R.~Li, X.~Li, J.~Xu, F.~Jiang, Z.~Jia, D.~Shao, L.~Pan, Energy-aware
  decision-making for dynamic task migration in {MEC}-based unmanned aerial
  vehicle delivery system, Concurrency and Computation Practice and Experience
  (2020).
\newblock \href {https://doi.org/10.1002/cpe.6092}
  {\path{doi:10.1002/cpe.6092}}.

\bibitem{MMessous2020}
M.~A. Messous, H.~Hellwagner, S.-M. Senouci, D.~Emini, D.~Schnieders, Edge
  computing for visual navigation and mapping in a {UAV} network, in: ICC 2020
  - 2020 IEEE International Conference on Communications (ICC), 2020, pp. 1--6.
\newblock \href {https://doi.org/10.1109/ICC40277.2020.9149087}
  {\path{doi:10.1109/ICC40277.2020.9149087}}.

\bibitem{DCallegaro2018}
D.~Callegaro, M.~Levorato, Optimal computation offloading in edge-assisted
  {UAV} systems, in: 2018 IEEE Global Communications Conference (GLOBECOM),
  2018, pp. 1--6.
\newblock \href {https://doi.org/10.1109/GLOCOM.2018.8648099}
  {\path{doi:10.1109/GLOCOM.2018.8648099}}.

\bibitem{JChen2021}
J.~Chen, Q.~Wu, Y.~Xu, N.~Qi, T.~Fang, L.~Jia, C.~Dong, A multi-leader
  multi-follower stackelberg game for coalition-based {UAV} {MEC} networks,
  IEEE Wireless Communications Letters 10~(11) (2021) 2350--2354.
\newblock \href {https://doi.org/10.1109/LWC.2021.3100113}
  {\path{doi:10.1109/LWC.2021.3100113}}.

\bibitem{MLiwang2021}
M.~Liwang, Z.~Gao, X.~Wang, Let’s trade in the future! a futures-enabled fast
  resource trading mechanism in edge computing-assisted {UAV} networks, IEEE
  Journal on Selected Areas in Communications 39~(11) (2021) 3252--3270.
\newblock \href {https://doi.org/10.1109/JSAC.2021.3088657}
  {\path{doi:10.1109/JSAC.2021.3088657}}.

\bibitem{LFan2019}
L.~Fan, W.~Yan, X.~Chen, Z.~Chen, Q.~Shi, An energy efficient design for {UAV}
  communication with mobile edge computing, China Communications 16~(1) (2019)
  26--36.
\newblock \href {https://doi.org/10.12676/j.cc.2019.01.003}
  {\path{doi:10.12676/j.cc.2019.01.003}}.

\bibitem{ZZhu2020}
Z.~Zhu, L.~Qian, J.~Shen, L.~Huang, Y.~Wu, Joint optimisation of {UAV} grouping
  and energy consumption in {MEC}-enabled {UAV} communication networks, IET
  Communications 14~(16) (2020) 2723--2730.
\newblock \href {https://doi.org/10.1049/iet-com.2019.1179}
  {\path{doi:10.1049/iet-com.2019.1179}}.

\bibitem{XCao2018}
X.~Cao, J.~Xu, R.~Zhang, Mobile edge computing for cellular-connected {UAV}:
  Computation offloading and trajectory optimization, in: 2018 IEEE 19th
  International Workshop on Signal Processing Advances in Wireless
  Communications (SPAWC), 2018, pp. 1--5.
\newblock \href {https://doi.org/10.1109/SPAWC.2018.8445936}
  {\path{doi:10.1109/SPAWC.2018.8445936}}.

\bibitem{MDai2020}
M.~Dai, Z.~Su, J.~Li, J.~Zhou, An energy-efficient edge offloading scheme for
  {UAV}-assisted internet of things, in: 2020 IEEE 40th International
  Conference on Distributed Computing Systems (ICDCS), 2020, pp. 1293--1297.
\newblock \href {https://doi.org/10.1109/ICDCS47774.2020.00167}
  {\path{doi:10.1109/ICDCS47774.2020.00167}}.

\bibitem{YRen2020}
Y.~Ren, Z.~Xie, Z.~Ding, X.~Sun, J.~Xia, Y.~Tian, Computation offloading game
  in multiple unmanned aerial vehicle-enabled mobile edge computing networks,
  IET Communications 15 (2020) 1392--1401.
\newblock \href {https://doi.org/10.1049/cmu2.12102}
  {\path{doi:10.1049/cmu2.12102}}.

\bibitem{RDuan2019}
R.~Duan, J.~Wang, C.~Jiang, Y.~Ren, L.~Hanzo, The transmit-energy vs
  computation-delay trade-off in gateway-selection for heterogenous cloud aided
  multi-{UAV} systems, IEEE Transactions on Communications 67~(4) (2019)
  3026--3039.
\newblock \href {https://doi.org/10.1109/TCOMM.2018.2889672}
  {\path{doi:10.1109/TCOMM.2018.2889672}}.

\bibitem{MMessous2019}
M.-A. Messous, S.-M. Senouci, H.~Sedjelmaci, S.~Cherkaoui, A game theory based
  efficient computation offloading in an {UAV} network, IEEE Transactions on
  Vehicular Technology 68~(5) (2019) 4964--4974.
\newblock \href {https://doi.org/10.1109/TVT.2019.2902318}
  {\path{doi:10.1109/TVT.2019.2902318}}.

\bibitem{PCao2019}
P.~Cao, Y.~Liu, C.~Yang, S.~Xie, K.~Xie, {MEC}-driven {UAV}-enabled routine
  inspection scheme in wind farm under wind influence, IEEE Access 7 (2019)
  179252--179265.
\newblock \href {https://doi.org/10.1109/ACCESS.2019.2958680}
  {\path{doi:10.1109/ACCESS.2019.2958680}}.

\bibitem{YYu2021}
Y.~Yu, X.~Bu, K.~Yang, H.~Yang, X.~Gao, Z.~Han, {UAV}-aided low latency
  multi-access edge computing, IEEE Transactions on Vehicular Technology 70~(5)
  (2021) 4955--4967.
\newblock \href {https://doi.org/10.1109/TVT.2021.3072065}
  {\path{doi:10.1109/TVT.2021.3072065}}.

\bibitem{XLiu2020}
X.~Liu, B.~Lai, L.~Gou, C.~Lin, M.~Zhou, Joint resource optimization for
  {UAV}-enabled multichannel internet of things based on intelligent fog
  computing, IEEE Transactions on Network Science and Engineering 8~(4) (2021)
  2814--2824.
\newblock \href {https://doi.org/10.1109/TNSE.2020.3027098}
  {\path{doi:10.1109/TNSE.2020.3027098}}.

\bibitem{FGuo2019}
F.~Guo, H.~Zhang, H.~Ji, X.~Li, V.~C. Leung, Joint trajectory and computation
  offloading optimization for {UAV}-assisted {MEC} with {NOMA}, in: IEEE
  INFOCOM 2019 - IEEE Conference on Computer Communications Workshops (INFOCOM
  WKSHPS), 2019, pp. 1--6.
\newblock \href {https://doi.org/10.1109/INFOCOMWKSHPS47286.2019.9093764}
  {\path{doi:10.1109/INFOCOMWKSHPS47286.2019.9093764}}.

\bibitem{QWang2021}
Q.~Wang, A.~Gao, Y.~Hu, Joint power and {Q}o{E} optimization scheme for
  multi-{UAV} assisted offloading in mobile computing, IEEE Access 9 (2021)
  21206--21217.
\newblock \href {https://doi.org/10.1109/ACCESS.2021.3055335}
  {\path{doi:10.1109/ACCESS.2021.3055335}}.

\bibitem{SZheng2020}
S.~Zheng, Z.~Ren, X.~Hou, H.~Zhang, Optimal communication-computing-caching for
  maximizing revenue in {UAV}-aided mobile edge computing, in: GLOBECOM 2020 -
  2020 IEEE Global Communications Conference, 2020, pp. 1--6.
\newblock \href {https://doi.org/10.1109/GLOBECOM42002.2020.9322229}
  {\path{doi:10.1109/GLOBECOM42002.2020.9322229}}.

\bibitem{LZhang2021}
L.~Zhang, N.~Ansari, Optimizing the operation cost for {UAV}-aided mobile edge
  computing, IEEE Transactions on Vehicular Technology 70~(6) (2021)
  6085--6093.
\newblock \href {https://doi.org/10.1109/TVT.2021.3076980}
  {\path{doi:10.1109/TVT.2021.3076980}}.

\bibitem{TZhang2020}
T.~Zhang, Y.~Xu, J.~Loo, D.~Yang, L.~Xiao, Joint computation and communication
  design for {UAV}-assisted mobile edge computing in {I}o{T}, IEEE Transactions
  on Industrial Informatics 16~(8) (2020) 5505--5516.
\newblock \href {https://doi.org/10.1109/TII.2019.2948406}
  {\path{doi:10.1109/TII.2019.2948406}}.

\bibitem{XHu2019}
X.~Hu, K.-K. Wong, K.~Yang, Z.~Zheng, {UAV}-assisted relaying and edge
  computing: Scheduling and trajectory optimization, IEEE Transactions on
  Wireless Communications 18~(10) (2019) 4738--4752.
\newblock \href {https://doi.org/10.1109/TWC.2019.2928539}
  {\path{doi:10.1109/TWC.2019.2928539}}.

\bibitem{LZhang2020}
L.~Zhang, N.~Ansari, Latency-aware {I}o{T} service provisioning in {UAV}-aided
  mobile-edge computing networks, IEEE Internet of Things Journal 7~(10) (2020)
  10573--10580.
\newblock \href {https://doi.org/10.1109/JIOT.2020.3005117}
  {\path{doi:10.1109/JIOT.2020.3005117}}.

\bibitem{ZYu2020}
Z.~Yu, Y.~Gong, S.~Gong, Y.~Guo, Joint task offloading and resource allocation
  in {UAV}-enabled mobile edge computing, IEEE Internet of Things Journal 7~(4)
  (2020) 3147--3159.
\newblock \href {https://doi.org/10.1109/JIOT.2020.2965898}
  {\path{doi:10.1109/JIOT.2020.2965898}}.

\bibitem{YZhu2020}
Y.~Zhu, S.~Wang, X.~Liu, H.~Tong, C.~Yin, Joint task and resource allocation in
  {SDN}-based {UAV}-assisted cellular networks, in: 2020 IEEE/CIC International
  Conference on Communications in China (ICCC), 2020, pp. 430--435.
\newblock \href {https://doi.org/10.1109/ICCC49849.2020.9238969}
  {\path{doi:10.1109/ICCC49849.2020.9238969}}.

\bibitem{LZhao2020}
L.~{Zhao}, K.~{Yang}, Z.~{Tan}, X.~{Li}, S.~{Sharma}, Z.~{Liu}, A novel cost
  optimization strategy for {SDN}-enabled {UAV}-assisted vehicular computation
  offloading, IEEE Transactions on Intelligent Transportation Systems 22~(6)
  (2021) 3664--3674.
\newblock \href {https://doi.org/10.1109/TITS.2020.3024186}
  {\path{doi:10.1109/TITS.2020.3024186}}.

\bibitem{XHu2020}
X.~Hu, K.-K. Wong, Y.~Zhang, Wireless-powered edge computing with cooperative
  {UAV}: Task, time scheduling and trajectory design, IEEE Transactions on
  Wireless Communications 19~(12) (2020) 8083--8098.
\newblock \href {https://doi.org/10.1109/TWC.2020.3019097}
  {\path{doi:10.1109/TWC.2020.3019097}}.

\bibitem{JZhang2020}
J.~{Zhang}, Y.~{Wu}, G.~{Min}, F.~{Hao}, L.~{Cui}, Balancing energy consumption
  and reputation gain of {UAV} scheduling in edge computing, IEEE Transactions
  on Cognitive Communications and Networking 6~(4) (2020) 1204--1217.
\newblock \href {https://doi.org/10.1109/TCCN.2020.3004592}
  {\path{doi:10.1109/TCCN.2020.3004592}}.

\bibitem{JrWang2020}
J.~Wang, K.~Liu, J.~Pan, Online {UAV}-mounted edge server dispatching for
  mobile-to-mobile edge computing, IEEE Internet of Things Journal 7~(2) (2020)
  1375--1386.
\newblock \href {https://doi.org/10.1109/JIOT.2019.2954798}
  {\path{doi:10.1109/JIOT.2019.2954798}}.

\bibitem{YWang2020}
Y.~Wang, Z.-Y. Ru, K.~Wang, P.-Q. Huang, Joint deployment and task scheduling
  optimization for large-scale mobile users in multi-{UAV}-enabled mobile edge
  computing, IEEE Transactions on Cybernetics 50~(9) (2020) 3984--3997.
\newblock \href {https://doi.org/10.1109/TCYB.2019.2935466}
  {\path{doi:10.1109/TCYB.2019.2935466}}.

\bibitem{NHMotlagh2016}
N.~Hossein~Motlagh, T.~Taleb, O.~Arouk, Low-altitude unmanned aerial
  vehicles-based internet of things services: Comprehensive survey and future
  perspectives, IEEE Internet of Things Journal 3~(6) (2016) 899--922.
\newblock \href {https://doi.org/10.1109/JIOT.2016.2612119}
  {\path{doi:10.1109/JIOT.2016.2612119}}.

\bibitem{LYang2020}
L.~Yang, H.~Yao, J.~Wang, C.~Jiang, A.~Benslimane, Y.~Liu, Multi-{UAV}-enabled
  load-balance mobile-edge computing for {I}o{T} networks, IEEE Internet of
  Things Journal 7~(8) (2020) 6898--6908.
\newblock \href {https://doi.org/10.1109/JIOT.2020.2971645}
  {\path{doi:10.1109/JIOT.2020.2971645}}.

\bibitem{CJiang2020}
C.~Jiang, Y.~Li, R.~Su, Z.~Xiao, F.~Yan, A load balancing based resource
  allocation algorithm in {UAV}-aided {MEC} systems, in: 2020 IEEE 6th
  International Conference on Computer and Communications (ICCC), 2020, pp.
  519--523.
\newblock \href {https://doi.org/10.1109/ICCC51575.2020.9345215}
  {\path{doi:10.1109/ICCC51575.2020.9345215}}.

\bibitem{HEl-Sayed2019}
H.~{El-Sayed}, M.~{Chaqfa}, S.~{Zeadally}, D.~{Puthal}, A traffic-aware
  approach for enabling unmanned aerial vehicles ({UAV}s) in smart city
  scenarios, IEEE Access 7 (2019) 86297--86305.
\newblock \href {https://doi.org/10.1109/ACCESS.2019.2922213}
  {\path{doi:10.1109/ACCESS.2019.2922213}}.

\bibitem{XWang2021}
X.~Wang, L.~Duan, Economic analysis of unmanned aerial vehicle ({UAV}) provided
  mobile services, IEEE Transactions on Mobile Computing 20~(5) (2021)
  1804--1816.
\newblock \href {https://doi.org/10.1109/TMC.2020.2973088}
  {\path{doi:10.1109/TMC.2020.2973088}}.

\bibitem{SSun2021}
S.~Sun, G.~Zhang, H.~Mei, K.~Wang, K.~Yang, Optimizing multi-{UAV} deployment
  in 3-{D} space to minimize task completion time in {UAV}-enabled mobile edge
  computing systems, IEEE Communications Letters 25~(2) (2021) 579--583.
\newblock \href {https://doi.org/10.1109/LCOMM.2020.3029144}
  {\path{doi:10.1109/LCOMM.2020.3029144}}.

\bibitem{ZLiao2021}
Z.~Liao, Y.~Ma, J.~Huang, J.~Wang, J.~Wang, {HOTSPOT}: A {UAV}-assisted dynamic
  mobility-aware offloading for mobile edge computing in 3{D} space, IEEE
  Internet of Things Journal 8~(13) (2021) 10940--10952.
\newblock \href {https://doi.org/10.1109/JIOT.2021.3051214}
  {\path{doi:10.1109/JIOT.2021.3051214}}.

\bibitem{RIslambouli2019}
R.~Islambouli, S.~Sharafeddine, Optimized 3{D} deployment of {UAV}-mounted
  cloudlets to support latency-sensitive services in {I}o{T} networks, IEEE
  Access 7 (2019) 172860--172870.
\newblock \href {https://doi.org/10.1109/ACCESS.2019.2956150}
  {\path{doi:10.1109/ACCESS.2019.2956150}}.

\bibitem{NAnsari202048}
N.~Ansari, L.~Zhang, Flexible backhaul-aware {DBS}-aided {H}et{N}et with {IBFD}
  communications, ICT Express 6~(1) (2020) 48--56.
\newblock \href {https://doi.org/10.1016/j.icte.2019.08.003}
  {\path{doi:10.1016/j.icte.2019.08.003}}.

\bibitem{LSun2021}
L.~Sun, L.~Wan, X.~Wang, Learning-based resource allocation strategy for
  industrial {I}o{T} in {UAV}-enabled {MEC} systems, IEEE Transactions on
  Industrial Informatics 17~(7) (2021) 5031--5040.
\newblock \href {https://doi.org/10.1109/TII.2020.3024170}
  {\path{doi:10.1109/TII.2020.3024170}}.

\bibitem{CTang2020}
C.~Tang, C.~Zhu, X.~Wei, J.~J. P.~C. Rodrigues, M.~Guizani, W.~Jia, {UAV}
  placement optimization for internet of medical things, in: 2020 International
  Wireless Communications and Mobile Computing (IWCMC), 2020, pp. 752--757.
\newblock \href {https://doi.org/10.1109/IWCMC48107.2020.9148581}
  {\path{doi:10.1109/IWCMC48107.2020.9148581}}.

\bibitem{YZheng2020}
Y.~Zheng, B.~Yang, C.~Chen, Joint optimization of the deployment and resource
  allocation of {UAV}s in vehicular edge computing and networks, in: 2020 IEEE
  92nd Vehicular Technology Conference (VTC2020-Fall), 2020, pp. 1--6.
\newblock \href {https://doi.org/10.1109/VTC2020-Fall49728.2020.9348819}
  {\path{doi:10.1109/VTC2020-Fall49728.2020.9348819}}.

\bibitem{XYu2020}
X.~Yu, X.~Dong, X.~Yang, C.~Chen, L.~Ruan, F.~Song, Y.~Gong, Air–ground
  integrated deployment for {UAV}-enabled mobile edge computing: A hierarchical
  game approach, IET Communications 14~(15) (2020) 2491--2499.
\newblock \href {https://doi.org/10.1049/iet-com.2019.1209}
  {\path{doi:10.1049/iet-com.2019.1209}}.

\bibitem{WYou2021}
W.~You, C.~Dong, X.~Cheng, X.~Zhu, Q.~Wu, G.~Chen, Joint optimization of area
  coverage and mobile-edge computing with clustering for {FANET}s, IEEE
  Internet of Things Journal 8~(2) (2021) 695--707.
\newblock \href {https://doi.org/10.1109/JIOT.2020.3006891}
  {\path{doi:10.1109/JIOT.2020.3006891}}.

\bibitem{JWang2020}
J.~Wang, C.~Jin, Q.~Tang, N.~Xiong, G.~Srivastava, Intelligent ubiquitous
  network accessibility for wireless-powered {MEC} in {UAV}-assisted {B5G},
  IEEE Transactions on Network Science and Engineering 8~(4) (2021) 2801--2813.
\newblock \href {https://doi.org/10.1109/TNSE.2020.3029048}
  {\path{doi:10.1109/TNSE.2020.3029048}}.

\bibitem{LXie2021}
L.~Xie, X.~Cao, J.~Xu, R.~Zhang, {UAV}-enabled wireless power transfer: A
  tutorial overview, IEEE Transactions on Green Communications and Networking
  5~(4) (2021) 2042--2064.
\newblock \href {https://doi.org/10.1109/TGCN.2021.3093718}
  {\path{doi:10.1109/TGCN.2021.3093718}}.

\bibitem{FZhou2018-C}
F.~Zhou, Y.~Wu, H.~Sun, Z.~Chu, {UAV}-enabled mobile edge computing: Offloading
  optimization and trajectory design, in: 2018 IEEE International Conference on
  Communications (ICC), 2018, pp. 1--6.
\newblock \href {https://doi.org/10.1109/ICC.2018.8422277}
  {\path{doi:10.1109/ICC.2018.8422277}}.

\bibitem{FZhou2018}
F.~{Zhou}, Y.~{Wu}, R.~Q. {Hu}, Y.~{Qian}, Computation rate maximization in
  {UAV}-enabled wireless-powered mobile-edge computing systems, IEEE Journal on
  Selected Areas in Communications 36~(9) (2018) 1927--1941.
\newblock \href {https://doi.org/10.1109/JSAC.2018.2864426}
  {\path{doi:10.1109/JSAC.2018.2864426}}.

\bibitem{YDu2019}
Y.~Du, K.~Yang, K.~Wang, G.~Zhang, Y.~Zhao, D.~Chen, Joint resources and
  workflow scheduling in {UAV}-enabled wirelessly-powered {MEC} for {I}o{T}
  systems, IEEE Transactions on Vehicular Technology 68~(10) (2019)
  10187--10200.
\newblock \href {https://doi.org/10.1109/TVT.2019.2935877}
  {\path{doi:10.1109/TVT.2019.2935877}}.

\bibitem{YLiu2020}
Y.~Liu, K.~Xiong, Q.~Ni, P.~Fan, K.~B. Letaief, {UAV}-assisted wireless powered
  cooperative mobile edge computing: Joint offloading, {CPU} control, and
  trajectory optimization, IEEE Internet of Things Journal 7~(4) (2020)
  2777--2790.
\newblock \href {https://doi.org/10.1109/JIOT.2019.2958975}
  {\path{doi:10.1109/JIOT.2019.2958975}}.

\bibitem{WLu2020}
W.~Lu, X.~Xu, F.~Lu, H.~Peng, B.~Li, G.~Liu, Y.~Gong, Resource optimization in
  anti-interference {UAV} powered cooperative mobile edge computing network,
  Physical Communication 42 (2020) 101128.
\newblock \href {https://doi.org/10.1016/j.phycom.2020.101128}
  {\path{doi:10.1016/j.phycom.2020.101128}}.

\bibitem{WLu2020IET}
W.~Lu, X.~Xu, Q.~Ye, B.~Li, H.~Peng, S.~Hu, Y.~Gong, Power optimization in
  {UAV}-assisted wireless powered cooperative mobile edge computing systems,
  IET Communications 14~(15) (2020) 2516--2523.
\newblock \href {https://doi.org/10.1049/iet-com.2019.1063}
  {\path{doi:10.1049/iet-com.2019.1063}}.

\bibitem{YLiu2020I}
Y.~Liu, M.~Qiu, J.~Hu, H.~Yu, Incentive {UAV}-enabled mobile edge computing
  based on microwave power transmission, IEEE Access 8 (2020) 28584--28593.
\newblock \href {https://doi.org/10.1109/ACCESS.2020.2971962}
  {\path{doi:10.1109/ACCESS.2020.2971962}}.

\bibitem{YLiu2021-1}
Y.~Liu, J.~Zhou, D.~Tian, Z.~Sheng, X.~Duan, G.~Qu, V.~C.~M. Leung, Joint
  communication and computation resource scheduling of a {UAV}-assisted mobile
  edge computing system for platooning vehicles, IEEE Transactions on
  Intelligent Transportation Systems (2021) 1--16\href
  {https://doi.org/10.1109/TITS.2021.3082539}
  {\path{doi:10.1109/TITS.2021.3082539}}.

\bibitem{WLu2021}
W.~Lu, Y.~Ding, Y.~Gao, S.~Hu, Y.~Wu, N.~Zhao, Y.~Gong, Resource and trajectory
  optimization for secure communications in dual-{UAV}-{MEC} systems, IEEE
  Transactions on Industrial Informatics 18~(4) (2022) 2704--2713.
\newblock \href {https://doi.org/10.1109/TII.2021.3087726}
  {\path{doi:10.1109/TII.2021.3087726}}.

\bibitem{TBai2019}
T.~Bai, J.~Wang, Y.~Ren, L.~Hanzo, Energy-efficient computation offloading for
  secure {UAV}-edge-computing systems, IEEE Transactions on Vehicular
  Technology 68~(6) (2019) 6074--6087.
\newblock \href {https://doi.org/10.1109/TVT.2019.2912227}
  {\path{doi:10.1109/TVT.2019.2912227}}.

\bibitem{DHan2020}
D.~Han, T.~Shi, Secrecy capacity maximization for a {UAV}-assisted {MEC}
  system, China Communications 17~(10) (2020) 64--81.
\newblock \href {https://doi.org/10.23919/JCC.2020.10.005}
  {\path{doi:10.23919/JCC.2020.10.005}}.

\bibitem{YZhou2020}
Y.~Zhou, C.~Pan, P.~L. Yeoh, K.~Wang, M.~Elkashlan, B.~Vucetic, Y.~Li, Secure
  communications for {UAV}-enabled mobile edge computing systems, IEEE
  Transactions on Communications 68~(1) (2020) 376--388.
\newblock \href {https://doi.org/10.1109/TCOMM.2019.2947921}
  {\path{doi:10.1109/TCOMM.2019.2947921}}.

\bibitem{PAmos2021}
P.~Amos, P.~Li, W.~Wu, B.~Wang, Computation efficiency maximization for secure
  {UAV}-enabled mobile edge computing networks, Physical Communication 46
  (2021) 101284.
\newblock \href {https://doi.org/https://doi.org/10.1016/j.phycom.2021.101284}
  {\path{doi:https://doi.org/10.1016/j.phycom.2021.101284}}.

\bibitem{YXu2021}
Y.~Xu, T.~Zhang, D.~Yang, Y.~Liu, M.~Tao, Joint resource and trajectory
  optimization for security in {UAV}-assisted {MEC} systems, IEEE Transactions
  on Communications 69~(1) (2021) 573--588.
\newblock \href {https://doi.org/10.1109/TCOMM.2020.3025910}
  {\path{doi:10.1109/TCOMM.2020.3025910}}.

\bibitem{SLai2021}
S.~Lai, R.~Zhao, S.~Tang, J.~Xia, F.~Zhou, L.~Fan, Intelligent secure mobile
  edge computing for beyond 5{G} wireless networks, Physical Communication 45
  (2021) 101283.
\newblock \href {https://doi.org/https://doi.org/10.1016/j.phycom.2021.101283}
  {\path{doi:https://doi.org/10.1016/j.phycom.2021.101283}}.

\bibitem{SLi2021}
S.~Li, B.~Duo, M.~D. Renzo, M.~Tao, X.~Yuan, Robust secure {UAV} communications
  with the aid of reconfigurable intelligent surfaces, IEEE Transactions on
  Wireless Communications 20~(10) (2021) 6402--6417.
\newblock \href {https://doi.org/10.1109/TWC.2021.3073746}
  {\path{doi:10.1109/TWC.2021.3073746}}.

\bibitem{SLi2020}
S.~Li, B.~Duo, X.~Yuan, Y.-C. Liang, M.~Di~Renzo, Reconfigurable intelligent
  surface assisted {UAV} communication: Joint trajectory design and passive
  beamforming, IEEE Wireless Communications Letters 9~(5) (2020) 716--720.
\newblock \href {https://doi.org/10.1109/LWC.2020.2966705}
  {\path{doi:10.1109/LWC.2020.2966705}}.

\bibitem{HMei2021}
H.~Mei, K.~Yang, J.~Shen, Q.~Liu, Joint trajectory-task-cache optimization with
  phase-shift design of {RIS}-assisted {UAV} for {MEC}, IEEE Wireless
  Communications Letters 10~(7) (2021) 1586--1590.
\newblock \href {https://doi.org/10.1109/LWC.2021.3074990}
  {\path{doi:10.1109/LWC.2021.3074990}}.

\bibitem{XWang2020}
X.~Wang, Y.~Zhang, R.~Shen, Y.~Xu, F.-C. Zheng, {DRL}-based energy-efficient
  resource allocation frameworks for uplink {NOMA} systems, IEEE Internet of
  Things Journal 7~(8) (2020) 7279--7294.
\newblock \href {https://doi.org/10.1109/JIOT.2020.2982699}
  {\path{doi:10.1109/JIOT.2020.2982699}}.

\bibitem{TRodrigues2020}
T.~K. Rodrigues, K.~Suto, H.~Nishiyama, J.~Liu, N.~Kato, Machine learning meets
  computation and communication control in evolving edge and cloud: Challenges
  and future perspective, IEEE Communications Surveys \& Tutorials 22~(1)
  (2020) 38--67.
\newblock \href {https://doi.org/10.1109/COMST.2019.2943405}
  {\path{doi:10.1109/COMST.2019.2943405}}.

\bibitem{NCLuong2019}
N.~C. Luong, D.~T. Hoang, S.~Gong, D.~Niyato, P.~Wang, Y.-C. Liang, D.~I. Kim,
  Applications of deep reinforcement learning in communications and networking:
  A survey, IEEE Communications Surveys \& Tutorials 21~(4) (2019) 3133--3174.
\newblock \href {https://doi.org/10.1109/COMST.2019.2916583}
  {\path{doi:10.1109/COMST.2019.2916583}}.

\bibitem{XDiao2018}
X.~Diao, J.~Zheng, Y.~Cai, X.~Dong, X.~Zhang, Joint user clustering, resource
  allocation and power control for {NOMA}-based mobile edge computing, in: 2018
  10th International Conference on Wireless Communications and Signal
  Processing (WCSP), 2018, pp. 1--6.
\newblock \href {https://doi.org/10.1109/WCSP.2018.8555861}
  {\path{doi:10.1109/WCSP.2018.8555861}}.

\bibitem{AKSangaiah2019}
A.~K. Sangaiah, D.~V. Medhane, T.~Han, M.~S. Hossain, G.~Muhammad, Enforcing
  position-based confidentiality with machine learning paradigm through mobile
  edge computing in real-time industrial informatics, IEEE Transactions on
  Industrial Informatics 15~(7) (2019) 4189--4196.
\newblock \href {https://doi.org/10.1109/TII.2019.2898174}
  {\path{doi:10.1109/TII.2019.2898174}}.

\bibitem{AShakarami2020}
A.~Shakarami, M.~Ghobaei-Arani, A.~Shahidinejad, A survey on the computation
  offloading approaches in mobile edge computing: A machine learning-based
  perspective, Computer Networks 182 (2020) 107496.
\newblock \href {https://doi.org/https://doi.org/10.1016/j.comnet.2020.107496}
  {\path{doi:https://doi.org/10.1016/j.comnet.2020.107496}}.

\bibitem{MWang2020}
M.~Wang, S.~Shi, S.~Gu, X.~Gu, X.~Qin, {Q}-learning based computation
  offloading for multi-{UAV}-enabled cloud-edge computing networks, IET
  Communications 14~(15) (2020) 2481--2490.
\newblock \href {https://doi.org/https://doi.org/10.1049/iet-com.2019.1184}
  {\path{doi:https://doi.org/10.1049/iet-com.2019.1184}}.

\bibitem{SZhu2020}
S.~Zhu, L.~Gui, N.~Cheng, Q.~Zhang, F.~Sun, X.~Lang, {UAV}-enabled computation
  migration for complex missions: A reinforcement learning approach, IET
  Communications 14~(15) (2020) 2472--2480.
\newblock \href {https://doi.org/10.1049/iet-com.2019.1188}
  {\path{doi:10.1049/iet-com.2019.1188}}.

\bibitem{AAsheralieva2019}
A.~Asheralieva, D.~Niyato, Hierarchical game-theoretic and reinforcement
  learning framework for computational offloading in {UAV}-enabled mobile edge
  computing networks with multiple service providers, IEEE Internet of Things
  Journal 6~(5) (2019) 8753--8769.
\newblock \href {https://doi.org/10.1109/JIOT.2019.2923702}
  {\path{doi:10.1109/JIOT.2019.2923702}}.

\bibitem{GFaraci2020}
G.~Faraci, C.~Grasso, G.~Schembra, Design of a 5{G} network slice extension
  with {MEC} {UAV}s managed with reinforcement learning, IEEE Journal on
  Selected Areas in Communications 38~(10) (2020) 2356--2371.
\newblock \href {https://doi.org/10.1109/JSAC.2020.3000416}
  {\path{doi:10.1109/JSAC.2020.3000416}}.

\bibitem{GFragkos2020}
G.~Fragkos, N.~Kemp, E.~E. Tsiropoulou, S.~Papavassiliou, Artificial
  intelligence empowered {UAV}s data offloading in mobile edge computing, in:
  ICC 2020 - 2020 IEEE International Conference on Communications (ICC), 2020,
  pp. 1--7.
\newblock \href {https://doi.org/10.1109/ICC40277.2020.9149115}
  {\path{doi:10.1109/ICC40277.2020.9149115}}.

\bibitem{YLiao2020}
Y.~{Liao}, L.~{Shou}, Q.~{Yu}, Q.~{Ai}, Q.~{Liu}, An intelligent computation
  demand response framework for {II}o{T}-{MEC} interactive networks, IEEE
  Networking Letters 2~(3) (2020) 154--158.
\newblock \href {https://doi.org/10.1109/LNET.2020.3001178}
  {\path{doi:10.1109/LNET.2020.3001178}}.

\bibitem{AFeriani2021}
A.~Feriani, E.~Hossain, Single and multi-agent deep reinforcement learning for
  {AI}-enabled wireless networks: A tutorial, IEEE Communications Surveys \&
  Tutorials 23~(2) (2021) 1226--1252.
\newblock \href {https://doi.org/10.1109/COMST.2021.3063822}
  {\path{doi:10.1109/COMST.2021.3063822}}.

\bibitem{XMa2020}
X.~Ma, C.~Yin, X.~Liu, Machine learning based joint offloading and trajectory
  design in {UAV} based {MEC} system for {I}o{T} devices, in: 2020 IEEE 6th
  International Conference on Computer and Communications (ICCC), 2020, pp.
  902--909.
\newblock \href {https://doi.org/10.1109/ICCC51575.2020.9345069}
  {\path{doi:10.1109/ICCC51575.2020.9345069}}.

\bibitem{MWang2020-C}
M.~Wang, S.~Shi, S.~Gu, N.~Zhang, X.~Gu, Intelligent resource allocation in
  {UAV}-enabled mobile edge computing networks, in: 2020 IEEE 92nd Vehicular
  Technology Conference (VTC2020-Fall), 2020, pp. 1--5.
\newblock \href {https://doi.org/10.1109/VTC2020-Fall49728.2020.9348573}
  {\path{doi:10.1109/VTC2020-Fall49728.2020.9348573}}.

\bibitem{FJiang2020}
F.~{Jiang}, K.~{Wang}, L.~{Dong}, C.~{Pan}, W.~{Xu}, K.~{Yang},
  Deep-learning-based joint resource scheduling algorithms for hybrid {MEC}
  networks, IEEE Internet of Things Journal 7~(7) (2020) 6252--6265.
\newblock \href {https://doi.org/10.1109/JIOT.2019.2954503}
  {\path{doi:10.1109/JIOT.2019.2954503}}.

\bibitem{MMukherjee2020}
M.~Mukherjee, V.~Kumar, A.~Lat, M.~Guo, R.~Matam, Y.~Lv, Distributed deep
  learning-based task offloading for {UAV}-enabled mobile edge computing, in:
  IEEE INFOCOM 2020 - IEEE Conference on Computer Communications Workshops
  (INFOCOM WKSHPS), 2020, pp. 1208--1212.
\newblock \href {https://doi.org/10.1109/INFOCOMWKSHPS50562.2020.9162899}
  {\path{doi:10.1109/INFOCOMWKSHPS50562.2020.9162899}}.

\bibitem{LWang20211}
L.~Wang, K.~Wang, C.~Pan, W.~Xu, N.~Aslam, L.~Hanzo, Multi-agent deep
  reinforcement learning-based trajectory planning for multi-{UAV} assisted
  mobile edge computing, IEEE Transactions on Cognitive Communications and
  Networking 7~(1) (2021) 73--84.
\newblock \href {https://doi.org/10.1109/TCCN.2020.3027695}
  {\path{doi:10.1109/TCCN.2020.3027695}}.

\bibitem{FXu2020}
F.~{Xu}, F.~{Yang}, C.~{Zhao}, S.~{Wu}, Deep reinforcement learning based joint
  edge resource management in maritime network, China Communications 17~(5)
  (2020) 211--222.
\newblock \href {https://doi.org/10.23919/JCC.2020.05.016}
  {\path{doi:10.23919/JCC.2020.05.016}}.

\bibitem{RKrisna2020}
R.~K.~A. Sakir, M.~R. Ramli, J.-M. Lee, D.-S. Kim, {UAV}-assisted real-time
  data processing using deep {Q}-network for industrial internet of things, in:
  2020 International Conference on Artificial Intelligence in Information and
  Communication (ICAIIC), 2020, pp. 208--211.
\newblock \href {https://doi.org/10.1109/ICAIIC48513.2020.9065203}
  {\path{doi:10.1109/ICAIIC48513.2020.9065203}}.

\bibitem{HWang2020}
H.~Wang, H.~Ke, W.~Sun, Unmanned-aerial-vehicle-assisted computation offloading
  for mobile edge computing based on deep reinforcement learning, IEEE Access 8
  (2020) 180784--180798.
\newblock \href {https://doi.org/10.1109/ACCESS.2020.3028553}
  {\path{doi:10.1109/ACCESS.2020.3028553}}.

\bibitem{QLiu2021}
Q.~Liu, L.~Shi, L.~Sun, J.~Li, M.~Ding, F.~Shu, Path planning for {UAV}-mounted
  mobile edge computing with deep reinforcement learning, IEEE Transactions on
  Vehicular Technology 69~(5) (2020) 5723--5728.
\newblock \href {https://doi.org/10.1109/TVT.2020.2982508}
  {\path{doi:10.1109/TVT.2020.2982508}}.

\bibitem{LWang2021}
L.~{Wang}, K.~{Wang}, C.~{Pan}, W.~{Xu}, N.~{Aslam}, A.~{Nallanathan}, Deep
  reinforcement learning based dynamic trajectory control for {UAV}-assisted
  mobile edge computing, IEEE Transactions on Mobile Computing (2021)
  1--15\href {https://doi.org/10.1109/TMC.2021.3059691}
  {\path{doi:10.1109/TMC.2021.3059691}}.

\bibitem{HPeng2021}
H.~Peng, X.~Shen, Multi-agent reinforcement learning based resource management
  in {MEC}- and {UAV}-assisted vehicular networks, IEEE Journal on Selected
  Areas in Communications 39~(1) (2021) 131--141.
\newblock \href {https://doi.org/10.1109/JSAC.2020.3036962}
  {\path{doi:10.1109/JSAC.2020.3036962}}.

\bibitem{AMSeid2021}
A.~M. {Seid}, G.~O. {Boateng}, S.~{Anokye}, T.~{Kwantwi}, G.~{Sun}, G.~{Liu},
  Collaborative computation offloading and resource allocation in multi-{UAV}
  assisted {I}o{T} networks: A deep reinforcement learning approach, IEEE
  Internet of Things Journal 8~(15) (2021) 12203--12218.
\newblock \href {https://doi.org/10.1109/JIOT.2021.3063188}
  {\path{doi:10.1109/JIOT.2021.3063188}}.

\bibitem{YNie2021}
Y.~Nie, J.~Zhao, F.~Gao, F.~R. Yu, Semi-distributed resource management in
  {UAV}-aided {MEC} systems: A multi-agent federated reinforcement learning
  approach, IEEE Transactions on Vehicular Technology 70~(12) (2021)
  13162--13173.
\newblock \href {https://doi.org/10.1109/TVT.2021.3118446}
  {\path{doi:10.1109/TVT.2021.3118446}}.

\bibitem{DRahbari2021}
D.~Rahbari, M.~M. Alam, Y.~L. Moullec, M.~Jenihhin, Fast and fair computation
  offloading management in a swarm of drones using a rating-based federated
  learning approach, IEEE Access 9 (2021) 113832--113849.
\newblock \href {https://doi.org/10.1109/ACCESS.2021.3104117}
  {\path{doi:10.1109/ACCESS.2021.3104117}}.

\bibitem{TRen2021}
T.~Ren, J.~Niu, B.~Dai, X.~Liu, Z.~Hu, M.~Xu, M.~Guizani, Enabling efficient
  scheduling in large-scale {UAV}-assisted mobile edge computing via
  hierarchical reinforcement learning, IEEE Internet of Things Journal (2021)
  1--15\href {https://doi.org/10.1109/JIOT.2021.3071531}
  {\path{doi:10.1109/JIOT.2021.3071531}}.

\bibitem{VMnih2015}
V.~Mnih, K.~Kavukcuoglu, D.~Silver, A.~A. Rusu, J.~Veness, M.~G. Bellemare,
  A.~Graves, M.~Riedmiller, A.~K. Fidjeland, G.~Ostrovski, et~al., Human-level
  control through deep reinforcement learning, Nature 518~(7540) (2015) 529.
\newblock \href {https://doi.org/10.1038/nature14236}
  {\path{doi:10.1038/nature14236}}.

\bibitem{TPLillicrap2015}
T.~P. Lillicrap, J.~J. Hunt, A.~Pritzel, N.~Heess, T.~Erez, Y.~Tassa,
  D.~Silver, D.~Wierstra, Continuous control with deep reinforcement learning,
  arXiv preprint arXiv:1509.02971 (2015).
\newblock \href {https://doi.org/10.48550/arXiv.1509.02971}
  {\path{doi:10.48550/arXiv.1509.02971}}.

\bibitem{VRKonda2000}
V.~R. Konda, J.~N. Tsitsiklis, Actor-critic algorithms, in: Advances in neural
  information processing systems, 2000, pp. 1008--1014.

\bibitem{MChen2021}
M.~Chen, Z.~Yang, W.~Saad, C.~Yin, H.~V. Poor, S.~Cui, A joint learning and
  communications framework for federated learning over wireless networks, IEEE
  Transactions on Wireless Communications 20~(1) (2021) 269--283.
\newblock \href {https://doi.org/10.1109/TWC.2020.3024629}
  {\path{doi:10.1109/TWC.2020.3024629}}.

\bibitem{LZang2022}
L.~Zang, X.~Zhang, B.~Guo, Federated deep reinforcement learning for online
  task offloading and resource allocation in {WPC-MEC} networks, IEEE Access 10
  (2022) 9856--9867.
\newblock \href {https://doi.org/10.1109/ACCESS.2022.3144415}
  {\path{doi:10.1109/ACCESS.2022.3144415}}.

\bibitem{ZYang2021}
Z.~Yang, M.~Chen, X.~Liu, Y.~Liu, Y.~Chen, S.~Cui, H.~V. Poor, {AI}-driven
  {UAV-NOMA-MEC} in next generation wireless networks, IEEE Wireless
  Communications 28~(5) (2021) 66--73.
\newblock \href {https://doi.org/10.1109/MWC.121.2100058}
  {\path{doi:10.1109/MWC.121.2100058}}.

\bibitem{QVPham2022}
Q.-V. Pham, M.~Le, T.~Huynh-The, Z.~Han, W.-J. Hwang, Energy-efficient
  federated learning over {UAV}-enabled wireless powered communications, IEEE
  Transactions on Vehicular Technology (2022) 1--1\href
  {https://doi.org/10.1109/TVT.2022.3150004}
  {\path{doi:10.1109/TVT.2022.3150004}}.

\bibitem{SLi2020-C}
S.~Li, B.~Liu, H.~Chen, Z.~Huang, A domain adaptation method for object
  detection in {UAV} based on semi-supervised learning, in: 2020 17th
  International Computer Conference on Wavelet Active Media Technology and
  Information Processing (ICCWAMTIP), 2020, pp. 138--141.
\newblock \href {https://doi.org/10.1109/ICCWAMTIP51612.2020.9317405}
  {\path{doi:10.1109/ICCWAMTIP51612.2020.9317405}}.

\bibitem{DPan2020}
D.~Pan, L.~Nie, W.~Kang, Z.~Song, {UAV} anomaly detection using active learning
  and improved {S3VM} model, in: 2020 International Conference on Sensing,
  Measurement \& Data Analytics in the era of Artificial Intelligence (ICSMD),
  2020, pp. 253--258.
\newblock \href {https://doi.org/10.1109/ICSMD50554.2020.9261709}
  {\path{doi:10.1109/ICSMD50554.2020.9261709}}.

\bibitem{BKellenberger2019}
B.~Kellenberger, D.~Marcos, S.~Lobry, D.~Tuia, Half a percent of labels is
  enough: Efficient animal detection in {UAV} imagery using deep {CNN}s and
  active learning, IEEE Transactions on Geoscience and Remote Sensing 57~(12)
  (2019) 9524--9533.
\newblock \href {https://doi.org/10.1109/TGRS.2019.2927393}
  {\path{doi:10.1109/TGRS.2019.2927393}}.

\bibitem{XWang2022}
X.~Wang, Z.~Ning, L.~Guo, S.~Guo, X.~Gao, G.~Wang, Online learning for
  distributed computation offloading in wireless powered mobile edge computing
  networks, IEEE Transactions on Parallel and Distributed Systems 33~(8) (2022)
  1841--1855.
\newblock \href {https://doi.org/10.1109/TPDS.2021.3129618}
  {\path{doi:10.1109/TPDS.2021.3129618}}.

\bibitem{ZSun2020}
Z.~Sun, M.~R. Nakhai, An online learning algorithm for distributed task
  offloading in multi-access edge computing, IEEE Transactions on Signal
  Processing 68 (2020) 3090--3102.
\newblock \href {https://doi.org/10.1109/TSP.2020.2991383}
  {\path{doi:10.1109/TSP.2020.2991383}}.

\bibitem{DRen2022}
D.~Ren, X.~Gui, K.~Zhang, Adaptive request scheduling and service caching for
  {MEC}-assisted {I}o{T} networks: An online learning approach, IEEE Internet
  of Things Journal (2022) 1--1\href
  {https://doi.org/10.1109/JIOT.2022.3157677}
  {\path{doi:10.1109/JIOT.2022.3157677}}.

\bibitem{QCui2019}
Q.~Cui, Z.~Gong, W.~Ni, Y.~Hou, X.~Chen, X.~Tao, P.~Zhang, Stochastic online
  learning for mobile edge computing: Learning from changes, IEEE
  Communications Magazine 57~(3) (2019) 63--69.
\newblock \href {https://doi.org/10.1109/MCOM.2019.1800644}
  {\path{doi:10.1109/MCOM.2019.1800644}}.

\bibitem{NCheng2019}
N.~Cheng, F.~Lyu, W.~Quan, C.~Zhou, H.~He, W.~Shi, X.~Shen,
  Space/aerial-assisted computing offloading for {I}o{T} applications: A
  learning-based approach, IEEE Journal on Selected Areas in Communications
  37~(5) (2019) 1117--1129.
\newblock \href {https://doi.org/10.1109/JSAC.2019.2906789}
  {\path{doi:10.1109/JSAC.2019.2906789}}.

\bibitem{BMao2021}
B.~Mao, F.~Tang, Y.~Kawamoto, N.~Kato, Optimizing computation offloading in
  satellite-{UAV}-served 6{G} {I}o{T}: A deep learning approach, IEEE Network
  35~(4) (2021) 102--108.
\newblock \href {https://doi.org/10.1109/MNET.011.2100097}
  {\path{doi:10.1109/MNET.011.2100097}}.

\bibitem{ZZhang2019}
Z.~Zhang, W.~Zhang, F.-H. Tseng, Satellite mobile edge computing: Improving
  {Q}o{S} of high-speed satellite-terrestrial networks using edge computing
  techniques, IEEE Netw. 33~(1) (2019) 70--76.
\newblock \href {https://doi.org/10.1109/MNET.2018.1800172}
  {\path{doi:10.1109/MNET.2018.1800172}}.

\bibitem{PSemasinghe2017}
P.~Semasinghe, S.~Maghsudi, E.~Hossain, Game theoretic mechanisms for resource
  management in massive wireless {I}o{T} systems, IEEE Communications Magazine
  55~(2) (2017) 121--127.
\newblock \href {https://doi.org/10.1109/MCOM.2017.1600568CM}
  {\path{doi:10.1109/MCOM.2017.1600568CM}}.

\bibitem{ASharma2020}
A.~Sharma, P.~Vanjani, N.~Paliwal, C.~M. Basnayaka, D.~N.~K. Jayakody, H.-C.
  Wang, P.~Muthuchidambaranathan, Communication and networking technologies for
  {UAV}s: A survey, Journal of Network and Computer Applications 168 (2020)
  102739.
\newblock \href {https://doi.org/https://doi.org/10.1016/j.jnca.2020.102739}
  {\path{doi:https://doi.org/10.1016/j.jnca.2020.102739}}.

\bibitem{HSedjelmaci2019}
H.~Sedjelmaci, A.~Boudguiga, I.~B. Jemaa, S.~M. Senouci, An efficient cyber
  defense framework for {UAV-E}dge computing network, Ad Hoc Networks 94 (2019)
  101970.
\newblock \href {https://doi.org/https://doi.org/10.1016/j.adhoc.2019.101970}
  {\path{doi:https://doi.org/10.1016/j.adhoc.2019.101970}}.

\bibitem{YSu2021}
Y.~{Su}, A trust based scheme to protect 5{G} {UAV} communication networks,
  IEEE Open Journal of the Computer Society (2021) 1--1\href
  {https://doi.org/10.1109/OJCS.2021.3058001}
  {\path{doi:10.1109/OJCS.2021.3058001}}.

\bibitem{PGope2020}
P.~{Gope}, B.~{Sikdar}, An efficient privacy-preserving authenticated key
  agreement scheme for edge-assisted internet of drones, IEEE Transactions on
  Vehicular Technology 69~(11) (2020) 13621--13630.
\newblock \href {https://doi.org/10.1109/TVT.2020.3018778}
  {\path{doi:10.1109/TVT.2020.3018778}}.

\bibitem{MYahuza2021}
M.~{Yahuza}, M.~Y.~I. {Idris}, A.~W.~A. {Wahab}, T.~{Nandy}, I.~B. {Ahmedy},
  R.~{Ramli}, An edge assisted secure lightweight authentication technique for
  safe communication on the internet of drones network, IEEE Access 9 (2021)
  31420--31440.
\newblock \href {https://doi.org/10.1109/ACCESS.2021.3060420}
  {\path{doi:10.1109/ACCESS.2021.3060420}}.

\bibitem{RGupta2021}
R.~Gupta, A.~Nair, S.~Tanwar, N.~Kumar, Blockchain-assisted secure {UAV}
  communication in 6{G} environment: Architecture, opportunities, and
  challenges, IET Communications 15 (2021) 1352--1367.
\newblock \href {https://doi.org/https://doi.org/10.1049/cmu2.12113}
  {\path{doi:https://doi.org/10.1049/cmu2.12113}}.

\bibitem{ZGuan2020}
Z.~Guan, H.~Lyu, D.~Li, Y.~Hei, T.~Wang, Blockchain: a distributed solution to
  {UAV}-enabled mobile edge computing, IET Communications 14~(15) (2020)
  2420--2426.
\newblock \href {https://doi.org/10.1049/iet-com.2019.1131}
  {\path{doi:10.1049/iet-com.2019.1131}}.

\bibitem{DHeeger2020}
D.~Heeger, M.~Garigan, E.~E. Tsiropoulou, J.~Plusquellic, Secure energy
  constrained {L}o{R}a mesh network, in: Proceedings of 19th International
  Conference on Ad-Hoc Networks and Wireless, ADHOC-NOW 2020, Bari, Italy,
  2020, pp. 228--240.
\newblock \href {https://doi.org/10.1007/978-3-030-61746-2_17}
  {\path{doi:10.1007/978-3-030-61746-2_17}}.

\bibitem{AIHentati2020}
A.~I. Hentati, L.~C. Fourati, Comprehensive survey of {UAV}s communication
  networks, Computer Standards \& Interfaces 72 (2020) 103451.
\newblock \href {https://doi.org/10.1016/j.csi.2020.103451}
  {\path{doi:10.1016/j.csi.2020.103451}}.

\bibitem{PKChittoor2021}
P.~K. Chittoor, B.~Chokkalingam, L.~Mihet-Popa, A review on {UAV} wireless
  charging: Fundamentals, applications, charging techniques and standards, IEEE
  Access 9 (2021) 69235--69266.
\newblock \href {https://doi.org/10.1109/ACCESS.2021.3077041}
  {\path{doi:10.1109/ACCESS.2021.3077041}}.

\bibitem{WLiu2021}
W.~Liu, L.~Zhang, N.~Ansari, Laser charging enabled {DBS} placement for
  downlink communications, IEEE Transactions on Network Science and Engineering
  8~(4) (2021) 3009--3018.
\newblock \href {https://doi.org/10.1109/TNSE.2021.3118328}
  {\path{doi:10.1109/TNSE.2021.3118328}}.

\bibitem{WLiu2022}
W.~Liu, S.~Zhang, N.~Ansari, Joint laser charging and {DBS} placement for
  drone-assisted edge computing, IEEE Transactions on Vehicular Technology
  71~(1) (2022) 780--789.
\newblock \href {https://doi.org/10.1109/TVT.2021.3126710}
  {\path{doi:10.1109/TVT.2021.3126710}}.

\bibitem{NAnsari2019}
N.~Ansari, Q.~Fan, X.~Sun, L.~Zhang, Soarnet, IEEE Wireless Communications
  26~(6) (2019) 37--43.
\newblock \href {https://doi.org/10.1109/MWC.001.1900126}
  {\path{doi:10.1109/MWC.001.1900126}}.

\bibitem{JOuyang2018}
J.~Ouyang, Y.~Che, J.~Xu, K.~Wu, Throughput maximization for laser-powered
  {UAV} wireless communication systems, in: 2018 IEEE International Conference
  on Communications Workshops (ICC Workshops), 2018, pp. 1--6.
\newblock \href {https://doi.org/10.1109/ICCW.2018.8403572}
  {\path{doi:10.1109/ICCW.2018.8403572}}.

\bibitem{BGalkin2019}
B.~Galkin, J.~Kibilda, L.~A. DaSilva, {UAV}s as mobile infrastructure:
  Addressing battery lifetime, IEEE Communications Magazine 57~(6) (2019)
  132--137.
\newblock \href {https://doi.org/10.1109/MCOM.2019.1800545}
  {\path{doi:10.1109/MCOM.2019.1800545}}.

\bibitem{ABera2020}
A.~Bera, S.~Misra, C.~Chatterjee, {Q}o{E} analysis in cache-enabled multi-{UAV}
  networks, IEEE Transactions on Vehicular Technology 69~(6) (2020) 6680--6687.
\newblock \href {https://doi.org/10.1109/TVT.2020.2985933}
  {\path{doi:10.1109/TVT.2020.2985933}}.

\bibitem{MChen2017}
M.~Chen, M.~Mozaffari, W.~Saad, C.~Yin, M.~Debbah, C.~S. Hong, Caching in the
  sky: Proactive deployment of cache-enabled unmanned aerial vehicles for
  optimized quality-of-experience, IEEE Journal on Selected Areas in
  Communications 35~(5) (2017) 1046--1061.
\newblock \href {https://doi.org/10.1109/JSAC.2017.2680898}
  {\path{doi:10.1109/JSAC.2017.2680898}}.

\bibitem{JJi2020}
J.~Ji, K.~Zhu, D.~Niyato, R.~Wang, Joint cache placement, flight trajectory,
  and transmission power optimization for multi-{UAV} assisted wireless
  networks, IEEE Transactions on Wireless Communications 19~(8) (2020)
  5389--5403.
\newblock \href {https://doi.org/10.1109/TWC.2020.2992926}
  {\path{doi:10.1109/TWC.2020.2992926}}.

\bibitem{YLiu2021}
Y.~Liu, J.~Nie, X.~Li, S.~H. Ahmed, W.~Y.~B. Lim, C.~Miao, Federated learning
  in the sky: Aerial-ground air quality sensing framework with {UAV} swarms,
  IEEE Internet of Things Journal 8~(12) (2021) 9827--9837.
\newblock \href {https://doi.org/10.1109/JIOT.2020.3021006}
  {\path{doi:10.1109/JIOT.2020.3021006}}.

\bibitem{CDong2021}
C.~Dong, Y.~Shen, Y.~Qu, K.~Wang, J.~Zheng, Q.~Wu, F.~Wu, {UAV}s as an
  intelligent service: Boosting edge intelligence for air-ground integrated
  networks, IEEE Network 35~(4) (2021) 167--175.
\newblock \href {https://doi.org/10.1109/MNET.011.2000651}
  {\path{doi:10.1109/MNET.011.2000651}}.

\bibitem{ZLi2021}
Z.~Li, M.~Chen, Z.~Yang, J.~Zhao, Y.~Wang, J.~Shi, C.~Huang, Energy efficient
  reconfigurable intelligent surface enabled mobile edge computing networks
  with {NOMA}, IEEE Transactions on Cognitive Communications and Networking
  7~(2) (2021) 427--440.
\newblock \href {https://doi.org/10.1109/TCCN.2021.3068750}
  {\path{doi:10.1109/TCCN.2021.3068750}}.

\end{thebibliography}

\end{sloppypar}
\end{document}